\documentclass[fleqn,usenatbib]{mnras}

\usepackage{newtxtext,newtxmath}

\usepackage[T1]{fontenc}
\usepackage{longtable}
\usepackage{enumitem}
\DeclareRobustCommand{\VAN}[3]{#2}
\let\VANthebibliography\thebibliography
\def\thebibliography{\DeclareRobustCommand{\VAN}[3]{##3}\VANthebibliography}

\usepackage{graphicx}	
\usepackage{amsmath}	
\usepackage{xcolor}
\usepackage{subfig}

\newcommand{\kepler}{\emph{Kepler}~}

\newcommand{\feh}{\ensuremath{[\textup{Fe}/\textup{H}]}~}

\newcommand{\teff}{\ensuremath{T_{\textup{eff}}}~}
\newcommand{\logg}{\ensuremath{\log g}~}

\newcommand{\numax}{\ensuremath{\nu_\textup{max}}~}

\newcommand{\rl}{$R_{g}$~}
\newcommand{\rb}{$R_{b}$~}

\newcommand{\alfa}{$\alpha$}

\title[Chemical clocks]{Tracing the Milky Way: Calibrating chemical ages with high-precision \kepler data}

\author[G. Casali et al.]{
G. Casali,$^{1,2,3,4}$
\thanks{E-mail: giada.casali@anu.edu.au}
J. Montalb\'an,$^{3,5}$
A. Miglio,$^{3}$
L. Casagrande,$^{1,2}$
L. Magrini,$^{6}$
C. Chiappini,$^{7}$
A. Bragaglia,$^{4}$
\newauthor
M. Matteuzzi,$^{3,4}$
K. Brogaard,$^{3,8}$
A. Stokholm,$^{3,5,8}$
V. Grisoni,$^{9,3}$
M. Tailo,$^{3,10,11}$
E. Willett$^{5}$
\\
$^{1}$Research School of Astronomy and Astrophysics, The Australian National University, Canberra, ACT 2611, Australia\\
$^{2}$ARC Centre of Excellence for All Sky Astrophysics in 3 Dimensions (ASTRO 3D), Australia\\
$^{3}$Dipartimento di Fisica e Astronomia, Universit\`a di Bologna, Via Gobetti 93/2, I-40129 Bologna, Italy\\
$^{4}$INAF – Osservatorio di Astrofisica e Scienza dello Spazio, Via P. Gobetti 93/3, 40129 Bologna, Italy\\
$^{5}$School of Physics \& Astronomy, University of Birmingham, Edgbaston, Birmingham, B15 2TT, UK\\
$^{6}$INAF – Osservatorio Astrofisico di Arcetri, Largo E. Fermi 5, 50125 Firenze, Italy\\
$^{7}$Leibniz-Institut fur Astrophysik Potsdam (AIP), An der Sternwarte 16, D-14482 Potsdam, Germany\\
$^{8}$Stellar Astrophysics Centre, Aarhus University, Ny Munkegade 120, Bldg. 1520, 8000 Aarhus C, Denmark\\
$^{9}$INAF, Osservatorio Astronomico di Trieste, via G.B. Tiepolo 11, I-34131, Trieste, Italy\\
$^{10}$INAF-Osservatorio Astronomico di Padova, Vicolo dell'Osservatorio 5, Padova, IT-35122, Italy\\
$^{11}$INAF, Observatory of Rome, Via Frascati 33, 00077, Monte Porzio Catone (RN), Italy\\
}

\date{Accepted XXX. Received YYY; in original form ZZZ}

\pubyear{\the\year{}}

\begin{document}
\label{firstpage}
\pagerange{\pageref{firstpage}--\pageref{lastpage}}
\maketitle

\begin{abstract}
Chemical clocks offer a powerful tool for estimating stellar ages from spectroscopic surveys. We present a new detailed spectroscopic analysis of 68 \kepler red giant stars to provide a suite of high-precision abundances along with asteroseismic ages with better than 10 percent precision from individual mode frequencies. We obtained several chemical clocks as ratios between s-process elements (Y, Zr, Ba, La, Ce) and \alfa-elements (Mg, Ca, Si, Al, Ti). Our data show that [Ce/Mg] and [Zr/Ti] display a remarkably tight correlation with stellar ages, with abundance dispersions of 0.08 and 0.01 dex respectively and
below 3 Gyr in ages, across the entire Galactic chronochemical history. 
While improving the precision floor of spectroscopic surveys is critical for broadening the scope and applicability of chemical clocks, the intrinsic accuracy of our relations -- enabled by high-resolution chemical abundances and stellar ages in our sample --
allows us to draw meaningful conclusions about age trends across stellar populations. By applying our relations to the APOGEE and \emph{Gaia}-ESO surveys, we are able to differentiate the low- and high-\alfa~sequences in age, recover the age-metallicity relation, observe the disc flaring of the Milky Way, and identify a population of old metal-rich stars. 
\end{abstract}




\section{Introduction}
In the era of the {\em Gaia} satellite \citep{gaiaedr3}, astrometric data coupled with high-resolution spectra give constraints to models of the Milky Way formation and assembly history. An important piece of information is provided by our ability to infer stellar ages and consequently to sketch a Galactic timeline. Asteroseismology can come to our aid giving us the possibility to infer precise ages of field stars, which are known to be very difficult to obtain \citep[see][for a review]{soderblom10}. 
Through asteroseismology, we are able to detect solar-like pulsations in thousands of G-K giants using data collected by the COnvection ROtation and planetary Transits \citep[CoRoT,][]{baglin06}, \emph{Kepler} \citep{gilliland10,Borucki2010}, K2 \citep{howell14}, and Transiting Exoplanet Survey Satellite \citep[TESS,][]{ricker14} missions. The pulsation frequencies are directly linked to the stellar structure and thus provide tight constraints on stellar properties (radius, mass, age) and evolutionary state \citep[see][for a review]{chaplin13}. In fact, through two global seismic parameters -- the average large frequency separation ($\Delta \nu$) and the frequency of maximum oscillations power ($\nu_{\rm max}$) -- stellar parameters can be deduced. 
Another method to get the highest levels of precision from asteroseismology is obtained when comparing observed individual mode frequencies (tens of constraints, compared to the two provided by the global seismic parameters) to stellar models \citep[i.e.][Montalbán in prep.]{huber13,lillobox14,lebreton14,davies16,silvaaguirre17,montalban21}.
These frequencies can be measured in long time series, such as \kepler (4.5 yrs compared to dozen/hundreds days for the other missions). 
Indeed, \emph{Kepler} light curves have allowed us to reach an unprecedented precision in stellar age for isolated field stars, similar to what can be done for stellar clusters \citep[$\sim$ 20\% using \numax and $\Delta \nu$, $\sim$ 11\% using individual mode frequencies, see][respectively]{miglio21,montalban21}. 

Starting from high-quality spectra for a sample of stars with well-known ages ("calibrators", in this case stars with ages from asteroseismology), it is possible to obtain empirical relations between abundance ratios and stellar ages.
The best abundance ratios are those composed of two elements produced on different timescales due to their different nucleosynthetic sites -- e.g., a slow neutron-capture element (or s-process element) and an $\alpha$-element -- resulting in a strong correlation with stellar age.
In Galactic chemical evolution, abundances of different elements are produced by stars of different masses, thus with different timescales. Indeed, s-process elements are mainly ejected in the interstellar medium (ISM) from low- and intermediate-mass asymptotic giant branch (AGB) stars with timescales from about 0.5 to 7 Gyr, whereas $\alpha$-elements are mainly ejected in the ISM from Type II supernovae with very short timescale less than 50 Myr.  
These abundance ratios are called chemical clocks and the most studied ones are [Y/Mg] and [Y/Al] \citep[][and references therein]{dasilva12,nissen15,feltzing17,slumstrup17,delgado19,casali20,casamiquela21,viscasillas22,shejeelammal24}. 
Although they are very promising, and show a clear relationship between age and abundance ratio for solar twins \citep{nissen15,spina16,spina18,jofre20,nissen20}, recent works have shown that these empirical relations differ slightly in the inner and outer regions of the Galactic disc with respect to the solar vicinity 
and depend on metallicity \citep{feltzing17,delgado19,casali20,magrini21,viscasillas22}. 

In the present work, we aim to calibrate chemical clocks with a well-controlled high-quality sample of stars and transfer their age information to field stars ("test sample") in the large spectroscopic surveys, such as APOGEE DR17 \citep{apogeedr17} and \emph{Gaia}-ESO \citep{randich22}. To achieve this goal, we observed 68 giants in the \emph{Kepler} field with the HARPS-N \citep{harps-n} and FIES \citep{fies} spectrographs to carry out a detailed chemical inventory (also complementing the APOGEE DR17 information with n-capture elements) coupled with asteroseismic ages determined using individual mode frequencies from Montalbán et al. (in prep), {where stellar ages for a total of $\sim 5000$ \emph{Kepler} stars will be presented.} 
In a companion paper, we will focus on the full set of chemo-age relations computed with the sample of 68 stars and compare with Galactic chemical evolutionary models (Casali et al. in prep).

The paper is structured as follows. In Sect.~\ref{sec:datasample}, we describe the data sample, its high-precision spectroscopy analysis, the age inference, and the comparison with the APOGEE survey; calibration of chemical clocks and applications of the empirical relations are in Sect.~\ref{sec:clocks} and Sect.~\ref{sec:appl}, respectively; finally, in Sect.~\ref{sec:conclusions}, we summarise and conclude.

\section{Observations and data sample}
\label{sec:datasample}
We selected 68 stars among the \emph{Kepler} red giant branch (RGB) stars with high-quality power spectra\footnote{The entire sample of \kepler RGB stars with high-quality power spectra will be published in Montalbàn et al. in prep.} from the \citet{miglio21}'s catalogue cross-matched with the SDSS data release 17 of the high-resolution spectroscopic survey, Apache Point Observatory Galactic Evolution Experiment \citep[APOGEE,][]{majewski17,apogeedr17}. The selection was made taking stars in the high- and low-$\alpha$ sequences of APOGEE DR17, selecting the same number of stars in the two sequences over the metallicity range [Fe/H] = [$-0.8$, $0.4$] dex. 
Moreover, we selected stars bright enough (G < 11) to be observed with high-resolution spectrographs on 2-4 meter telescopes. 

We obtained spectra with two different instruments: HARPS-N ($R \sim 115,000$, $\Delta \lambda = 383-693$~nm) at the Telescopio Nazionale Galileo (TNG) and FIES ($R \sim 67,000$, $\Delta \lambda = 370-830 $~nm) at the Nordic Optical Telescope (NOT), both at the Roque de los Muchachos Observatory (La Palma, Spain). The observations were conducted from 19 to 22 July 2022 and from 25 to 27 July 2023 (programmes A45TAC\_22, A47TAC\_9), respectively at the TNG and from 24 to 26 June 2022 at the NOT (proposal number: 65-006). The spectra were acquired with total exposure times ranging from 1200 to 3600 seconds depending on the star brightness, in order to reach a signal-to-noise ratio per pixel at blue wavelengths of $S/N > 40$. Exposure times longer than 1800s were usually split into two or three sub-exposures to reduce the contamination of cosmic rays and to avoid saturation. 

The stars analysed in the present work are listed in Table~\ref{tab:info} with their coordinates, \emph{Gaia} magnitudes, parallax, and proper motions. 
Four stars were observed using both instruments to check for possible offsets in atmospheric parameters and chemical abundances caused by the use of two different spectrographs. 
All spectra are available from TNG and NOT Archives.

\begin{table*}
\caption{Parameters of the \emph{Kepler} stars.}
\begin{center}
\setlength{\tabcolsep}{5pt}
\selectfont
\scriptsize
\begin{tabular}{lcccccccc}
\hline
\hline
Name & Telescope & \emph{Gaia} DR3 ID   &   R.A.        & Dec.    & Plx       &  $\mu_{\alpha}$   &  $\mu_{\delta}$   & Gmag   \\ 
 (KIC)    &     &           &  (J2000)      &         & (mas)     & (mas~yr$^{-1}$)   &  (mas~yr$^{-1}$)  &        \\
\hline

1433803  & NOT      & 2051703785658841600 & 291.72570630879 & 37.03846643288 & 2.5376 & 1.832   & -12.493 & 10.15  \\
2451509  & NOT      & 2051665607692314368 & 293.24321361009 & 37.70252480197 & 1.667  & -1.873  & -9.830  & 9.93  \\
2970584  & TNG      & 2099962209991763456 & 286.03837277092 & 38.16971098852 & 1.973  & 7.064   & -15.797 & 10.01  \\
3429738  & TNG      & 2099600161429446016 & 287.19501623686 & 38.59784869632 & 1.874  & -18.696 & -11.068 & 10.02  \\
3539408  & TNG      & 2052888058462055296 & 289.95602718206 & 38.68070416077 & 2.2985 & -9.392  & -5.065  & 9.64  \\
3644223  & NOT      & 2052939086972190208 & 291.07699505523 & 38.78791593017 & 3.1637 & -8.748  & -28.62  & 9.54  \\
3661494  & TNG      & 2076159569968802176 & 295.33964205499 & 38.72007372105 & 1.6322 & -22.057 & -7.288  & 10.57  \\
3744043  & TNG      & 2052864114022602496 & 290.6144515805  & 38.84814303874 & 2.5287 & 10.405  & 25.379  & 9.65  \\
4055294  & NOT      & 2052965234733414528 & 291.04521477091 & 39.12651402293 & 2.1343 & -10.799 & -10.518 & 9.80  \\
4648485  & TNG      & 2101044507393877248 & 289.8882549336  & 39.76055768539 & 2.2062 & 3.844   & -5.397  & 10.53  \\
4756219  & TNG      & 2076460526911309440 & 294.40519631193 & 39.80229068803 & 1.1785 & 14.91   & 25.146  & 11.00  \\
4826087  & NOT      & 2101241629211118720 & 288.91716240894 & 39.96424275903 & 1.7006 & -9.6    & -9.419  & 9.51  \\
4913049  & TNG      & 2101299211835208704 & 288.49796516747 & 40.08652722798 & 2.9383 & 4.729   & 12.322  & 10.30  \\
4931389  & TNG      & 2052513438539100544 & 293.90812945332 & 40.07863655336 & 3.3937 & 5.606   & 19.098  & 9.68  \\
5265256  & TNG      & 2101334224404935808 & 288.78444311507 & 40.49287806719 & 1.3298 & -5.758  & -24.946 & 10.62  \\
5769244  & TNG      & 2103697147912507904 & 283.41902409105 & 41.03654428312 & 2.9417 & -25.193 & -24.086 & 9.86  \\
5882005  & NOT      & 2077382948448694144 & 293.58754916997 & 41.10299865842 & 1.7129 & -6.663  & -26.876 & 10.19  \\
5940060  & TNG      & 2103703160866757376 & 283.84136373493 & 41.23187541068 & 2.0936 & -6.442  & -29.312 & 9.68  \\
6365511  & TNG      & 2101681880540210816 & 292.29333546278 & 41.78788591373 & 1.1798 & -1.29   & -25.558 & 11.25  \\
6547007  & TNG      & 2075411524106352896 & 298.23841960779 & 41.97024284079 & 1.2814 & 8.739   & -4.842  & 10.45  \\
6851499  & TNG      & 2102506548618822272 & 288.09390243154 & 42.39708323493 & 1.2974 & -8.761  & -23.784 & 10.66  \\
6859803  & NOT      & 2101942288694401792 & 290.80359448096 & 42.33279512806 & 2.7956 & 25.759  & 41.3    & 9.81  \\
6964342  & NOT      & 2077220426885950976 & 296.0893851256  & 42.43766636329 & 1.667  & 14.996  & 29.675  & 9.84  \\
7429055  & TNG      & 2105552608144587008 & 286.30448783205 & 43.03668783654 & 1.1228 & -7.816  & 9.118   & 11.19  \\
7430868  & TNG      & 2105587513340280832 & 287.07512031919 & 43.00396515398 & 1.6869 & -12.007 & 1.624   & 10.22  \\
7450230  & TNG      & 2077839520649382912 & 293.39148078301 & 43.03586846178 & 5.0844 & -16.225 & -30.856 & 9.06  \\
7533995  & TNG      & 2078201569212490112 & 293.63876550384 & 43.10556031286 & 2.3541 & -14.546 & -13.262 & 10.11  \\
7617227  & TNG      & 2077985893136622720 & 294.65772917773 & 43.2050581937  & 2.246  & 7.274   & 28.134  & 9.62  \\
7812552  & TNG      & 2102960333383028864 & 288.21198411368 & 43.50628926117 & 3.0691 & 2.044   & -10.235 & 10.09  \\
8129047  & TNG      & 2076056903066401408 & 300.79512348024 & 43.98263371667 & 1.4667 & 3.739   & 10.518  & 10.70  \\
8493969  & TNG      & 2126240056861504768 & 291.31898259395 & 44.52162539813 & 1.8747 & -18.969 & -1.276  & 10.59  \\
8587329  & TNG      & 2076151186191153408 & 300.15781414731 & 44.65790976271 & 1.901  & -4.434  & -21.296 & 10.38  \\
8590920  & TNG      & 2082106038083789184 & 301.2334322159  & 44.66624917466 & 1.3475 & -2.408  & -10.732 & 11.01  \\
8612241  & NOT      & 2106313740768003712 & 286.26347122477 & 44.70791149712 & 1.3996 & 6.049   & -22.101 & 10.24  \\
8737032  & NOT      & 2106790314639185280 & 283.86388902736 & 44.98524940721 & 1.2746 & 6.929   & 9.98    & 10.19  \\
9080175  & TNG      & 2130239221168800128 & 288.03397469718 & 45.49335362678 & 1.6128 & -6.065  & -8.657  & 10.15  \\
9145955  & NOT      & 2130219769257640576 & 287.8855801582  & 45.52287564413 & 2.4141 & 3.138   & -15.988 & 9.77  \\
9535399  & TNG, NOT & 2080029640428099840 & 295.3782447566  & 46.15220017926 & 2.2502 & 7.537   & -9.335  & 9.63  \\
9640480  & TNG      & 2130416929734202624 & 287.71772206553 & 46.39464244871 & 1.4604 & -9.38   & 9.414   & 10.70  \\
9711269  & TNG, NOT & 2128306485885973248 & 292.24848338049 & 46.41193805565 & 2.138  & 0.928   & 25.083  & 9.77  \\
9772366  & TNG      & 2128291127083154432 & 292.59732866502 & 46.58594066557 & 1.9758 & 4.309   & -13.709 & 9.66  \\
9777293  & TNG      & 2128141627862366720 & 294.54782801418 & 46.59345892176 & 1.9066 & -9.328  & 9.874   & 10.99  \\
9783226  & NOT      & 2080331907350333312 & 296.68965799804 & 46.58398730699 & 1.6824 & 0.844   & 11.227  & 9.84  \\
9967700  & TNG      & 2086305725822665728 & 297.91512418488 & 46.87070339907 & 1.9044 & -5.736  & 5.194   & 10.30  \\
10095427 & TNG      & 2085570771021453056 & 298.69412782257 & 47.08869166931 & 2.2811 & 8.92    & -47.646 & 9.61  \\
10351820 & NOT      & 2080579160028012928 & 296.33947586705 & 47.41981620496 & 1.7124 & 11.351  & -2.045  & 9.86  \\
10513837 & NOT      & 2119801557282726912 & 280.86796292697 & 47.70362889589 & 3.7792 & 4.718   & -1.304  & 9.21  \\
10550429 & TNG      & 2086419800156327680 & 297.34535863072 & 47.78287096973 & 2.556  & -5.297  & -25.731 & 9.57  \\
10659842 & NOT      & 2127945639913049984 & 289.44254327873 & 47.94136786497 & 1.7343 & 10.002  & 13.189  & 9.90  \\
10722175 & NOT      & 2130984827490239232 & 288.09718661588 & 48.06239066917 & 2.8962 & 2.505   & 28.877  & 9.46  \\
10793771 & NOT      & 2129414106412064256 & 290.4112575946  & 48.1503155865  & 1.2973 & -0.953  & -17.183 & 10.25  \\
10801063 & TNG      & 2128852943165070336 & 293.40401419551 & 48.12821469224 & 2.6771 & -28.996 & -32.222 & 10.37  \\
10992711 & TNG      & 2086863212577271040 & 298.79955322383 & 48.40574133545 & 1.6143 & 13.523  & 40.395  & 9.95  \\
11020211 & TNG      & 2131166276968591360 & 286.72887458747 & 48.50281492322 & 1.9772 & -37.815 & 59.779  & 10.41  \\
11029423 & TNG      & 2129338961663941504 & 291.71024475923 & 48.59500213454 & 2.7315 & -9.554  & 21.707  & 9.46  \\
11087371 & TNG, NOT & 2129006870493887744 & 293.46990960596 & 48.61222842243 & 2.8327 & -7.734  & -1.006  & 9.76  \\
11228549 & TNG      & 2143769635160629760 & 282.56542385642 & 48.90544475372 & 1.1368 & -3.458  & 4.498   & 10.95  \\
11250139 & TNG      & 2134800338999043584 & 294.63588634634 & 48.95780220302 & 1.1997 & 6.304   & -17.961 & 10.69  \\
11358669 & NOT      & 2134772129653795328 & 295.7471194045  & 49.19190636441 & 2.1599 & 21.089  & 29.626  & 9.74  \\
11453721 & TNG      & 2129947988025131520 & 291.02538198849 & 49.38136639365 & 1.9252 & 0.654   & -26.499 & 9.94  \\
11496569 & NOT      & 2132105435996154368 & 285.71381737169 & 49.49053613947 & 1.8927 & -12.441 & -10.541 & 9.55  \\
11550492 & TNG      & 2131326599508772352 & 286.74955720864 & 49.54020004945 & 3.1245 & -4.51   & 19.454  & 9.26  \\
11702195 & TNG      & 2132142166556843264 & 285.6532396511  & 49.81256772245 & 1.3476 & 5.714   & 18.753  & 10.63  \\
11775511 & TNG      & 2087252680210082432 & 297.52129742582 & 49.97923069817 & 1.519  & -3.225  & -11.811 & 10.33  \\
11819363 & NOT      & 2135151083206715008 & 294.55092213007 & 50.08866204207 & 2.0492 & -9.251  & -4.202  & 9.39  \\
12115227 & TNG      & 2135229973166427008 & 293.99161760649 & 50.63598157364 & 2.3219 & -13.407 & 6.422   & 9.95  \\
12122151 & NOT      & 2135411834961721728 & 297.08500386861 & 50.60769726425 & 1.6858 & 0.516   & -7.605  & 9.69  \\
12647227 & TNG, NOT & 2139238517681314944 & 290.55928721926 & 51.78082445724 & 1.7849 & 4.43    & -1.281  & 9.41  \\

\hline
\end{tabular}

\end{center}
\label{tab:info}
\end{table*}

\subsection{Spectral analysis}
\label{sec:spectral}
We performed a spectral analysis based on the measurements of the equivalent widths (EWs) to determine the atmospheric parameters and abundances of our sample stars. The EWs of the spectral absorption lines were measured using the Daospec \citep{stetson08} tool through an automatic wrapper DOOp \cite[Daospec Output Optimiser pipeline,][]{cantat14}. The linelist of atomic transitions used in this work was the master list prepared for the analysis of the stellar spectra for the {\em Gaia}-ESO survey \citep{heiter15, heiter21}. This linelist includes three flags that provide information about the quality of the transition probabilities and the blending properties, such as “Y” (yes), “N” (no), and “U” (undetermined). These flags are assigned on the basis of the accuracy of the $\log~gf$ (line oscillator strength) and the quality of the line profiles (at the resolving power $R \sim 47,000$). In our analysis, we considered all lines, except those with “N” for the $\log~gf$ values. Moreover, we selected lines with EWs between \ensuremath{20-120} m\AA~for iron and between $10-120$ m\AA~for the other elements to avoid saturated and overly weak lines. Regarding neutron-capture elements, we considered only lines with "Y" for the $\log~gf$ values.

The EWs measured through DOOp were passed to MOOG in the automatic form: Fast Automatic MOOG Analysis \citep[FAMA,][]{magrini13}. FAMA uses MOOG in its 2017 version \citep{sneden12} and a grid of MARCS model atmospheres \citep{gustafsson08} to determine atmospheric parameters and abundances. It iteratively searches for the effective temperature $T_{\rm eff}$ by minimising the slope between iron abundance and the excitation potential and the microturbulence $\xi$ by minimising the trend between iron abundance and the reduced EWs, $\log(EW/\lambda)$. The $\log~g$ is fixed to the seismic value computed using the scaling relation:
\begin{equation}
\log~g = \log~g_{\odot} + \log(\nu_{\rm max}/\nu_{\rm max,\odot}) + 1/2 \log(T_{\rm eff}/T_{\rm eff, \odot})
\end{equation}
where $\log~g_{\odot} = 4.44$ dex, $\nu_{\rm max,\odot}$ = 3090 $\mu$Hz and $T_{\rm eff, \odot}$ = 5777 K \citep{handberg17,huber11}.
The evaluation of the uncertainties on the final atmospheric parameters is described by \citet{magrini13}.
Finally, the elemental abundances computed by FAMA are: Na, Mg, Al, Si, Ca, Sc, Ti ({\sc i} and {\sc ii}), V, Cr ({\sc i} and {\sc ii}), Mn, Fe ({\sc i} and {\sc ii}), Co, Ni, Cu, Zn, Sr, Y, Zr ({\sc i} and {\sc ii}), Ba, La, Ce, Pr, Nd, Sm and Eu (the n-capture elements are not available in the APOGEE survey for these stars, except Ce  -- for this reason, we observed stars with previously measured APOGEE abundances using spectrographs that include the blue spectral region, which contains key absorption lines of neutron-capture elements).
The Solar abundance scale adopted in this work is from \citet{asplund09}. 
When only one line is detected, as is the case for Ba, the observational error is estimated through the uncertainty on the EW measured by DOOp.
No systematic offsets were found between the FIES and HARPS-N spectra for the 4 stars observed by both instruments, for this reason we use the average of the two values.
In the Tables~\ref{tab:parametersage} and \ref{tab:abu} we present the results of our spectral analysis\footnote{In this work, we present the abundances of elements used in our analysis and compared with the APOGEE survey: Na, Mg, Al, Si, Ca, Ti ({\sc i} and {\sc ii}), V, Cr, Fe ({\sc i} and {\sc ii}), Co, Ni, Y, Zr {\sc i}, Ba, La, Ce.}.  

 \begin{table*}
\caption{Atmospheric parameters derived from spectral analysis, along with \logg obtained from asteroseismology, for the \emph{Kepler} stars.}
\begin{center}
\scriptsize
\begin{tabular}{lcccc}
\hline
\hline
  Name       &     T$_{\rm eff}$   &      logg (seismo)         &        [Fe/H]        &        $\xi$  \\
             &     (K)             &      (dex)         &        (dex)         &        (dex)  \\
\hline 
  1433803 &   $4683 \pm   89$ & $3.079 \pm  0.004$ &       $ 0.14 \pm  0.01$ &  $1.40 \pm  0.07$ \\ 
  2451509 &   $4942 \pm   76$ & $2.820 \pm  0.004$ &       $-0.15 \pm  0.04$ &  $1.49 \pm  0.07$ \\ 
  2970584 &   $4694 \pm  109$ & $2.698 \pm  0.005$ &       $-0.33 \pm  0.03$ &  $1.36 \pm  0.09$ \\ 
  3429738 &   $4664 \pm  119$ & $2.651 \pm  0.005$ &       $-0.41 \pm  0.03$ &  $1.33 \pm  0.08$ \\ 
  3539408 &   $5002 \pm  110$ & $2.971 \pm  0.004$ &       $-0.23 \pm  0.05$ &  $1.18 \pm  0.09$ \\ 
  3644223 &   $4958 \pm   80$ & $3.127 \pm  0.003$ &       $-0.21 \pm  0.03$ &  $1.29 \pm  0.09$ \\ 
  3661494 &   $4903 \pm   90$ & $2.852 \pm  0.003$ &       $-0.38 \pm  0.04$ &  $1.14 \pm  0.08$ \\ 
  3744043 &   $5023 \pm  100$ & $2.960 \pm  0.003$ &       $-0.36 \pm  0.05$ &  $1.18 \pm  0.08$ \\ 
  4055294 &   $5021 \pm   94$ & $2.989 \pm  0.004$ &       $ 0.17 \pm  0.03$ &  $1.45 \pm  0.08$ \\ 
  4648485 &   $4754 \pm  110$ & $3.056 \pm  0.003$ &       $-0.08 \pm  0.02$ &  $1.16 \pm  0.10$ \\ 
  4756219 &   $4597 \pm  142$ & $2.602 \pm  0.010$ &       $-0.25 \pm  0.01$ &  $1.29 \pm  0.11$ \\ 
  4826087 &   $4802 \pm   94$ & $2.559 \pm  0.010$ &       $-0.10 \pm  0.04$ &  $1.53 \pm  0.07$ \\ 
  4913049 &   $4944 \pm  136$ & $3.246 \pm  0.003$ &       $-0.07 \pm  0.04$ &  $0.99 \pm  0.12$ \\ 
  4931389 &   $5007 \pm  107$ & $3.218 \pm  0.004$ &       $ 0.11 \pm  0.03$ &  $1.07 \pm  0.10$ \\ 
  5265256 &   $4639 \pm  120$ & $2.619 \pm  0.005$ &       $-0.26 \pm  0.02$ &  $1.34 \pm  0.10$ \\ 
  5769244 &   $4756 \pm  106$ & $3.046 \pm  0.003$ &       $ 0.08 \pm  0.02$ &  $1.01 \pm  0.09$ \\ 
  5882005 &   $4843 \pm   70$ & $2.818 \pm  0.004$ &       $-0.16 \pm  0.03$ &  $1.34 \pm  0.06$ \\ 
  5940060 &   $4683 \pm   93$ & $2.615 \pm  0.008$ &       $-0.45 \pm  0.04$ &  $1.32 \pm  0.07$ \\ 
  6365511 &   $4845 \pm   74$ & $2.789 \pm  0.003$ &       $-0.62 \pm  0.04$ &  $1.31 \pm  0.07$ \\ 
  6547007 &   $4772 \pm  103$ & $2.475 \pm  0.005$ &       $-0.82 \pm  0.06$ &  $1.44 \pm  0.08$ \\ 
  6851499 &   $4671 \pm  110$ & $2.612 \pm  0.006$ &       $-0.39 \pm  0.03$ &  $1.31 \pm  0.08$ \\ 
  6859803 &   $5059 \pm  104$ & $3.146 \pm  0.003$ &       $-0.17 \pm  0.04$ &  $1.18 \pm  0.08$ \\ 
  6964342 &   $4604 \pm   88$ & $2.507 \pm  0.011$ &       $-0.09 \pm  0.02$ &  $1.55 \pm  0.07$ \\ 
  7429055 &   $4550 \pm   84$ & $2.669 \pm  0.006$ &       $-0.09 \pm  0.02$ &  $1.20 \pm  0.05$ \\ 
  7430868 &   $4567 \pm   89$ & $2.627 \pm  0.006$ &       $-0.34 \pm  0.02$ &  $1.24 \pm  0.07$ \\ 
  7450230 &   $4909 \pm  122$ & $3.238 \pm  0.003$ &       $-0.13 \pm  0.03$ &  $0.99 \pm  0.12$ \\ 
  7533995 &   $4656 \pm   76$ & $2.929 \pm  0.003$ &       $-0.08 \pm  0.03$ &  $1.12 \pm  0.04$ \\ 
  7617227 &   $4866 \pm  117$ & $2.780 \pm  0.004$ &       $-0.33 \pm  0.04$ &  $1.18 \pm  0.11$ \\ 
  7812552 &   $5106 \pm   76$ & $3.235 \pm  0.003$ &       $-0.52 \pm  0.04$ &  $1.06 \pm  0.05$ \\ 
  8129047 &   $4635 \pm   69$ & $2.682 \pm  0.004$ &       $-0.08 \pm  0.02$ &  $1.10 \pm  0.05$ \\ 
  8493969 &   $4830 \pm   91$ & $2.942 \pm  0.003$ &       $-0.34 \pm  0.04$ &  $1.07 \pm  0.06$ \\ 
  8587329 &   $4357 \pm  111$ & $2.718 \pm  0.005$ &       $ 0.14 \pm  0.00$ &  $1.13 \pm  0.09$ \\ 
  8590920 &   $4460 \pm   77$ & $2.785 \pm  0.006$ &       $ 0.18 \pm  0.05$ &  $1.26 \pm  0.05$ \\ 
  8612241 &   $5029 \pm  103$ & $2.811 \pm  0.005$ &       $-0.23 \pm  0.06$ &  $1.53 \pm  0.07$ \\ 
  8737032 &   $4711 \pm   71$ & $2.535 \pm  0.004$ &       $-0.21 \pm  0.02$ &  $1.54 \pm  0.08$ \\ 
  9080175 &   $4700 \pm  106$ & $2.586 \pm  0.005$ &       $-0.55 \pm  0.04$ &  $1.33 \pm  0.09$ \\ 
  9145955 &   $5064 \pm   95$ & $3.027 \pm  0.004$ &       $-0.30 \pm  0.04$ &  $1.32 \pm  0.09$ \\ 
  9535399 &   $4549 \pm   62$ & $2.653 \pm  0.005$ &       $ 0.12 \pm  0.02$ &  $1.08 \pm  0.03$ \\ 
  9640480 &   $4764 \pm   75$ & $2.581 \pm  0.004$ &       $-0.40 \pm  0.02$ &  $1.17 \pm  0.09$ \\ 
  9711269 &   $4656 \pm  119$ & $2.709 \pm  0.004$ &       $ 0.07 \pm  0.02$ &  $1.09 \pm  0.07$ \\ 
  9772366 &   $4691 \pm  105$ & $2.780 \pm  0.005$ &       $ 0.14 \pm  0.02$ &  $1.11 \pm  0.08$ \\ 
  9777293 &   $5006 \pm   73$ & $3.171 \pm  0.003$ &       $-0.61 \pm  0.04$ &  $1.23 \pm  0.06$ \\ 
  9783226 &   $4762 \pm   75$ & $2.625 \pm  0.003$ &       $-0.02 \pm  0.03$ &  $1.38 \pm  0.07$ \\ 
  9967700 &   $4976 \pm   76$ & $2.944 \pm  0.003$ &       $-0.42 \pm  0.04$ &  $1.15 \pm  0.07$ \\ 
 10095427 &   $4682 \pm  103$ & $2.662 \pm  0.004$ &       $-0.55 \pm  0.04$ &  $1.26 \pm  0.08$ \\ 
 10351820 &   $4788 \pm   77$ & $2.658 \pm  0.006$ &       $-0.12 \pm  0.03$ &  $1.40 \pm  0.08$ \\ 
 10513837 &   $5024 \pm  112$ & $3.199 \pm  0.003$ &       $ 0.04 \pm  0.04$ &  $1.31 \pm  0.08$ \\ 
 10550429 &   $4980 \pm  118$ & $3.030 \pm  0.004$ &       $ 0.02 \pm  0.04$ &  $1.06 \pm  0.10$ \\ 
 10659842 &   $4973 \pm   79$ & $2.784 \pm  0.004$ &       $-0.22 \pm  0.04$ &  $1.32 \pm  0.08$ \\ 
 10722175 &   $4764 \pm   86$ & $2.957 \pm  0.003$ &       $ 0.11 \pm  0.03$ &  $1.39 \pm  0.07$ \\ 
 10793771 &   $4956 \pm  105$ & $2.806 \pm  0.003$ &       $ 0.05 \pm  0.04$ &  $1.48 \pm  0.09$ \\ 
 10801063 &   $4948 \pm   80$ & $3.207 \pm  0.002$ &       $-0.35 \pm  0.03$ &  $0.95 \pm  0.06$ \\ 
 10992711 &   $4571 \pm  119$ & $2.444 \pm  0.004$ &       $-0.44 \pm  0.03$ &  $1.38 \pm  0.10$ \\ 
 11020211 &   $4891 \pm   83$ & $2.937 \pm  0.003$ &       $-0.40 \pm  0.03$ &  $1.07 \pm  0.07$ \\ 
 11029423 &   $4616 \pm  116$ & $2.779 \pm  0.004$ &       $-0.16 \pm  0.02$ &  $1.21 \pm  0.08$ \\ 
 11087371 &   $4739 \pm   72$ & $3.056 \pm  0.004$ &       $ 0.14 \pm  0.02$ &  $1.07 \pm  0.05$ \\ 
 11228549 &   $4890 \pm   89$ & $2.649 \pm  0.004$ &       $-0.83 \pm  0.05$ &  $1.42 \pm  0.08$ \\ 
 11250139 &   $4675 \pm   98$ & $2.506 \pm  0.005$ &       $-0.53 \pm  0.04$ &  $1.35 \pm  0.07$ \\ 
 11358669 &   $4899 \pm   61$ & $2.896 \pm  0.003$ &       $ 0.11 \pm  0.02$ &  $1.41 \pm  0.07$ \\ 
 11453721 &   $4615 \pm  107$ & $2.685 \pm  0.003$ &       $-0.42 \pm  0.04$ &  $1.25 \pm  0.09$ \\ 
 11496569 &   $4894 \pm  117$ & $2.730 \pm  0.003$ &       $-0.03 \pm  0.05$ &  $1.46 \pm  0.09$ \\ 
 11550492 &   $4801 \pm   86$ & $2.875 \pm  0.003$ &       $-0.07 \pm  0.03$ &  $1.03 \pm  0.08$ \\ 
 11702195 &   $4662 \pm   97$ & $2.619 \pm  0.005$ &       $-0.51 \pm  0.04$ &  $1.29 \pm  0.07$ \\ 
 11775511 &   $4756 \pm   73$ & $2.574 \pm  0.004$ &       $-0.32 \pm  0.03$ &  $1.20 \pm  0.07$ \\ 
 11819363 &   $4868 \pm   84$ & $2.611 \pm  0.005$ &       $-0.37 \pm  0.04$ &  $1.33 \pm  0.10$ \\ 
 12115227 &   $4903 \pm   78$ & $2.898 \pm  0.003$ &       $-0.77 \pm  0.04$ &  $1.32 \pm  0.08$ \\ 
 12122151 &   $4739 \pm  122$ & $2.610 \pm  0.008$ &       $-0.15 \pm  0.04$ &  $1.52 \pm  0.10$ \\ 
 12647227 &   $4803 \pm   96$ & $2.491 \pm  0.007$ &       $-0.54 \pm  0.05$ &  $1.29 \pm  0.09$ \\ 
 \hline
\end{tabular}
\label{tab:parametersage}
\end{center}
\end{table*}

\begin{table*}
\caption{Chemical abundances for the sample of \emph{Kepler} stars.}
\begin{center}
\scriptsize
\begin{tabular}{cccccccccc}
\hline
\hline
  Name      & Na/H            & Mg/H           & Al/H           & Si/H           & Ca/H           & TiI/H          &  TiII/H           &  V/H         &  Cr/H            \\
\hline
  1433803 &  $6.70 \pm  0.03$ &  $7.97 \pm 0.11$ &  $6.75 \pm 0.10$ &  $7.72 \pm 0.07$ &  $6.43 \pm 0.14$ &  $5.11 \pm 0.09$ &  $5.15 \pm  0.20$ &  $4.34 \pm 0.08$ &  $5.83 \pm  0.10$  \\
  2451509 &  $6.26 \pm  0.07$ &  $7.56 \pm 0.06$ &  $6.36 \pm 0.06$ &  $7.35 \pm 0.03$ &  $6.23 \pm 0.07$ &  $4.92 \pm 0.07$ &  $4.85 \pm  0.10$ &  $4.01 \pm 0.11$ &  $5.58 \pm  0.04$  \\
  2970584 &  $6.23 \pm  0.08$ &  $7.59 \pm 0.20$ &  $6.54 \pm 0.02$ &  $7.26 \pm 0.01$ &  $6.24 \pm 0.08$ &  $5.00 \pm 0.07$ &  $4.86 \pm  0.11$ &  $4.09 \pm 0.08$ &  $5.47 \pm  0.06$  \\
  3429738 &  $6.10 \pm  0.06$ &  $7.62 \pm 0.01$ &  $6.40 \pm 0.12$ &  $7.26 \pm 0.01$ &  $6.10 \pm 0.07$ &  $4.83 \pm 0.06$ &  $4.83 \pm  0.08$ &  $3.92 \pm 0.09$ &  $5.31 \pm  0.06$  \\
  3539408 &  $6.12 \pm  0.04$ &  $7.47 \pm 0.02$ &  $6.21 \pm 0.01$ &  $7.21 \pm 0.02$ &  $6.14 \pm 0.04$ &  $4.80 \pm 0.05$ &  $4.82 \pm  0.07$ &  $3.83 \pm 0.08$ &  $5.41 \pm  0.05$  \\
  3644223 &  $6.21 \pm  0.10$ &  $7.57 \pm 0.08$ &  $6.33 \pm 0.04$ &  $7.35 \pm 0.02$ &  $6.13 \pm 0.04$ &  $4.83 \pm 0.06$ &  $4.87 \pm  0.12$ &  $3.88 \pm 0.11$ &  $5.49 \pm  0.07$  \\
  3661494 &  $5.96 \pm  0.04$ &  $7.40 \pm 0.02$ &  $6.23 \pm 0.04$ &  $7.12 \pm 0.01$ &  $6.03 \pm 0.03$ &  $4.72 \pm 0.06$ &  $4.75 \pm  0.07$ &  $3.74 \pm 0.09$ &  $5.25 \pm  0.05$  \\
  3744043 &  $5.98 \pm  0.05$ &  $7.29 \pm 0.03$ &  $6.14 \pm 0.07$ &  $7.10 \pm 0.01$ &  $6.00 \pm 0.05$ &  $4.72 \pm 0.06$ &  $4.68 \pm  0.07$ &  $3.73 \pm 0.08$ &  $5.28 \pm  0.04$  \\
  4055294 &  $6.79 \pm  0.06$ &  $7.95 \pm 0.07$ &  $6.80 \pm 0.04$ &  $7.62 \pm 0.03$ &  $6.55 \pm 0.12$ &  $5.30 \pm 0.09$ &  $5.11 \pm  0.17$ &  $4.51 \pm 0.09$ &  $5.92 \pm  0.07$  \\
  4648485 &  $6.34 \pm  0.07$ &  $7.78 \pm 0.01$ &  $6.61 \pm 0.06$ &  $7.47 \pm 0.02$ &  $6.33 \pm 0.04$ &  $5.03 \pm 0.06$ &  $5.08 \pm  0.13$ &  $4.18 \pm 0.09$ &  $5.64 \pm  0.06$  \\
  4756219 &  $6.20 \pm  0.07$ &  $7.66 \pm 0.02$ &  $6.48 \pm 0.04$ &  $7.30 \pm 0.01$ &  $6.18 \pm 0.04$ &  $4.85 \pm 0.06$ &  $4.88 \pm  0.13$ &  $3.97 \pm 0.08$ &  $5.47 \pm  0.06$  \\
  4826087 &  $6.38 \pm  0.08$ &  $7.73 \pm 0.11$ &  $6.47 \pm 0.01$ &  $7.40 \pm 0.05$ &  $6.36 \pm 0.10$ &  $5.00 \pm 0.07$ &  $4.90 \pm  0.12$ &  $4.12 \pm 0.07$ &  $5.66 \pm  0.07$  \\
  4913049 &  $6.35 \pm  0.08$ &  $7.85 \pm 0.05$ &  $6.65 \pm 0.05$ &  $7.34 \pm 0.01$ &  $6.37 \pm 0.07$ &  $5.18 \pm 0.07$ &  $5.07 \pm  0.11$ &  $4.38 \pm 0.12$ &  $5.68 \pm  0.07$  \\
  4931389 &  $6.57 \pm  0.05$ &  $7.80 \pm 0.02$ &  $6.62 \pm 0.10$ &  $7.44 \pm 0.03$ &  $6.45 \pm 0.03$ &  $5.21 \pm 0.07$ &  $5.07 \pm  0.14$ &  $4.47 \pm 0.11$ &  $5.84 \pm  0.07$  \\
  5265256 &  $6.12 \pm  0.06$ &  $7.62 \pm 0.04$ &  $6.42 \pm 0.01$ &  $7.28 \pm 0.02$ &  $6.16 \pm 0.03$ &  $4.82 \pm 0.06$ &  $4.85 \pm  0.10$ &  $3.93 \pm 0.08$ &  $5.47 \pm  0.05$  \\
  5769244 &  $6.54 \pm  0.06$ &  $7.96 \pm 0.04$ &  $6.74 \pm 0.07$ &  $7.48 \pm 0.03$ &  $6.47 \pm 0.04$ &  $5.21 \pm 0.08$ &  $5.16 \pm  0.17$ &  $4.40 \pm 0.10$ &  $5.83 \pm  0.08$  \\
  5882005 &  $6.23 \pm  0.08$ &  $7.62 \pm 0.05$ &  $6.41 \pm 0.03$ &  $7.42 \pm 0.03$ &  $6.20 \pm 0.07$ &  $4.88 \pm 0.07$ &  $4.90 \pm  0.12$ &  $3.98 \pm 0.11$ &  $5.50 \pm  0.07$  \\
  5940060 &  $6.08 \pm  0.05$ &  $7.59 \pm 0.01$ &  $6.33 \pm 0.17$ &  $7.21 \pm 0.01$ &  $6.05 \pm 0.07$ &  $4.84 \pm 0.06$ &  $4.77 \pm  0.09$ &  $3.90 \pm 0.08$ &  $5.29 \pm  0.06$  \\
  6365511 &  $5.85 \pm  0.06$ &  $7.34 \pm 0.11$ &  $6.20 \pm 0.03$ &  $7.06 \pm 0.01$ &  $5.92 \pm 0.03$ &  $4.65 \pm 0.06$ &  $4.64 \pm  0.07$ &  $3.62 \pm 0.06$ &  $5.08 \pm  0.04$  \\
  6547007 &  $5.60 \pm  0.03$ &  $7.24 \pm 0.04$ &  $5.96 \pm 0.10$ &  $6.95 \pm 0.01$ &  $5.68 \pm 0.04$ &  $4.35 \pm 0.05$ &  $4.47 \pm  0.05$ &  $3.29 \pm 0.06$ &  $4.81 \pm  0.05$  \\
  6851499 &  $6.06 \pm  0.08$ &  $7.62 \pm 0.01$ &  $6.37 \pm 0.04$ &  $7.21 \pm 0.02$ &  $6.05 \pm 0.06$ &  $4.76 \pm 0.06$ &  $4.80 \pm  0.09$ &  $3.85 \pm 0.09$ &  $5.32 \pm  0.06$  \\
  6859803 &  $6.24 \pm  0.12$ &  $7.57 \pm 0.09$ &  $6.31 \pm 0.03$ &  $7.38 \pm 0.02$ &  $6.21 \pm 0.09$ &  $4.90 \pm 0.08$ &  $4.93 \pm  0.09$ &  $3.95 \pm 0.15$ &  $5.44 \pm  0.08$  \\
  6964342 &  $6.52 \pm  0.07$ &  $7.79 \pm 0.09$ &  $6.60 \pm 0.08$ &  $7.47 \pm 0.01$ &  $6.37 \pm 0.10$ &  $5.01 \pm 0.08$ &  $4.89 \pm  0.16$ &  $4.19 \pm 0.09$ &  $5.75 \pm  0.08$  \\
  7429055 &  $6.43 \pm  0.08$ &  $7.83 \pm 0.04$ &  $6.69 \pm 0.07$ &  $7.41 \pm 0.01$ &  $6.35 \pm 0.04$ &  $5.04 \pm 0.07$ &  $5.08 \pm  0.15$ &  $4.21 \pm 0.07$ &  $5.63 \pm  0.06$  \\
  7430868 &  $6.17 \pm  0.06$ &  $7.69 \pm 0.02$ &  $6.20 \pm 0.10$ &  $7.30 \pm 0.01$ &  $6.15 \pm 0.06$ &  $4.86 \pm 0.06$ &  $4.92 \pm  0.11$ &  $3.95 \pm 0.08$ &  $5.38 \pm  0.07$  \\
  7450230 &  $6.27 \pm  0.09$ &  $7.71 \pm 0.04$ &  $6.54 \pm 0.04$ &  $7.31 \pm 0.01$ &  $6.28 \pm 0.04$ &  $5.05 \pm 0.06$ &  $5.03 \pm  0.11$ &  $4.22 \pm 0.11$ &  $5.59 \pm  0.06$  \\
  7533995 &  $6.37 \pm  0.06$ &  $7.83 \pm 0.03$ &  $6.66 \pm 0.06$ &  $7.43 \pm 0.01$ &  $6.33 \pm 0.04$ &  $5.04 \pm 0.06$ &  $5.08 \pm  0.14$ &  $4.22 \pm 0.08$ &  $5.63 \pm  0.07$  \\
  7617227 &  $6.07 \pm  0.05$ &  $7.42 \pm 0.04$ &  $6.25 \pm 0.04$ &  $7.14 \pm 0.03$ &  $6.03 \pm 0.02$ &  $4.69 \pm 0.05$ &  $4.74 \pm  0.08$ &  $3.75 \pm 0.08$ &  $5.31 \pm  0.05$  \\
  7812552 &  $5.97 \pm  0.06$ &  $7.52 \pm 0.10$ &  $6.14 \pm 0.01$ &  $7.05 \pm 0.02$ &  $6.01 \pm 0.02$ &  $4.75 \pm 0.05$ &  $4.71 \pm  0.05$ &  $3.66 \pm 0.07$ &  $5.16 \pm  0.05$  \\
  8129047 &  $6.44 \pm  0.05$ &  $7.82 \pm 0.01$ &  $6.67 \pm 0.04$ &  $7.40 \pm 0.03$ &  $6.41 \pm 0.06$ &  $5.17 \pm 0.07$ &  $5.07 \pm  0.17$ &  $4.35 \pm 0.10$ &  $5.71 \pm  0.07$  \\
  8493969 &  $6.13 \pm  0.06$ &  $7.62 \pm 0.02$ &  $6.46 \pm 0.01$ &  $7.24 \pm 0.02$ &  $6.18 \pm 0.04$ &  $4.91 \pm 0.06$ &  $4.88 \pm  0.09$ &  $3.98 \pm 0.10$ &  $5.36 \pm  0.05$  \\
  8587329 &  $6.59 \pm  0.10$ &  $7.99 \pm 0.02$ &  $6.79 \pm 0.07$ &  $7.68 \pm 0.02$ &  $6.47 \pm 0.23$ &  $5.01 \pm 0.07$ &  $5.35 \pm  0.22$ &  $4.26 \pm 0.07$ &  $5.78 \pm  0.08$  \\
  8590920 &  $6.67 \pm  0.06$ &  $8.03 \pm 0.03$ &  $6.73 \pm 0.07$ &  $7.71 \pm 0.03$ &  $6.55 \pm 0.06$ &  $5.13 \pm 0.08$ &  $5.34 \pm  0.24$ &  $4.43 \pm 0.06$ &  $5.89 \pm  0.08$  \\
  8612241 &  $6.28 \pm  0.08$ &  $7.62 \pm 0.10$ &  $6.33 \pm 0.04$ &  $7.35 \pm 0.06$ &  $6.17 \pm 0.08$ &  $4.89 \pm 0.07$ &  $4.79 \pm  0.09$ &  $3.96 \pm 0.12$ &  $5.50 \pm  0.06$  \\
  8737032 &  $6.23 \pm  0.01$ &  $7.63 \pm 0.08$ &  $6.41 \pm 0.01$ &  $7.36 \pm 0.02$ &  $6.18 \pm 0.07$ &  $4.83 \pm 0.07$ &  $4.83 \pm  0.10$ &  $3.93 \pm 0.08$ &  $5.50 \pm  0.06$  \\
  9080175 &  $5.98 \pm  0.04$ &  $7.53 \pm 0.02$ &  $6.33 \pm 0.04$ &  $7.14 \pm 0.01$ &  $5.99 \pm 0.07$ &  $4.75 \pm 0.06$ &  $4.71 \pm  0.08$ &  $3.79 \pm 0.09$ &  $5.19 \pm  0.05$  \\
  9145955 &  $6.10 \pm  0.06$ &  $7.58 \pm 0.14$ &  $6.24 \pm 0.01$ &  $7.22 \pm 0.06$ &  $6.12 \pm 0.06$ &  $4.79 \pm 0.07$ &  $4.80 \pm  0.07$ &  $3.81 \pm 0.13$ &  $5.38 \pm  0.05$  \\
  9535399 &  $6.67 \pm  0.12$ &  $7.77 \pm 0.19$ &  $6.78 \pm 0.08$ &  $7.57 \pm 0.06$ &  $6.49 \pm 0.06$ &  $5.20 \pm 0.08$ &  $5.24 \pm  0.19$ &  $4.44 \pm 0.07$ &  $5.86 \pm  0.07$  \\
  9640480 &  $6.09 \pm  0.04$ &  $7.58 \pm 0.01$ &  $6.32 \pm 0.20$ &  $7.19 \pm 0.01$ &  $6.14 \pm 0.04$ &  $4.89 \pm 0.06$ &  $4.78 \pm  0.09$ &  $3.98 \pm 0.10$ &  $5.31 \pm  0.06$  \\
  9711269 &  $6.53 \pm  0.07$ &  $7.83 \pm 0.02$ &  $6.67 \pm 0.16$ &  $7.45 \pm 0.01$ &  $6.48 \pm 0.06$ &  $5.19 \pm 0.08$ &  $5.09 \pm  0.18$ &  $4.41 \pm 0.08$ &  $5.84 \pm  0.07$  \\
  9772366 &  $6.84 \pm  0.10$ &  $7.89 \pm 0.04$ &  $6.73 \pm 0.12$ &  $7.54 \pm 0.02$ &  $6.51 \pm 0.06$ &  $5.24 \pm 0.08$ &  $5.15 \pm  0.17$ &  $4.46 \pm 0.08$ &  $5.90 \pm  0.06$  \\
  9777293 &  $5.79 \pm  0.04$ &  $7.46 \pm 0.01$ &  $     --      $ &  $7.00 \pm 0.05$ &  $5.95 \pm 0.05$ &  $4.70 \pm 0.06$ &  $4.66 \pm  0.06$ &  $3.61 \pm 0.07$ &  $5.09 \pm  0.04$  \\
  9783226 &  $6.54 \pm  0.08$ &  $7.75 \pm 0.08$ &  $6.60 \pm 0.01$ &  $7.44 \pm 0.08$ &  $6.38 \pm 0.06$ &  $5.03 \pm 0.08$ &  $5.00 \pm  0.14$ &  $4.21 \pm 0.10$ &  $5.69 \pm  0.06$  \\
  9967700 &  $5.94 \pm  0.06$ &  $7.38 \pm 0.02$ &  $6.08 \pm 0.01$ &  $7.13 \pm 0.01$ &  $5.99 \pm 0.04$ &  $4.67 \pm 0.05$ &  $4.68 \pm  0.07$ &  $3.67 \pm 0.09$ &  $5.22 \pm  0.05$  \\
 10095427 &  $5.99 \pm  0.05$ &  $7.56 \pm 0.04$ &  $6.23 \pm 0.18$ &  $7.19 \pm 0.01$ &  $6.03 \pm 0.06$ &  $4.76 \pm 0.06$ &  $4.82 \pm  0.07$ &  $3.80 \pm 0.09$ &  $5.16 \pm  0.05$  \\
 10351820 &  $6.40 \pm  0.09$ &  $7.66 \pm 0.08$ &  $6.46 \pm 0.02$ &  $7.42 \pm 0.07$ &  $6.34 \pm 0.07$ &  $4.93 \pm 0.08$ &  $4.93 \pm  0.17$ &  $4.09 \pm 0.09$ &  $5.59 \pm  0.07$  \\
 10513837 &  $6.50 \pm  0.08$ &  $7.79 \pm 0.07$ &  $6.57 \pm 0.01$ &  $7.49 \pm 0.06$ &  $6.46 \pm 0.07$ &  $5.10 \pm 0.08$ &  $5.02 \pm  0.16$ &  $4.24 \pm 0.11$ &  $5.75 \pm  0.08$  \\
 10550429 &  $6.47 \pm  0.06$ &  $7.58 \pm 0.14$ &  $6.55 \pm 0.07$ &  $7.36 \pm 0.01$ &  $6.38 \pm 0.04$ &  $5.09 \pm 0.07$ &  $4.97 \pm  0.13$ &  $4.31 \pm 0.12$ &  $5.71 \pm  0.05$  \\
 10659842 &  $6.18 \pm  0.12$ &  $7.60 \pm 0.09$ &  $     --      $ &  $7.25 \pm 0.03$ &  $6.18 \pm 0.09$ &  $4.85 \pm 0.07$ &  $4.90 \pm  0.10$ &  $3.88 \pm 0.09$ &  $5.45 \pm  0.06$  \\
 10722175 &  $6.74 \pm  0.06$ &  $7.88 \pm 0.08$ &  $6.72 \pm 0.11$ &  $7.69 \pm 0.14$ &  $6.44 \pm 0.11$ &  $5.16 \pm 0.09$ &  $5.18 \pm  0.21$ &  $4.35 \pm 0.09$ &  $5.84 \pm  0.08$  \\
 10793771 &  $6.52 \pm  0.09$ &  $7.85 \pm 0.09$ &  $6.62 \pm 0.01$ &  $7.46 \pm 0.04$ &  $6.46 \pm 0.08$ &  $5.10 \pm 0.08$ &  $4.93 \pm  0.12$ &  $4.25 \pm 0.10$ &  $5.77 \pm  0.08$  \\
 10801063 &  $6.08 \pm  0.05$ &  $7.70 \pm 0.07$ &  $6.31 \pm 0.17$ &  $7.22 \pm 0.01$ &  $6.17 \pm 0.04$ &  $4.94 \pm 0.05$ &  $4.92 \pm  0.08$ &  $3.96 \pm 0.10$ &  $5.34 \pm  0.05$  \\
 10992711 &  $6.10 \pm  0.05$ &  $7.62 \pm 0.02$ &  $6.30 \pm 0.27$ &  $7.24 \pm 0.01$ &  $6.08 \pm 0.07$ &  $4.82 \pm 0.06$ &  $4.79 \pm  0.09$ &  $3.88 \pm 0.07$ &  $5.30 \pm  0.06$  \\
 11020211 &  $6.01 \pm  0.05$ &  $7.54 \pm 0.01$ &  $6.33 \pm 0.02$ &  $7.16 \pm 0.01$ &  $6.09 \pm 0.04$ &  $4.84 \pm 0.05$ &  $4.79 \pm  0.08$ &  $3.84 \pm 0.08$ &  $5.26 \pm  0.05$  \\
 11029423 &  $6.30 \pm  0.08$ &  $7.74 \pm 0.01$ &  $6.59 \pm 0.06$ &  $7.43 \pm 0.01$ &  $6.26 \pm 0.05$ &  $4.96 \pm 0.07$ &  $4.99 \pm  0.13$ &  $4.09 \pm 0.07$ &  $5.57 \pm  0.07$  \\
 11087371 &  $6.72 \pm  0.06$ &  $7.92 \pm 0.01$ &  $6.73 \pm 0.12$ &  $7.57 \pm 0.01$ &  $6.49 \pm 0.06$ &  $5.19 \pm 0.08$ &  $5.17 \pm  0.17$ &  $4.46 \pm 0.09$ &  $5.87 \pm  0.08$  \\
 11228549 &  $5.63 \pm  0.04$ &  $7.19 \pm 0.07$ &  $6.01 \pm 0.01$ &  $6.88 \pm 0.01$ &  $5.72 \pm 0.03$ &  $4.45 \pm 0.06$ &  $4.46 \pm  0.05$ &  $3.35 \pm 0.06$ &  $4.82 \pm  0.04$  \\
 11250139 &  $5.97 \pm  0.05$ &  $7.47 \pm 0.04$ &  $6.22 \pm 0.14$ &  $7.16 \pm 0.02$ &  $5.98 \pm 0.06$ &  $4.72 \pm 0.06$ &  $4.73 \pm  0.08$ &  $3.77 \pm 0.09$ &  $5.18 \pm  0.06$  \\
 11358669 &  $6.66 \pm  0.01$ &  $7.89 \pm 0.07$ &  $6.71 \pm 0.07$ &  $7.57 \pm 0.04$ &  $6.58 \pm 0.07$ &  $5.25 \pm 0.08$ &  $5.05 \pm  0.14$ &  $4.45 \pm 0.09$ &  $5.90 \pm  0.06$  \\
 11453721 &  $6.11 \pm  0.06$ &  $7.66 \pm 0.06$ &  $6.38 \pm 0.05$ &  $7.26 \pm 0.01$ &  $6.02 \pm 0.08$ &  $4.76 \pm 0.06$ &  $4.82 \pm  0.10$ &  $3.87 \pm 0.08$ &  $5.30 \pm  0.06$  \\
 11496569 &  $6.46 \pm  0.06$ &  $7.73 \pm 0.08$ &  $6.60 \pm 0.01$ &  $7.46 \pm 0.01$ &  $6.41 \pm 0.08$ &  $5.11 \pm 0.09$ &  $4.96 \pm  0.18$ &  $4.28 \pm 0.07$ &  $5.75 \pm  0.06$  \\
 11550492 &  $6.33 \pm  0.06$ &  $7.67 \pm 0.01$ &  $6.53 \pm 0.02$ &  $7.31 \pm 0.02$ &  $6.30 \pm 0.06$ &  $5.02 \pm 0.07$ &  $4.93 \pm  0.12$ &  $4.19 \pm 0.10$ &  $5.65 \pm  0.07$  \\
 11702195 &  $6.01 \pm  0.06$ &  $7.58 \pm 0.03$ &  $6.41 \pm 0.01$ &  $7.18 \pm 0.01$ &  $6.03 \pm 0.05$ &  $4.75 \pm 0.05$ &  $4.78 \pm  0.09$ &  $3.81 \pm 0.10$ &  $5.20 \pm  0.06$  \\
 11775511 &  $6.09 \pm  0.05$ &  $7.54 \pm 0.01$ &  $6.31 \pm 0.13$ &  $7.18 \pm 0.01$ &  $6.14 \pm 0.05$ &  $4.88 \pm 0.06$ &  $4.78 \pm  0.09$ &  $3.99 \pm 0.10$ &  $5.38 \pm  0.06$  \\
 11819363 &  $6.04 \pm  0.14$ &  $7.45 \pm 0.12$ &  $6.22 \pm 0.01$ &  $7.20 \pm 0.01$ &  $6.01 \pm 0.11$ &  $4.69 \pm 0.06$ &  $4.73 \pm  0.12$ &  $3.73 \pm 0.11$ &  $5.25 \pm  0.09$  \\
 12115227 &  $5.69 \pm  0.03$ &  $7.35 \pm 0.02$ &  $6.02 \pm 0.04$ &  $6.99 \pm 0.05$ &  $5.81 \pm 0.04$ &  $4.53 \pm 0.05$ &  $4.55 \pm  0.04$ &  $3.39 \pm 0.06$ &  $4.90 \pm  0.06$  \\
 12122151 &  $6.34 \pm  0.06$ &  $7.66 \pm 0.08$ &  $6.45 \pm 0.01$ &  $7.41 \pm 0.02$ &  $6.28 \pm 0.07$ &  $4.88 \pm 0.06$ &  $4.88 \pm  0.13$ &  $4.00 \pm 0.08$ &  $5.59 \pm  0.08$  \\
 12647227 &  $5.90 \pm  0.05$ &  $7.20 \pm 0.07$ &  $6.06 \pm 0.01$ &  $7.02 \pm 0.02$ &  $5.83 \pm 0.03$ &  $4.49 \pm 0.05$ &  $4.57 \pm  0.08$ &  $3.49 \pm 0.07$ &  $5.09 \pm  0.03$  \\
  \hline
\end{tabular}
\label{tab:abu}
\end{center}
\end{table*}

\begin{table*}
\ContinuedFloat
\caption{Continued}
\begin{center}
\scriptsize
\begin{tabular}{cccccccccc}
 \hline
 \hline
 Name &  FeI/H          & FeII/H         & Co/H           & Ni/H            &  Y/H          &  Zr/H           &  Ba/H            & La/H            & Ce/H          \\ 
 \hline
  1433803 &  $7.64 \pm 0.13$ &  $7.76 \pm 0.22$ &  $5.39 \pm 0.13$ &  $6.55 \pm  0.10$ &  $2.28 \pm 0.04$ &  $2.64 \pm 0.06$  &   $1.87 \pm 0.14$ &   $1.03 \pm 0.08$ &  $1.84 \pm 0.09$ \\
  2451509 &  $7.35 \pm 0.12$ &  $7.21 \pm 0.12$ &  $4.84 \pm 0.11$ &  $6.08 \pm  0.08$ &  $2.00 \pm 0.13$ &  $2.71 \pm 0.01$  &   $1.85 \pm 0.09$ &   $0.91 \pm 0.13$ &  $1.71 \pm 0.13$ \\
  2970584 &  $7.17 \pm 0.12$ &  $7.00 \pm 0.12$ &  $4.89 \pm 0.12$ &  $6.01 \pm  0.10$ &  $1.71 \pm 0.03$ &  $2.48 \pm 0.02$  &   $1.44 \pm 0.04$ &   $0.61 \pm 0.04$ &  $1.32 \pm 0.01$ \\
  3429738 &  $7.09 \pm 0.09$ &  $6.99 \pm 0.11$ &  $4.79 \pm 0.11$ &  $5.90 \pm  0.07$ &  $1.62 \pm 0.02$ &  $2.25 \pm 0.01$  &   $1.36 \pm 0.03$ &   $0.58 \pm 0.02$ &  $1.22 \pm 0.01$ \\
  3539408 &  $7.27 \pm 0.08$ &  $7.14 \pm 0.06$ &  $4.74 \pm 0.11$ &  $6.04 \pm  0.08$ &  $1.93 \pm 0.01$ &  $2.47 \pm 0.02$  &   $1.93 \pm 0.03$ &   $0.88 \pm 0.01$ &  $1.54 \pm 0.01$ \\
  3644223 &  $7.29 \pm 0.12$ &  $7.25 \pm 0.13$ &  $4.83 \pm 0.13$ &  $6.05 \pm  0.08$ &  $1.85 \pm 0.05$ &  $2.36 \pm 0.03$  &   $1.80 \pm 0.09$ &   $0.80 \pm 0.13$ &  $1.50 \pm 0.16$ \\
  3661494 &  $7.12 \pm 0.08$ &  $6.99 \pm 0.06$ &  $4.65 \pm 0.14$ &  $5.92 \pm  0.08$ &  $1.66 \pm 0.03$ &  $2.22 \pm 0.02$  &   $1.59 \pm 0.03$ &   $0.58 \pm 0.03$ &  $1.22 \pm 0.01$ \\
  3744043 &  $7.14 \pm 0.08$ &  $6.99 \pm 0.08$ &  $4.66 \pm 0.12$ &  $5.93 \pm  0.08$ &  $1.70 \pm 0.03$ &  $     --      $  &   $1.65 \pm 0.03$ &   $0.67 \pm 0.01$ &  $1.30 \pm 0.01$ \\
  4055294 &  $7.67 \pm 0.12$ &  $7.37 \pm 0.18$ &  $5.31 \pm 0.16$ &  $6.47 \pm  0.12$ &  $2.12 \pm 0.09$ &  $2.96 \pm 0.01$  &   $1.92 \pm 0.11$ &   $0.97 \pm 0.07$ &  $1.75 \pm 0.10$ \\
  4648485 &  $7.42 \pm 0.10$ &  $7.32 \pm 0.10$ &  $5.06 \pm 0.13$ &  $6.23 \pm  0.12$ &  $1.95 \pm 0.06$ &  $2.57 \pm 0.01$  &   $1.80 \pm 0.04$ &   $0.90 \pm 0.01$ &  $1.59 \pm 0.02$ \\
  4756219 &  $7.25 \pm 0.10$ &  $7.21 \pm 0.11$ &  $4.88 \pm 0.16$ &  $6.04 \pm  0.07$ &  $1.79 \pm 0.02$ &  $2.32 \pm 0.01$  &   $1.60 \pm 0.04$ &   $0.71 \pm 0.01$ &  $1.41 \pm 0.01$ \\
  4826087 &  $7.40 \pm 0.12$ &  $7.25 \pm 0.19$ &  $4.99 \pm 0.10$ &  $6.16 \pm  0.08$ &  $2.07 \pm 0.07$ &  $2.73 \pm 0.03$  &   $1.91 \pm 0.10$ &   $0.96 \pm 0.05$ &  $1.76 \pm 0.17$ \\
  4913049 &  $7.43 \pm 0.09$ &  $7.17 \pm 0.11$ &  $5.19 \pm 0.21$ &  $6.23 \pm  0.10$ &  $2.02 \pm 0.01$ &  $2.69 \pm 0.01$  &   $1.97 \pm 0.03$ &   $0.88 \pm 0.02$ &  $1.55 \pm 0.01$ \\
  4931389 &  $7.61 \pm 0.10$ &  $7.33 \pm 0.09$ &  $5.23 \pm 0.14$ &  $6.42 \pm  0.09$ &  $2.14 \pm 0.03$ &  $2.88 \pm 0.01$  &   $2.06 \pm 0.04$ &   $0.98 \pm 0.05$ &  $1.64 \pm 0.04$ \\
  5265256 &  $7.24 \pm 0.10$ &  $7.16 \pm 0.10$ &  $4.81 \pm 0.12$ &  $5.99 \pm  0.08$ &  $1.75 \pm 0.04$ &  $2.34 \pm 0.02$  &   $1.56 \pm 0.03$ &   $0.71 \pm 0.01$ &  $1.45 \pm 0.08$ \\
  5769244 &  $7.58 \pm 0.10$ &  $7.44 \pm 0.16$ &  $5.28 \pm 0.17$ &  $6.43 \pm  0.14$ &  $2.24 \pm 0.01$ &  $2.74 \pm 0.01$  &   $2.09 \pm 0.04$ &   $0.94 \pm 0.04$ &  $1.70 \pm 0.01$ \\
  5882005 &  $7.34 \pm 0.11$ &  $7.27 \pm 0.14$ &  $4.88 \pm 0.10$ &  $6.07 \pm  0.10$ &  $1.97 \pm 0.15$ &  $2.46 \pm 0.01$  &   $1.78 \pm 0.11$ &   $0.86 \pm 0.13$ &  $1.59 \pm 0.10$ \\
  5940060 &  $7.05 \pm 0.10$ &  $6.95 \pm 0.10$ &  $4.76 \pm 0.11$ &  $5.87 \pm  0.10$ &  $1.60 \pm 0.06$ &  $2.30 \pm 0.02$  &   $1.39 \pm 0.04$ &   $0.57 \pm 0.01$ &  $1.20 \pm 0.01$ \\
  6365511 &  $6.88 \pm 0.09$ &  $6.79 \pm 0.08$ &  $4.52 \pm 0.11$ &  $5.72 \pm  0.06$ &  $1.29 \pm 0.04$ &  $2.12 \pm 0.01$  &   $1.17 \pm 0.03$ &   $0.41 \pm 0.03$ &  $0.96 \pm 0.01$ \\
  6547007 &  $6.68 \pm 0.08$ &  $6.66 \pm 0.08$ &  $4.29 \pm 0.08$ &  $5.47 \pm  0.06$ &  $1.16 \pm 0.02$ &  $     --      $  &   $1.05 \pm 0.03$ &   $0.29 \pm 0.03$ &  $0.85 \pm 0.03$ \\
  6851499 &  $7.11 \pm 0.10$ &  $7.05 \pm 0.09$ &  $4.71 \pm 0.10$ &  $5.91 \pm  0.11$ &  $1.66 \pm 0.04$ &  $2.27 \pm 0.01$  &   $1.49 \pm 0.03$ &   $0.60 \pm 0.02$ &  $1.29 \pm 0.02$ \\
  6859803 &  $7.33 \pm 0.12$ &  $7.22 \pm 0.13$ &  $4.83 \pm 0.11$ &  $6.09 \pm  0.14$ &  $2.13 \pm 0.17$ &  $2.59 \pm 0.01$  &   $1.87 \pm 0.08$ &   $0.97 \pm 0.01$ &  $1.60 \pm 0.14$ \\
  6964342 &  $7.41 \pm 0.13$ &  $7.36 \pm 0.22$ &  $5.08 \pm 0.13$ &  $6.25 \pm  0.14$ &  $2.06 \pm 0.18$ &  $2.63 \pm 0.01$  &   $1.69 \pm 0.12$ &   $0.88 \pm 0.07$ &  $1.66 \pm 0.09$ \\
  7429055 &  $7.41 \pm 0.10$ &  $7.33 \pm 0.13$ &  $5.13 \pm 0.18$ &  $6.26 \pm  0.10$ &  $2.08 \pm 0.01$ &  $2.53 \pm 0.02$  &   $1.80 \pm 0.03$ &   $0.83 \pm 0.02$ &  $1.60 \pm 0.08$ \\
  7430868 &  $7.16 \pm 0.11$ &  $7.12 \pm 0.11$ &  $4.88 \pm 0.14$ &  $6.00 \pm  0.10$ &  $1.77 \pm 0.03$ &  $2.26 \pm 0.01$  &   $1.53 \pm 0.03$ &   $0.63 \pm 0.02$ &  $1.29 \pm 0.02$ \\
  7450230 &  $7.37 \pm 0.08$ &  $7.17 \pm 0.08$ &  $5.00 \pm 0.16$ &  $6.16 \pm  0.08$ &  $1.93 \pm 0.02$ &  $2.51 \pm 0.01$  &   $1.84 \pm 0.03$ &   $0.77 \pm 0.05$ &  $1.48 \pm 0.01$ \\
  7533995 &  $7.42 \pm 0.11$ &  $7.36 \pm 0.11$ &  $5.09 \pm 0.17$ &  $6.26 \pm  0.11$ &  $2.06 \pm 0.05$ &  $2.51 \pm 0.04$  &   $1.86 \pm 0.03$ &   $0.86 \pm 0.03$ &  $1.58 \pm 0.01$ \\
  7617227 &  $7.17 \pm 0.08$ &  $7.07 \pm 0.06$ &  $4.69 \pm 0.13$ &  $5.97 \pm  0.07$ &  $1.74 \pm 0.04$ &  $2.28 \pm 0.01$  &   $1.66 \pm 0.04$ &   $0.59 \pm 0.05$ &  $1.30 \pm 0.01$ \\
  7812552 &  $6.98 \pm 0.08$ &  $6.80 \pm 0.07$ &  $4.57 \pm 0.09$ &  $5.79 \pm  0.07$ &  $1.53 \pm 0.02$ &  $     --      $  &   $1.37 \pm 0.01$ &   $     --      $ &  $     --      $ \\
  8129047 &  $7.42 \pm 0.11$ &  $7.18 \pm 0.11$ &  $5.12 \pm 0.20$ &  $6.27 \pm  0.11$ &  $2.16 \pm 0.01$ &  $2.67 \pm 0.02$  &   $1.92 \pm 0.03$ &   $0.85 \pm 0.01$ &  $1.59 \pm 0.05$ \\
  8493969 &  $7.16 \pm 0.09$ &  $7.00 \pm 0.08$ &  $4.84 \pm 0.16$ &  $5.99 \pm  0.09$ &  $1.73 \pm 0.02$ &  $2.33 \pm 0.01$  &   $1.59 \pm 0.03$ &   $0.69 \pm 0.03$ &  $1.23 \pm 0.03$ \\
  8587329 &  $7.64 \pm 0.12$ &  $7.86 \pm 0.21$ &  $5.39 \pm 0.17$ &  $6.54 \pm  0.11$ &  $2.38 \pm 0.13$ &  $2.45 \pm 0.03$  &   $2.07 \pm 0.05$ &   $0.96 \pm 0.07$ &  $1.82 \pm 0.05$ \\
  8590920 &  $7.68 \pm 0.11$ &  $7.83 \pm 0.26$ &  $5.47 \pm 0.16$ &  $6.59 \pm  0.11$ &  $2.42 \pm 0.11$ &  $2.71 \pm 0.01$  &   $1.98 \pm 0.07$ &   $0.87 \pm 0.05$ &  $1.82 \pm 0.11$ \\
  8612241 &  $7.27 \pm 0.12$ &  $7.09 \pm 0.13$ &  $4.87 \pm 0.11$ &  $6.03 \pm  0.09$ &  $1.79 \pm 0.15$ &  $2.40 \pm 0.01$  &   $1.57 \pm 0.07$ &   $0.83 \pm 0.11$ &  $1.44 \pm 0.07$ \\
  8737032 &  $7.29 \pm 0.11$ &  $7.23 \pm 0.16$ &  $4.87 \pm 0.12$ &  $6.05 \pm  0.10$ &  $1.89 \pm 0.07$ &  $2.50 \pm 0.02$  &   $1.68 \pm 0.10$ &   $0.88 \pm 0.07$ &  $1.58 \pm 0.14$ \\
  9080175 &  $6.95 \pm 0.08$ &  $6.85 \pm 0.09$ &  $4.71 \pm 0.14$ &  $5.78 \pm  0.08$ &  $1.45 \pm 0.04$ &  $2.17 \pm 0.01$  &   $1.22 \pm 0.03$ &   $0.36 \pm 0.06$ &  $1.04 \pm 0.04$ \\
  9145955 &  $7.20 \pm 0.11$ &  $7.09 \pm 0.10$ &  $4.74 \pm 0.13$ &  $5.96 \pm  0.06$ &  $1.72 \pm 0.03$ &  $     --      $  &   $1.75 \pm 0.12$ &   $0.77 \pm 0.10$ &  $1.48 \pm 0.04$ \\
  9535399 &  $7.62 \pm 0.15$ &  $7.55 \pm 0.20$ &  $5.37 \pm 0.23$ &  $6.54 \pm  0.12$ &  $2.37 \pm 0.16$ &  $2.75 \pm 0.01$  &   $2.11 \pm 0.04$ &   $0.92 \pm 0.07$ &  $1.73 \pm 0.06$ \\
  9640480 &  $7.10 \pm 0.08$ &  $6.87 \pm 0.09$ &  $4.85 \pm 0.18$ &  $5.90 \pm  0.08$ &  $1.69 \pm 0.02$ &  $2.35 \pm 0.01$  &   $1.49 \pm 0.03$ &   $0.58 \pm 0.04$ &  $1.17 \pm 0.02$ \\
  9711269 &  $7.57 \pm 0.10$ &  $7.37 \pm 0.13$ &  $5.22 \pm 0.21$ &  $6.40 \pm  0.11$ &  $2.28 \pm 0.03$ &  $2.79 \pm 0.01$  &   $2.07 \pm 0.03$ &   $0.94 \pm 0.01$ &  $1.69 \pm 0.05$ \\
  9772366 &  $7.64 \pm 0.11$ &  $7.49 \pm 0.15$ &  $5.37 \pm 0.21$ &  $6.54 \pm  0.10$ &  $2.36 \pm 0.02$ &  $2.80 \pm 0.01$  &   $2.09 \pm 0.05$ &   $0.91 \pm 0.06$ &  $1.67 \pm 0.15$ \\
  9777293 &  $6.89 \pm 0.09$ &  $6.77 \pm 0.09$ &  $4.54 \pm 0.09$ &  $5.70 \pm  0.06$ &  $1.43 \pm 0.02$ &  $     --      $  &   $1.19 \pm 0.03$ &   $     --      $ &  $0.96 \pm 0.01$ \\
  9783226 &  $7.48 \pm 0.13$ &  $7.32 \pm 0.19$ &  $5.05 \pm 0.14$ &  $6.30 \pm  0.11$ &  $2.17 \pm 0.08$ &  $2.64 \pm 0.01$  &   $1.97 \pm 0.11$ &   $0.86 \pm 0.08$ &  $1.68 \pm 0.06$ \\
  9967700 &  $7.08 \pm 0.08$ &  $6.96 \pm 0.07$ &  $4.63 \pm 0.09$ &  $5.87 \pm  0.10$ &  $1.63 \pm 0.01$ &  $     --      $  &   $1.62 \pm 0.03$ &   $0.53 \pm 0.01$ &  $1.21 \pm 0.01$ \\
 10095427 &  $6.95 \pm 0.09$ &  $6.87 \pm 0.10$ &  $4.69 \pm 0.15$ &  $5.79 \pm  0.07$ &  $1.54 \pm 0.04$ &  $2.20 \pm 0.02$  &   $1.23 \pm 0.04$ &   $0.33 \pm 0.06$ &  $0.95 \pm 0.01$ \\
 10351820 &  $7.38 \pm 0.12$ &  $7.29 \pm 0.16$ &  $4.92 \pm 0.10$ &  $6.14 \pm  0.10$ &  $2.05 \pm 0.08$ &  $2.59 \pm 0.02$  &   $1.83 \pm 0.09$ &   $0.91 \pm 0.15$ &  $1.66 \pm 0.03$ \\
 10513837 &  $7.54 \pm 0.12$ &  $7.41 \pm 0.16$ &  $5.09 \pm 0.10$ &  $6.32 \pm  0.10$ &  $2.14 \pm 0.01$ &  $2.83 \pm 0.03$  &   $1.88 \pm 0.11$ &   $0.93 \pm 0.12$ &  $1.71 \pm 0.02$ \\
 10550429 &  $7.52 \pm 0.09$ &  $7.29 \pm 0.10$ &  $5.07 \pm 0.15$ &  $6.30 \pm  0.10$ &  $2.18 \pm 0.10$ &  $2.80 \pm 0.02$  &   $2.03 \pm 0.05$ &   $0.92 \pm 0.05$ &  $1.60 \pm 0.01$ \\
 10659842 &  $7.28 \pm 0.11$ &  $7.12 \pm 0.14$ &  $4.81 \pm 0.13$ &  $6.01 \pm  0.10$ &  $1.90 \pm 0.04$ &  $2.53 \pm 0.02$  &   $1.87 \pm 0.09$ &   $0.86 \pm 0.09$ &  $1.53 \pm 0.19$ \\
 10722175 &  $7.61 \pm 0.13$ &  $7.59 \pm 0.20$ &  $5.30 \pm 0.14$ &  $6.50 \pm  0.11$ &  $2.32 \pm 0.08$ &  $2.78 \pm 0.02$  &   $1.82 \pm 0.16$ &   $0.95 \pm 0.15$ &  $1.74 \pm 0.02$ \\
 10793771 &  $7.55 \pm 0.11$ &  $7.39 \pm 0.18$ &  $5.10 \pm 0.12$ &  $6.30 \pm  0.10$ &  $2.19 \pm 0.10$ &  $2.85 \pm 0.01$  &   $2.02 \pm 0.11$ &   $1.03 \pm 0.08$ &  $1.74 \pm 0.08$ \\
 10801063 &  $7.15 \pm 0.09$ &  $6.97 \pm 0.08$ &  $4.81 \pm 0.13$ &  $5.97 \pm  0.10$ &  $1.76 \pm 0.04$ &  $2.33 \pm 0.02$  &   $1.61 \pm 0.03$ &   $     --      $ &  $1.19 \pm 0.01$ \\
 10992711 &  $7.06 \pm 0.10$ &  $6.95 \pm 0.13$ &  $4.79 \pm 0.13$ &  $5.88 \pm  0.08$ &  $1.61 \pm 0.01$ &  $2.25 \pm 0.01$  &   $1.35 \pm 0.03$ &   $0.50 \pm 0.02$ &  $1.19 \pm 0.02$ \\
 11020211 &  $7.10 \pm 0.08$ &  $6.94 \pm 0.09$ &  $4.71 \pm 0.14$ &  $5.91 \pm  0.08$ &  $1.68 \pm 0.05$ &  $2.29 \pm 0.06$  &   $1.53 \pm 0.03$ &   $0.51 \pm 0.01$ &  $1.13 \pm 0.01$ \\
 11029423 &  $7.34 \pm 0.10$ &  $7.27 \pm 0.12$ &  $5.05 \pm 0.15$ &  $6.14 \pm  0.08$ &  $1.92 \pm 0.03$ &  $2.45 \pm 0.01$  &   $1.74 \pm 0.04$ &   $0.78 \pm 0.02$ &  $1.53 \pm 0.01$ \\
 11087371 &  $7.64 \pm 0.10$ &  $7.55 \pm 0.13$ &  $5.43 \pm 0.19$ &  $6.54 \pm  0.10$ &  $2.30 \pm 0.02$ &  $     --      $  &   $2.00 \pm 0.04$ &   $0.81 \pm 0.03$ &  $1.69 \pm 0.01$ \\
 11228549 &  $6.67 \pm 0.09$ &  $6.59 \pm 0.07$ &  $4.29 \pm 0.09$ &  $5.50 \pm  0.06$ &  $1.14 \pm 0.02$ &  $     --      $  &   $0.93 \pm 0.03$ &   $0.43 \pm 0.01$ &  $0.92 \pm 0.28$ \\
 11250139 &  $6.97 \pm 0.09$ &  $6.89 \pm 0.09$ &  $4.61 \pm 0.11$ &  $5.80 \pm  0.08$ &  $1.53 \pm 0.06$ &  $2.17 \pm 0.01$  &   $1.31 \pm 0.03$ &   $0.45 \pm 0.05$ &  $1.14 \pm 0.03$ \\
 11358669 &  $7.61 \pm 0.13$ &  $7.39 \pm 0.19$ &  $5.29 \pm 0.13$ &  $6.46 \pm  0.12$ &  $2.16 \pm 0.09$ &  $2.95 \pm 0.02$  &   $1.95 \pm 0.12$ &   $1.02 \pm 0.08$ &  $1.76 \pm 0.03$ \\
 11453721 &  $7.08 \pm 0.12$ &  $7.06 \pm 0.10$ &  $4.77 \pm 0.13$ &  $5.91 \pm  0.06$ &  $1.63 \pm 0.05$ &  $2.18 \pm 0.01$  &   $1.41 \pm 0.04$ &   $0.49 \pm 0.02$ &  $1.21 \pm 0.01$ \\
 11496569 &  $7.47 \pm 0.14$ &  $7.26 \pm 0.17$ &  $5.08 \pm 0.12$ &  $6.22 \pm  0.11$ &  $2.19 \pm 0.03$ &  $2.88 \pm 0.04$  &   $1.92 \pm 0.12$ &   $0.90 \pm 0.09$ &  $1.75 \pm 0.23$ \\
 11550492 &  $7.43 \pm 0.10$ &  $7.24 \pm 0.10$ &  $5.01 \pm 0.16$ &  $6.24 \pm  0.11$ &  $2.03 \pm 0.06$ &  $2.59 \pm 0.01$  &   $1.92 \pm 0.04$ &   $0.82 \pm 0.01$ &  $1.53 \pm 0.01$ \\
 11702195 &  $6.99 \pm 0.10$ &  $6.92 \pm 0.11$ &  $4.76 \pm 0.16$ &  $5.82 \pm  0.08$ &  $1.57 \pm 0.05$ &  $2.18 \pm 0.01$  &   $1.29 \pm 0.03$ &   $0.50 \pm 0.06$ &  $1.08 \pm 0.03$ \\
 11775511 &  $7.18 \pm 0.09$ &  $6.98 \pm 0.10$ &  $4.81 \pm 0.17$ &  $5.96 \pm  0.09$ &  $1.73 \pm 0.06$ &  $2.37 \pm 0.03$  &   $1.64 \pm 0.03$ &   $0.63 \pm 0.08$ &  $1.29 \pm 0.01$ \\
 11819363 &  $7.13 \pm 0.12$ &  $7.00 \pm 0.12$ &  $4.68 \pm 0.13$ &  $5.89 \pm  0.10$ &  $1.70 \pm 0.14$ &  $2.30 \pm 0.01$  &   $1.69 \pm 0.08$ &   $0.63 \pm 0.14$ &  $1.29 \pm 0.08$ \\
 12115227 &  $6.73 \pm 0.09$ &  $6.66 \pm 0.08$ &  $4.32 \pm 0.08$ &  $5.54 \pm  0.06$ &  $1.39 \pm 0.03$ &  $     --      $  &   $1.14 \pm 0.03$ &   $     --      $ &  $0.88 \pm 0.01$ \\
 12122151 &  $7.35 \pm 0.13$ &  $7.29 \pm 0.17$ &  $4.97 \pm 0.12$ &  $6.12 \pm  0.10$ &  $2.05 \pm 0.12$ &  $2.55 \pm 0.04$  &   $1.81 \pm 0.10$ &   $0.90 \pm 0.06$ &  $1.60 \pm 0.04$ \\
 12647227 &  $6.96 \pm 0.08$ &  $6.89 \pm 0.08$ &  $4.48 \pm 0.11$ &  $5.76 \pm  0.06$ &  $1.53 \pm 0.02$ &  $2.02 \pm 0.01$  &   $1.49 \pm 0.01$ &   $0.43 \pm 0.01$ &  $1.12 \pm 0.01$ \\
  \hline
\end{tabular}
\label{tab:abu}
\end{center}
\end{table*}

%

\subsection{Comparison with APOGEE DR17}
\label{sec:comp}
\begin{figure*}
\centering
\includegraphics[scale=0.5]{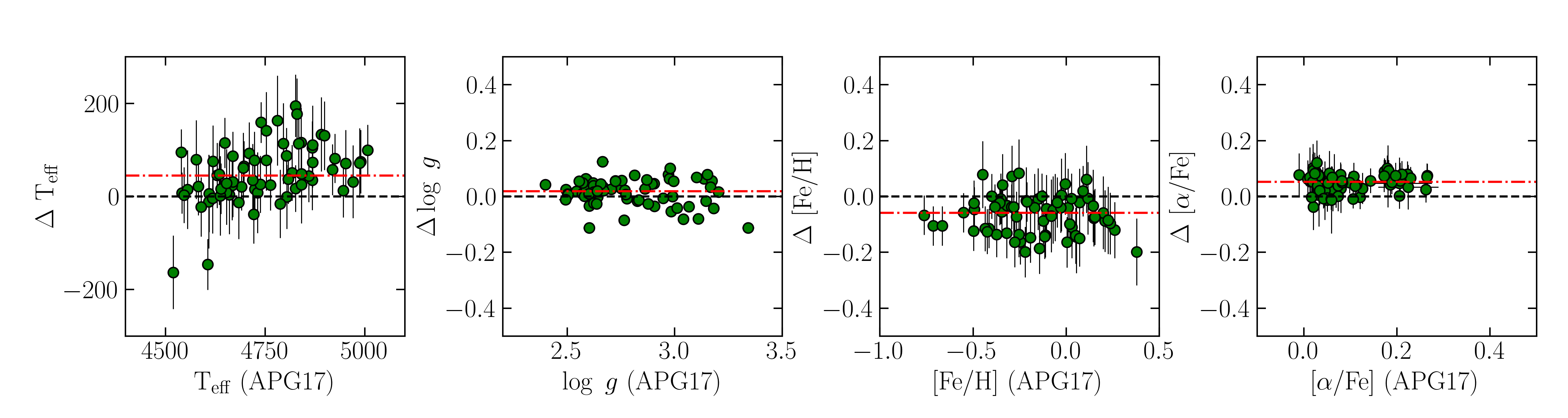}
\caption{Comparison between the atmospheric parameters and [$\alpha$/Fe] of APOGEE (APG17) and those found in this work for the sample of \kepler stars. The difference $\Delta$ on y-axis is (this work) - APG17. The red dash-dotted line is the average offset. \label{fig:comparisonapo}}
\end{figure*}

The atmospheric parameters and some abundances are available in APOGEE DR17 for our sample of stars. We compared our atmospheric parameters and [$\alpha$/Fe] with those available in APOGEE DR17. As we can see from Fig.~\ref{fig:comparisonapo}, \teff and [Fe/H] are in good agreement, without any systematic trend, but with an offset: our temperatures \teff are on average higher by $\sim 45$ K and our metallicities are on average lower by $\sim 0.05$ dex than those of APOGEE (similar offset are found by \citet{hegedus23} among APOGEE and the optical surveys GALAH and Gaia-ESO). The seismic $\log~g$ is in agreement with the $\log~g$ of APOGEE, which is calibrated on a set of surface gravities from asteroseismology and isochrones using \kepler stars \citep{apogeedr17}. Finally, [$\alpha$/Fe] shows an average-offset of 0.05 dex. The [$\alpha$/Fe] ratio is computed as $\rm 1/4([Ca/Fe] + [Mg/Fe] + [Si/Fe] + [Ti/Fe])$ in both cases.

\begin{figure*}
\centering
\includegraphics[scale=0.5]{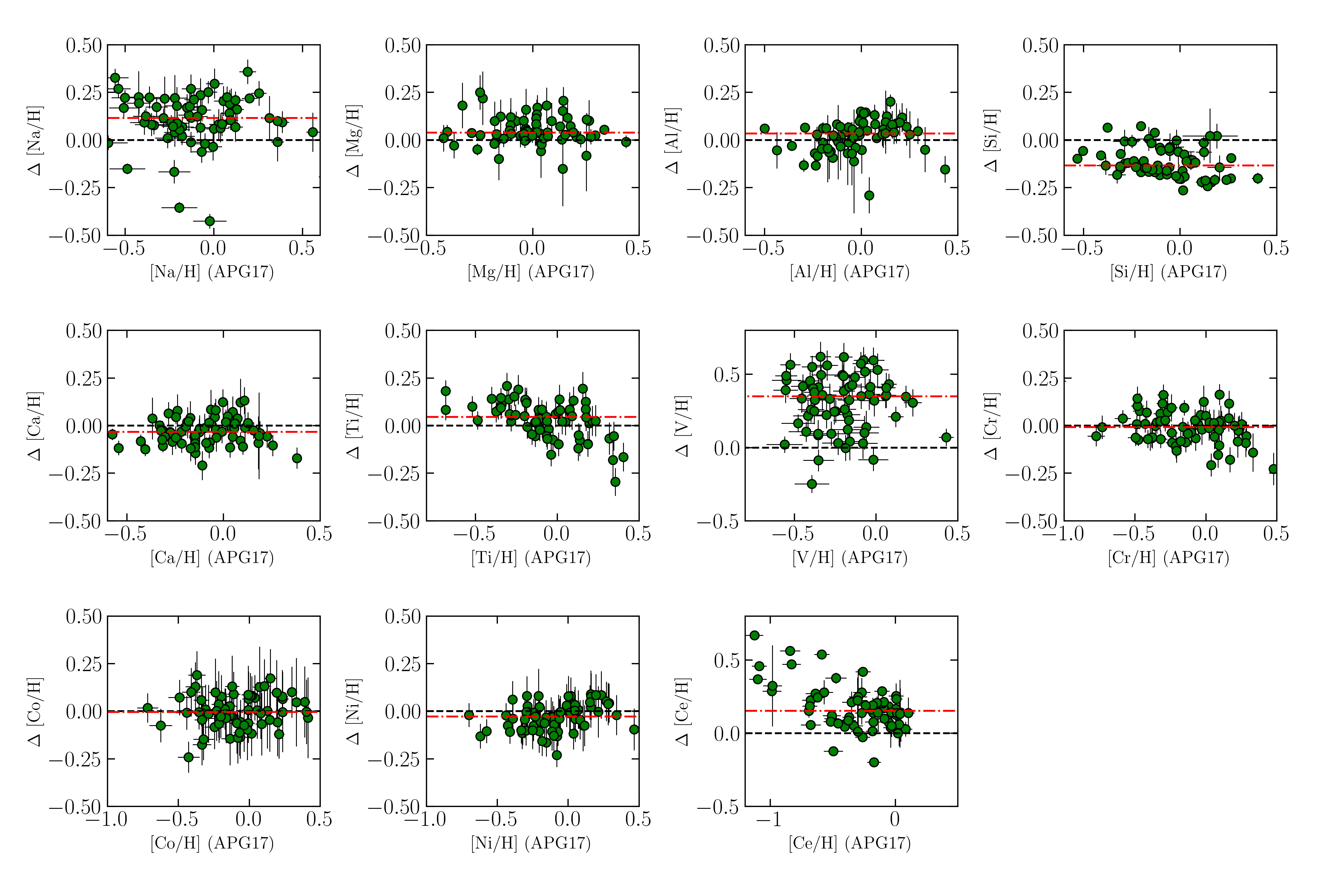}
\caption{Comparison between the elemental abundances of APOGEE (APG17) and those found in this work for the sample of \kepler stars. The difference $\Delta$ on y-axis is (this work) - APG17. The red dash-dotted line is the average offset. \label{fig:comparisonapoabu}}
\end{figure*}

Figure~\ref{fig:comparisonapoabu} shows the comparison between all the abundances in common with the APOGEE survey. 
The agreement is poor for Na, Si, V and Ce, whereas it is good for the other elements. APOGEE spectra contain two weak lines of sodium, close to telluric lines \citep{barbuy23}. Therefore, sodium is one of the least precisely determined element abundances in the APOGEE survey. Although Si is considered one of the most precisely measured elements in APOGEE, there is a large offset between our measurements and the APOGEE ones (offset of the order of 0.1 dex). This offset is related to the solar Si abundance reported by \citet{asplund09}. To investigate further, we analysed a solar spectrum collected with the HARPS spectrograph and we found excellent agreement with the \citet{asplund09} values for all elements in common (differences < 0.03 dex), except for silicon, where we found a difference of 0.07 dex (A(Si)$_\odot$ = 7.44).
By adopting this updated solar Si abundance, the offset with APOGEE measurements can be reduced to 0.06 dex. 
Vanadium abundances (0.3 dex of offset) have rather low precision in APOGEE and they should be used with caution \citep{jonsson20,hegedus23}. Finally, Ce in APOGEE DR17 \citep{cunha17,salessilva22,salessilva24} is lower by a factor of $0.1$ dex with respect to the abundances shown in this work, with a larger scatter at lower metallicity (\citet{hayes2022} found a metallicity-correlated offset for Ce at [Fe/H] < $-1$ dex, which lies outside the metallicity range considered in this work). This trend is already seen when the Ce of APOGEE is compared with the Ce from the optical survey {\em Gaia}-ESO \citep{casali23}.

\subsection{Age from asteroseismology}
\label{sec:ages}
The 4.5-year long \kepler light curves provide high-quality oscillation data, allowing for precise identification and characterisation of oscillation modes. 
The sample of stars, chosen from seismically classified RGBs in the \kepler field, was also manually inspected to ensure that any other potential signals in the light curve did not affect the extraction and characterization of oscillation modes.  
 Indeed, the presence of overlapping solar-like oscillation signals in the power spectral density, as well as harmonics from other classical pulsators in the field or signals due to stellar rotation, can interfere with the accurate identification of both oscillation modes and background components.
Seventy-five percent of the targets have seven or more identified radial modes. The frequencies of these modes, along with the frequency at maximum power ($\nu_{\rm max}$) and, when available, the asymptotic period spacing \citep[ referred to the regular spacing in period between gravity modes and derived from mixed dipolar modes by][]{vrard16}, 
are used as seismic observational constraints in the process of determining stellar parameters. For this purpose, we used the AIMS code \citep{aims,rendle19}, which implements a Bayesian inference approach to provide posterior probability distributions for the stellar parameters. Additional “classical” observational data included in the process are the effective temperature and the current metal abundance ([M/H]) in the photospheres of these stars. 
Given the constraints provided by asteroseismic frequencies, and the overall agreement of our stellar parameters with those in APOGEE shown in Sec.~\ref{sec:comp}, the impact of adopting one or the other set of "classical" stellar parameters is limited. For consistency with the more extensive catalogue of \kepler stars that will be presented in Montalbán et al. (in preparation), we have adopted effective temperature, [$\alpha$/M], and [M/H] from APOGEE-DR17. \citet{thomsen25} show that adopting different spectroscopic input values in the AIMS code does not significantly affect the inferred stellar ages. In their Table G.2, they report a difference of only 0.25$\sigma$ between the ages computed using atmospheric parameters from APOGEE and those derived from spectra collected with FIES and analysed in the same way as presented in this work\footnote{The differences in \teff\ and [Fe/H] between the two spectral analyses are 25~K and 0.05~dex, respectively. The APOGEE uncertainties were inflated to 60 K and 0.05 dex in the calculation as they are systematically underestimated in the survey, compared to the FIES uncertainties of 36~K and 0.07~dex.}.

Compared to the work presented in \cite{montalban21}, which focused on a low-metallicity sample, we have extended the range of chemical composition in the grid of models at the base of the method. In this study, we have adopted a grid computed with a helium enrichment law of $\Delta Y/\Delta Z=1.5$,  as adopted in several studies \cite[e.g.,][]{brogaard12, miglio21} and consistent 
with measurements \citep[e.g.,][]{casagrande07,verma19}.
For details on the impact of $\Delta Y/\Delta Z$ on the determination of stellar parameters, see Montalbán et al. (in preparation).

\begin{figure}
\centering
\includegraphics[trim={0.9cm 0cm 0cm 0cm}, clip,scale=0.4]{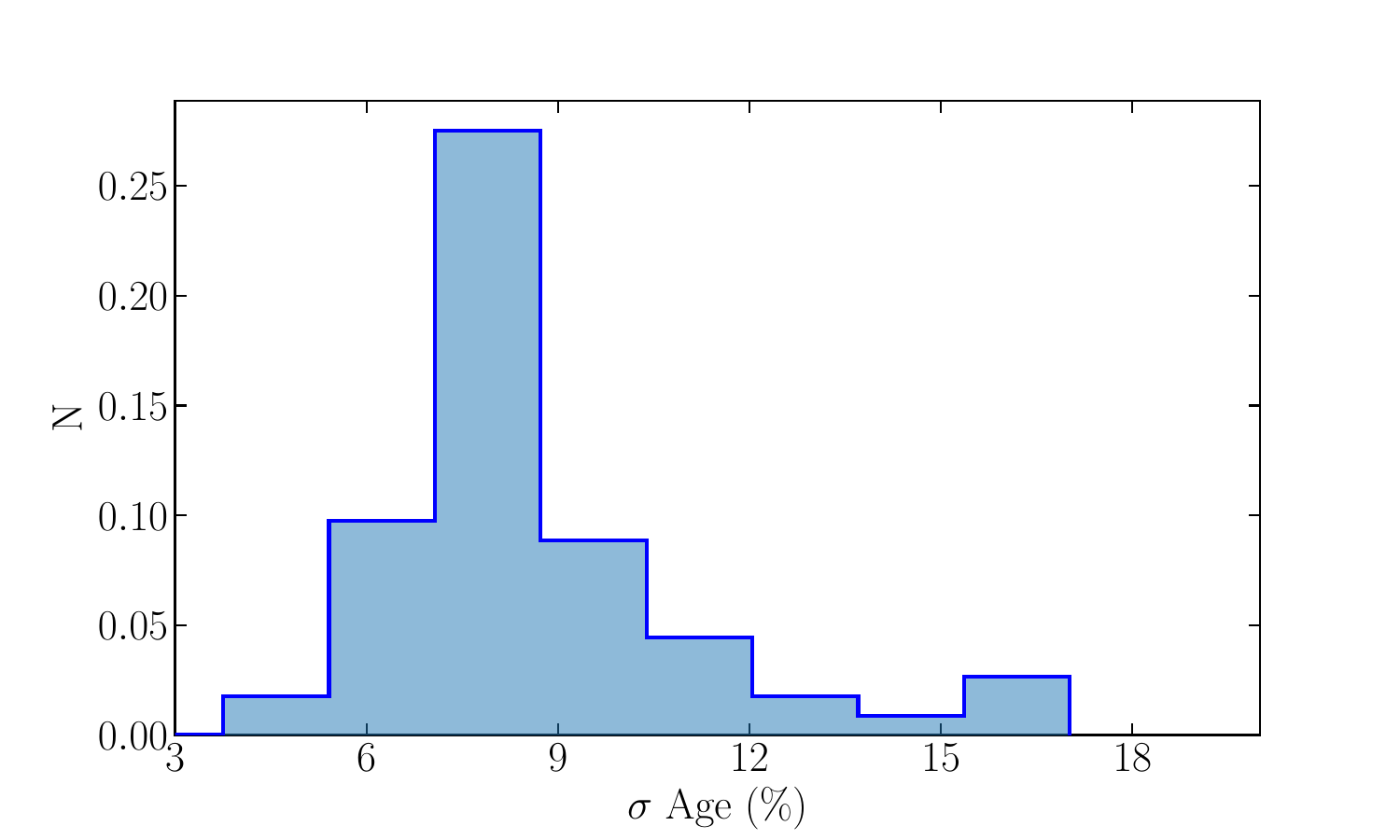}
\caption{Uncertainties distribution for the age computed using the individual mode frequencies for our \kepler sample. N is a probability density: each bin displays the number of stars per bin divided by the total number of stars and the bin width. \label{fig:agedist}}
\end{figure}

As will be shown in Montalbán et al. (in preparation), using individual frequencies as observational constraints instead of global seismic parameters allows us to reduce the internal uncertainty in estimating stellar ages. In the sample analysed in this paper, 50\% of the targets have an age uncertainty between 7\% and 9\% estimated using individual mode frequencies (see Fig.~\ref{fig:agedist}). This is lower than the typical value of 20-23\% reported in \cite{miglio21}, which was estimated using the Bayesian code PARAM \citep{dasilva2006,rodrigues2017} with global seismic parameters, APOGEE-DR14, and \emph{Gaia}-DR2 data as constraints. A recent update (Willett et al., submitted) using APOGEE-DR17 and \emph{Gaia}-DR3 has further decreased the typical age uncertainty with that method. As a result, 50\% of the stars in our sample have age values estimated by PARAM using global seismic parameters with an uncertainty between 10\% and 14.25\%.

\begin{figure}
\centering
\includegraphics[scale=0.4]{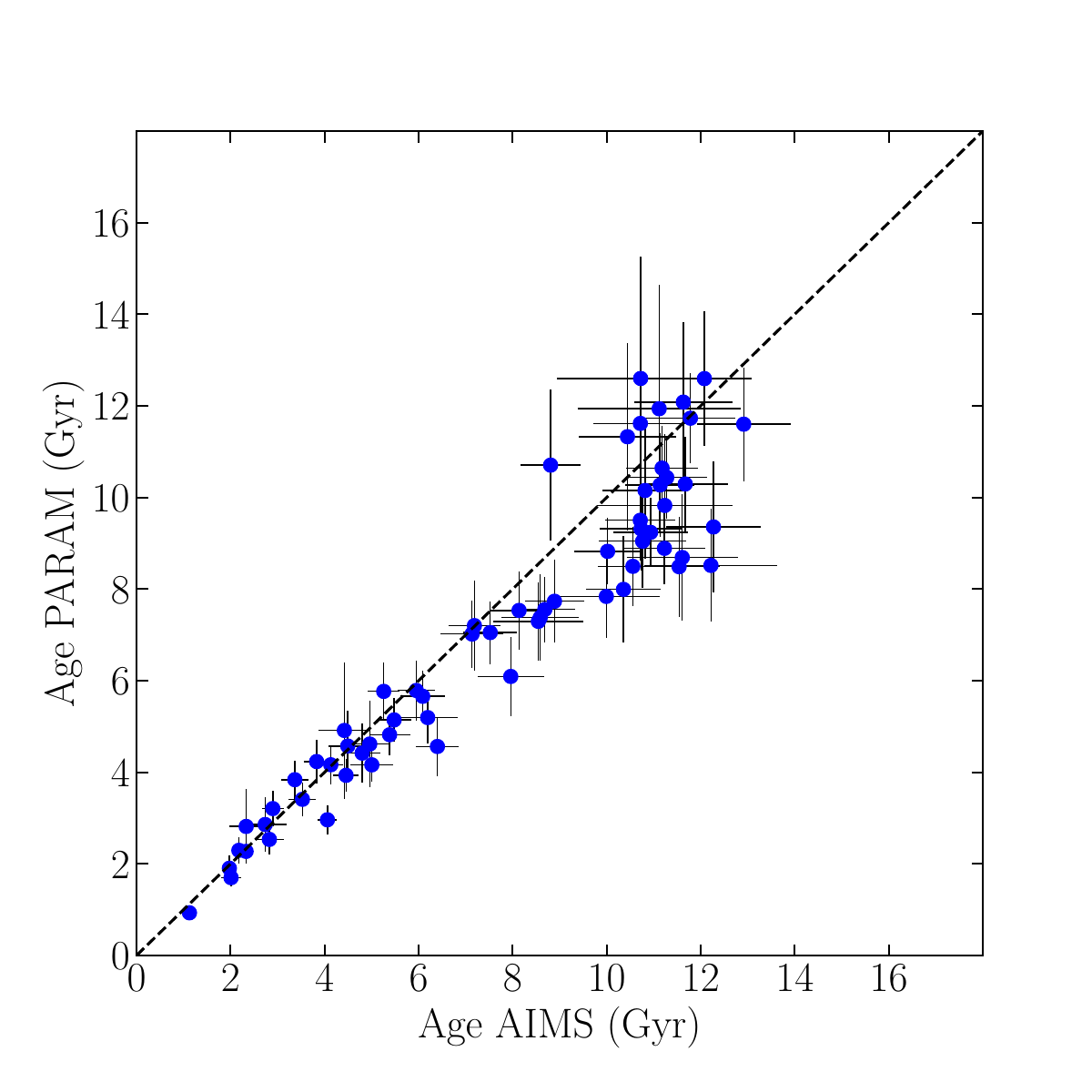}
\caption{Comparison between asteroseismic ages of our sample of \kepler stars computed using individual mode-frequencies (AIMS) and using $\nu_{\rm max}$ and $\Delta \nu$ by PARAM. 
\label{fig:agecomp}}
\end{figure}

The mean difference between ages derived with both methods, normalized by their uncertainties $(\tau_{\rm PARAM}-\tau_{\rm AIMS})/\left( \sqrt{\sigma_{\rm PARAM}^2+\sigma_{\rm AIMS}^2}\right)$ is $-0.72$. Indeed, PARAM ages tend to be smaller than  AIMS ones as we can see from Fig.~\ref{fig:agecomp}. For half of the sample, this difference falls between $-$1.2 and 0.07.

Recently, \citet{apokasc3} estimated a median fractional age uncertainty of $\sim 11.1\%$ for their gold sample of the APOKASC-3 catalogue using global seismic parameters \citep[comparisons of APOKASC-3 against APO-K2 sample are lately present in][]{apok2a,apok2b}. 
However, we found that APOKASC-3 yields younger ages compared to those computed via individual mode frequencies, with an average offset of 0.9 Gyr, increasing to up to 1.7 Gyr in the oldest age regime ($>10$ Gyr), showing a younger thick disc compared to its estimated average age of around 11 Gyr \citep{miglio21, mackereth21, queiroz23, gallart24}. This might be due to a systematic overestimation of mass in APOKASC3 with respect to individual mode frequencies, that yield mass consistent with well-studied eclipsing binaries present in literature \citep{thomsen25,jorgensen21,brogaard18,brogaard22}.

\subsection{Spatial distribution}
\label{sec:spatial}
Our stars are located in the solar neighbourhood ($7.5 < R_{GC} < 8.5$ kpc). Nonetheless, they could have come from  other regions of the Galactic disc, transiting into the solar vicinity, or could have migrated in previous epochs due to the change in their eccentricity through radial heating or in their angular momentum \citep{lynden72,sellwood02}. To better understand their provenance, we computed the guiding radius $R_{g}$, which is the radius of a circular orbit with specific angular momentum $L_{z}$ (estimated as $R_{g} = R_{GC} \cdot V /V_{LSR}$, where $V$ is the Galactocentric azimuthal velocity and $V_{LSR}$ is the circular velocity at the solar Galactocentric distance). Moreover, the adoption of $R_{g}$ instead of the Galactocentric radius $R_{GC}$ can mitigate the blurring effect due to epicyclic oscillations around the guiding radius \citep{schonrich09}. However, it cannot overcome the migrating effect due to churning, which can change $R_{g}$ due to interactions with spiral arms or bars \citep{sellwood02,binney08}.
The guiding radius is computed from the stellar orbits obtained using the {\sc GalPy} package of Python, in which the model {\sc MWpotential2014} for the gravitational potential of the Milky Way is assumed \citep{bovy15}. 
Through the astrometric information by {\em Gaia} DR3, distances from \citet{bailerjones21}\footnote{We also tested asteroseismic distances from PARAM: the use of PARAM distances in the calculation results in an average difference in guiding radius \rl of 0.002 kpc with standard deviation of 0.013 kpc.}, an assumed solar Galactocentric distance $R_{0}$ = 8 kpc, a height above the plane $z_{0}$ = 0.025 kpc \citep{juric08}, a circular velocity at the solar Galactocentric distance equal to $V_{LSR} = 220~ \rm km s^{-1} $, and the Sun's motion with respect to the local standard of rest [$U_{\odot}, V_{\odot}, W_{\odot}] = [11.1, 12.24, 7.25] ~ \rm km s^{-1}$ \citep{schonrich10}, we obtained the orbital parameters, among which the guiding radius $R_{g}$. 
The distribution of $R_{g}$ of the dataset is shown in Fig.~\ref{fig:radii}. It is peaked at 8~kpc with an extension to 4~kpc in the inner disc and more than 9~kpc in the outer disc.

\subsubsection{Birth radii}
To better constrain radial migration, recent works in literature \citep{ratcliffe23,lu24} estimated the birth radii (distances between the birth location of a star and the centre of the Milky Way) using stellar ages and metallicities. 
Knowing the birth radius can provide us with details on the evolution of the abundance gradient over time and, consequently, on the assembly history of our Galaxy, as well as the origin of the high- and low-\alfa~sequences. 

The empirical method explained in \cite{lu24} assumes a linear relation between the present day metallicity range and the birth radial metallicity gradient, inferred from cosmological simulations of disc formation. According to this approach, for any lookback time $\tau$, the metallicity can be written as a function of the metallicity gradient at that time $\nabla \rm [Fe/H] ~(\tau)$, birth radius and the metallicity at the Galactic centre [Fe/H]~(0, $\tau$). To estimate [Fe/H]~(0, $\tau$), \cite{lu24} used the upper boundary of the age-metallicity relation for stars currently found in the inner Milky Way disc.
While the use of birth radii can provide interesting insights, as by definition they are unaffected by radial migration, it should be kept in mind that their use strongly relies on the aforementioned assumptions used to derive them.

\begin{figure}
\centering
\includegraphics[trim={0.6cm 0cm 0cm 0cm}, clip,scale=0.4]{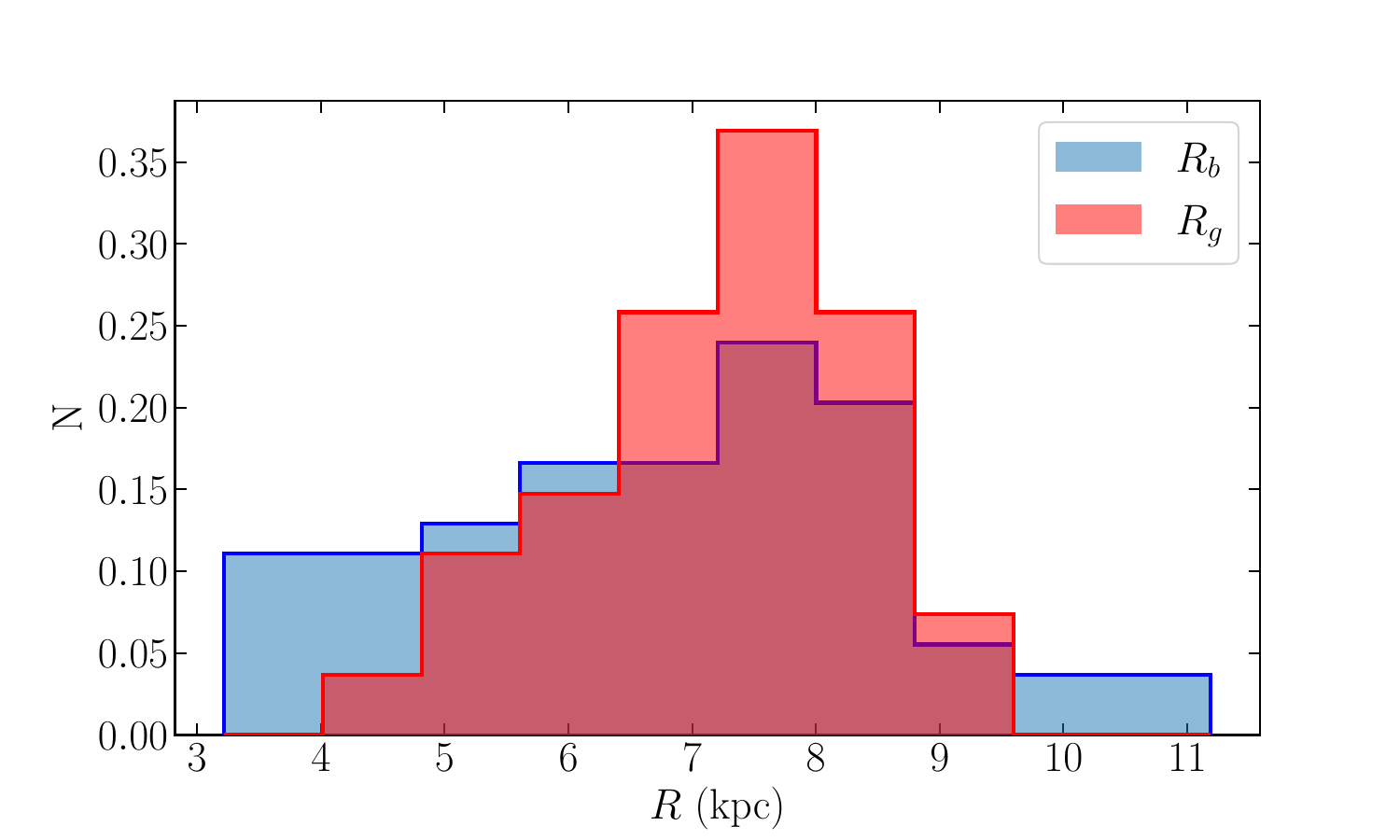}
\caption{Distribution of \rl (red) and \rb (blue) for the \kepler sample. N is a probability density as in Fig.~\ref{fig:agedist}. \label{fig:radii}}
\end{figure}

Figure~\ref{fig:radii} displays the radii distribution for \rl and $R_{b}$. 
\rb is computed using the equation 3 in \cite{lu24}, where we used metallicity and stellar age from our sample, while [Fe/H]~(0, $\tau$) from their Table~1.

The \rl distribution is peaked at 8~kpc, while the \rb distribution is more flat and widens at smaller and larger radii than \rl one. 
In the next sections, we will study the chemical clocks relations using different bins in \rl and $R_{b}$.

\section{Chemical clocks}
\label{sec:clocks}
\begin{figure*}
\centering
\includegraphics[scale=0.5]{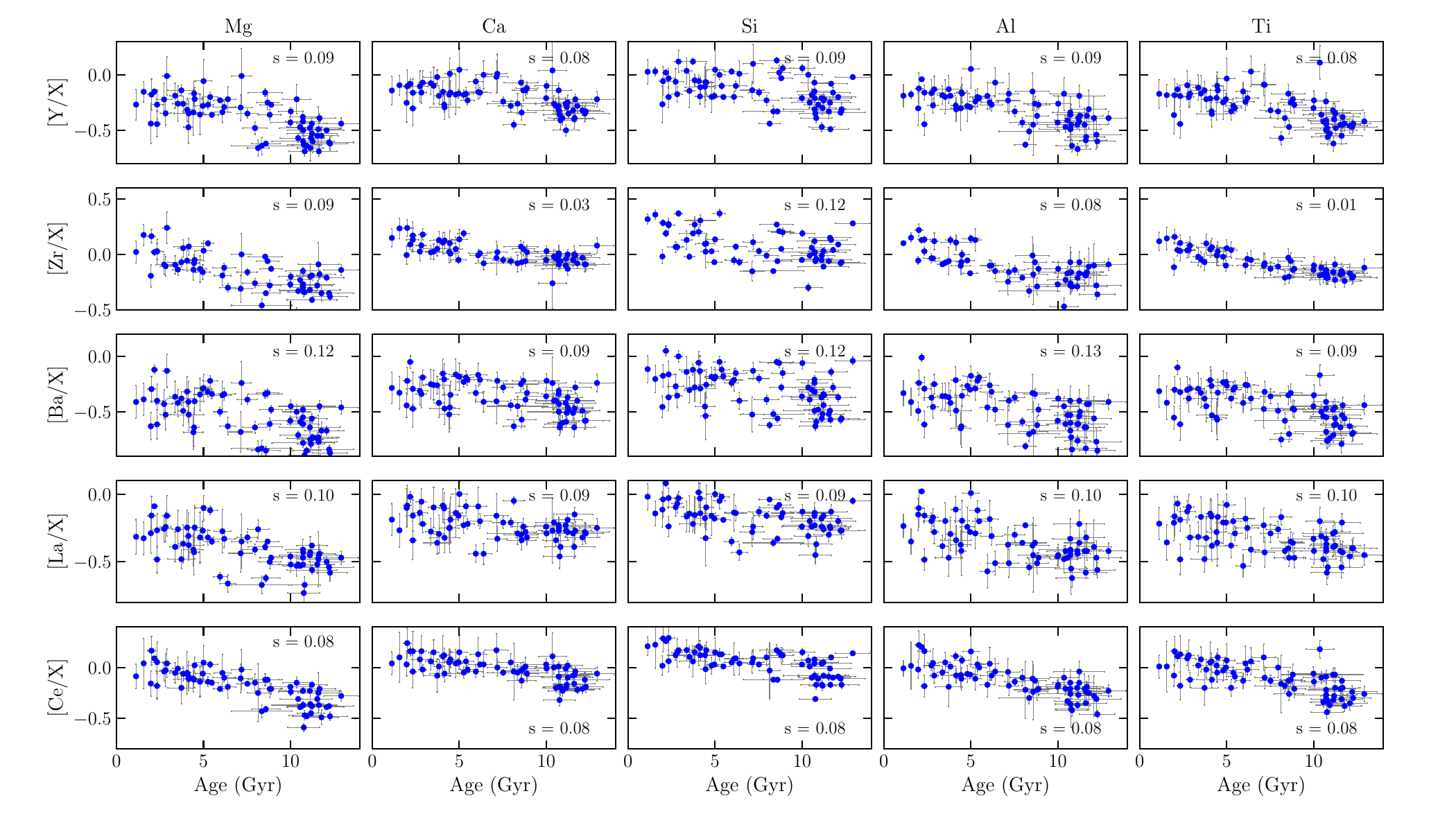}
\caption{Chemical clocks [s/\alfa] vs. stellar age for the \kepler sample. Different rows have different s-process elements, while different columns have different \alfa-elements. The scatter is indicated in each panel. \label{fig:chemclocks}}
\end{figure*}

The ratios of s-process elements to $\alpha$-elements are widely studied as age indicators because they show a steeper trend with age compared to [s/Fe] and [\alfa/Fe]. In Fig.~\ref{fig:chemclocks}, we show the trends of all chemical clocks that we can compose using the s- and \alfa-elements that we have at our disposal. They are respectively: Y, Zr, Ba, La, and Ce\footnote{We did not use Sr since it was measured for a few stars. Pr and Nd are considered s-process elements but their r-component is higher than the other s-process studied in this work.}; Mg, Ca, Si, Ti and the odd-Z Al (it has a similar behaviour to $\alpha$-elements). All of the abundance ratios display a decreasing trend with increasing stellar age.  This is expected due to the s-over-\alfa-elements ratio and their different nucleosynthetic origins.

Among the different abundance ratios, those composed by Zr and Ce show the smaller scatter with ages. Particularly, the tightest relation of Zr is with Ti with an intrinsic scatter of 0.01 dex. The trends including Ce have the same intrinsic scatter of $\sim$0.08 dex, consistent across all \alfa-elements. 
La and Ba are the element that reveal the largest scatter ($\sim 0.1$ dex) among the s-elements, whereas Y has a intrinsic scatter of $\sim 0.085$ dex. The large scatter observed for Ba is likely due to the measurement of a single spectral line, 5853.668 \AA, which may be saturated.

From this point onward, we focus on [Ce/Mg]\footnote{Despite Ce having a small scatter with all \alfa-elements in this study, we focus on [Ce/Mg] as Mg is the \alfa-element with the most measurements available in the APOGEE survey.} and [Zr/Ti] only, as they exhibit the least scatter among chemical clocks and since Zr and Ce are respectively two s-process elements of the first and second peak in the periodic table.  We selected one s-process element for each peak, as they exhibit different dependencies on metallicity and stellar age; consequently, the associated chemical clocks might also differ. In addition, Ce is the only s-process element available in the APOGEE survey, which is another reason to include it in this work. 

\begin{figure*}
\centering
\includegraphics[scale=0.5]{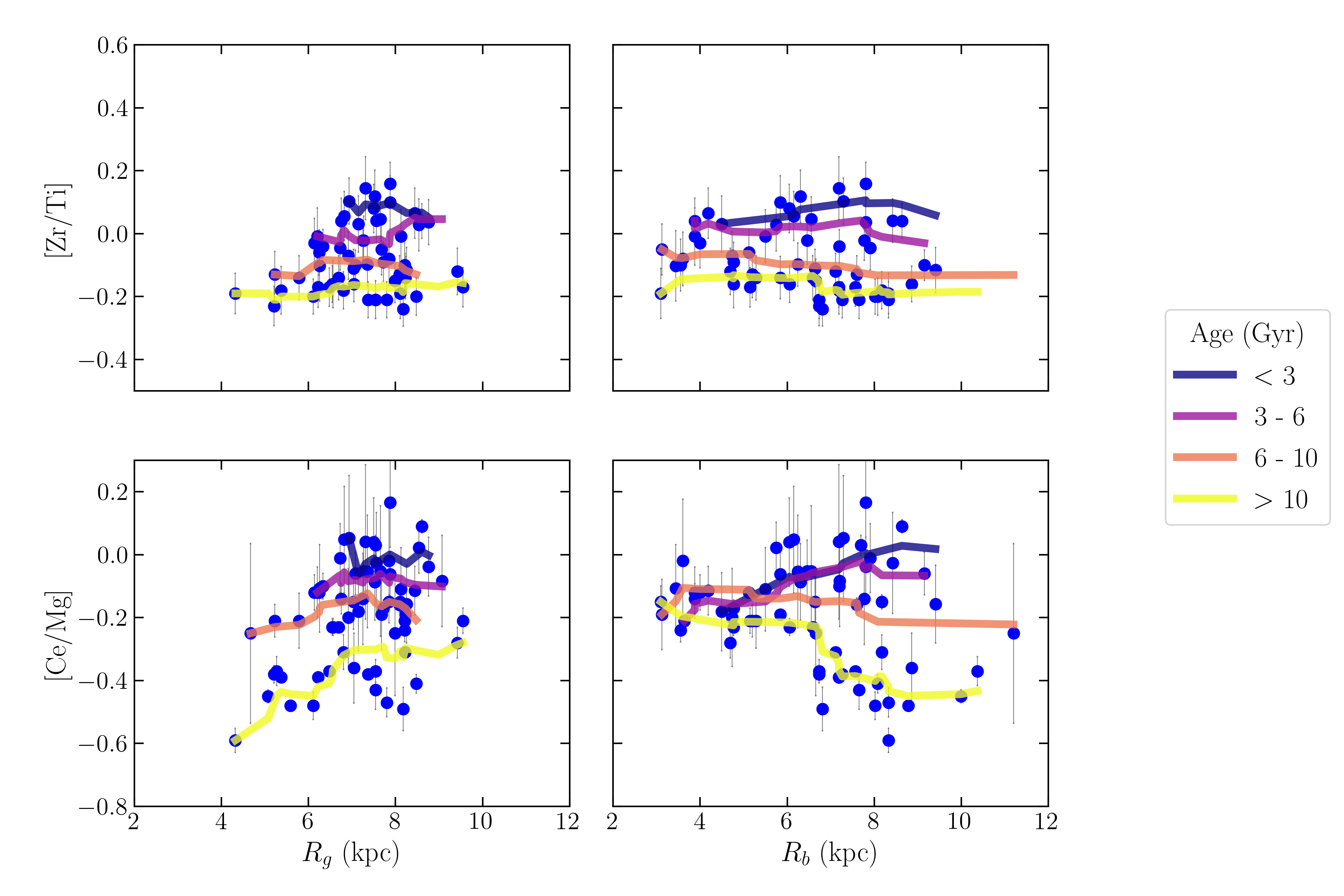}
\caption{Radial gradients of [Zr/Ti] (top panels) and [Ce/Mg] (bottom panels) for the \kepler sample. The lines are the running means of different mono-age populations, determined by calculating the average of [Zr/Ti] and [Ce/Mg] for different bins of \rl (left panels) and \rb (right panels). \label{fig:runningave1}}
\end{figure*}

Using the \rl and \rb estimations in Sec.~\ref{sec:spatial}, we examined the trends of [Zr/Ti] and [Ce/Mg] at different radii. In the Fig.~\ref{fig:runningave1}, we present [Zr/Ti] and [Ce/Mg] vs. radii, where the mono-age populations are highlighted using running means, which were obtained by averaging [Ce/Mg] and [Zr/Ti] for different radius and age bins.
The figure shows that the youngest mono-age populations have higher abundance ratios, while the oldest mono-age populations have lower content in each panel.  In addition, the oldest mono-age population is more spread out in \rl due to the radial migration. Since older stars have had more time to experience these effects, they are more likely to have drifted significantly from their original location. This results in a wider radial distribution for old stars compared to younger ones.  Moreover, the running means do not show different slopes at different age for [Zr/Ti]. Mono-age populations show flat abundance ratios as function of \rl and $R_{b}$, suggesting that spatial correlations are secondary, thus reinforcing the suitability of these ratios as chemical clocks. The situation is slightly different for [Ce/Mg] panels where the mono-age populations have a positive radial gradient, except for the oldest mono-aged populations in the \rb panel, corrisponging to the high-\alfa~sequence. A similar behaviour is observed in \citet{ratcliffe23} for [Ce/Mg] --  lesser extent for Y, an s-process element of the first peak like Zr -- where they interpret this radial gradient as an evidence of a rapid increase in the enrichment with Galactic radius. Moreover, the different behaviour of Ce with respect to Zr is due to the different yields dependence on metallicity (and as a consequence on radius because of the radial metallicity gradient).

\begin{figure*}
\centering
\includegraphics[scale=0.5]{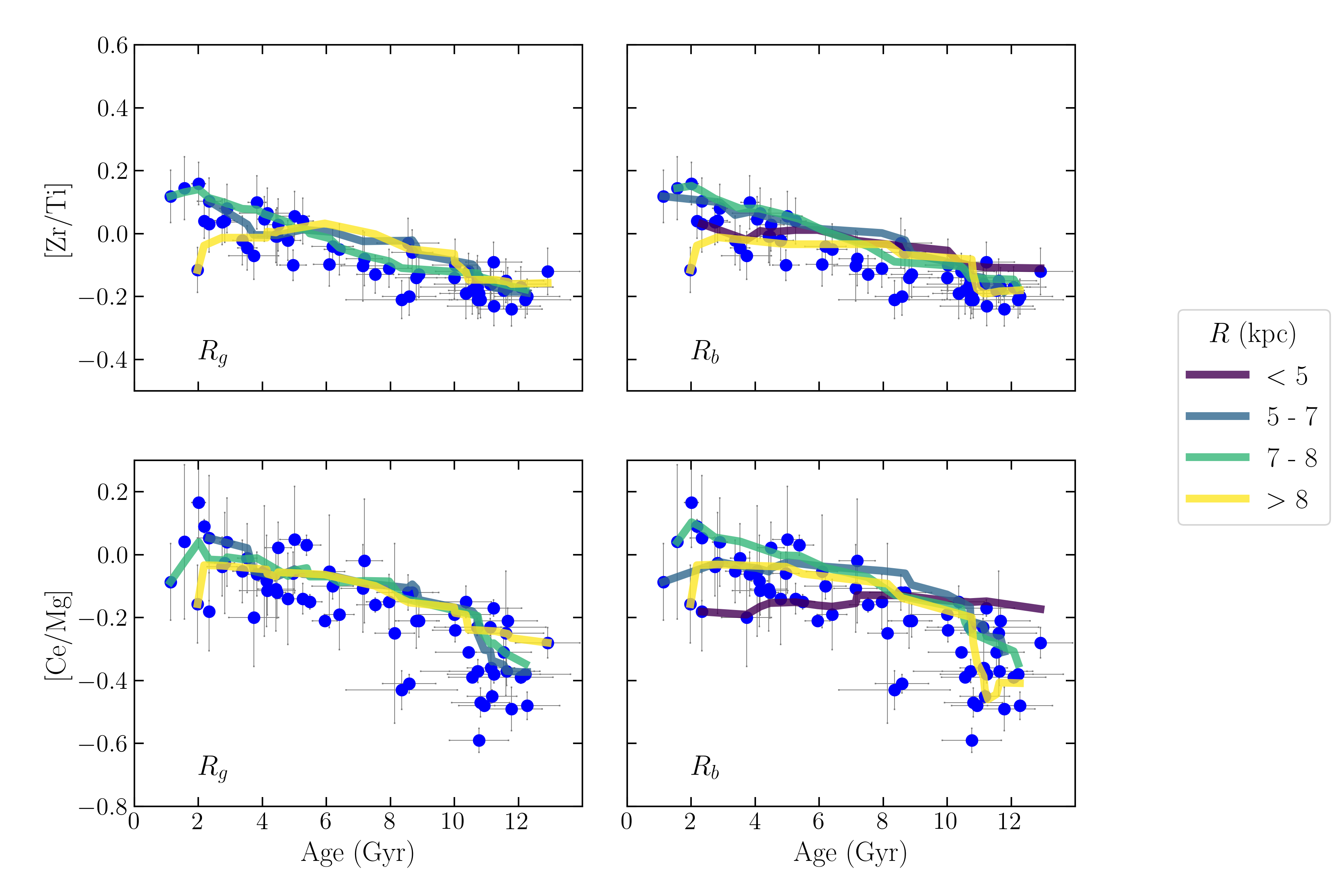}
\caption{[Zr/Ti] (top panels) and [Ce/Mg] (bottom panels) vs. stellar age. The lines are the running means of mono-\rl populations (left panels) and mono-\rb populations (right panels).  \label{fig:runningave2}}
\end{figure*}

Figure~\ref{fig:runningave2} shows the age–[Zr/Ti] and [Ce/Mg] relations for stars born at different locations in the Galaxy. 
Following \citet{casali23} and \citet{ratcliffe23} that studied chemical clocks composed by Ce, Ba and Y, we would expect a flattening for the most internal mono-\rb populations (less prominent for the mono-\rl populations) and an increase in the slopes for the mono-\rb populations at larger radii. This effect is more evident for the s-process of the second peak, such as Ce and Ba, while is less evident for element of the second peak, such as Y and Zr. 
In our work we found a similar behaviour: the abundance ratios are almost flat in the inner regions and steeper moving towards the outer regions for the \rb bins (we do not have the innermost \rl bin); this behaviour is more evident for [Ce/Mg] with respect to [Zr/Ti]. Indeed, Zr is a s-process of the first peak for which the flattening of the innermost mono-\rb population is less pronounced. Furthermore, Zr has a relevant r-component, which could make all slopes similar. 


\subsection{[s/\alfa]-[Fe/H]-age relations}
\label{sec:relations}

In this section, we investigate [s/$\alpha$]-[Fe/H]-age trends for the stars in our sample, dividing stars in \rl and $R_{b}$. The aim of this section is to use these relations to derive ages for a large sample of stars in spectroscopic surveys.
We focused on [Zr/Ti] and [Ce/Mg] only, since they are the ratios with the smallest scatter. 

We modelled the [s/$\alpha$] distributions at different \rl (or $R_{b}$) bins, also taking into account the \feh dependence as follows:

\begin{equation}
[s/\alpha] = m_{1} \cdot {\rm Age} + m_{2} \cdot {\rm [Fe/H]}+c
\end{equation}

where $m_{1}$, $m_{2}$ and $c$ are in Table~\ref{tab:mcmc_zrti_cemg}.
The bins are respectively \rl < 7~kpc, $7 \leq$ \rl $ \leq 8$~kpc, \rl > 8~kpc, and \rb < 6~kpc, $6 \leq$ \rb $ \leq 7$~kpc, \rb > 7~kpc. The different separation between the two radii is related to the different distribution for \rl and $R_{b}$. The bins were chosen by considering the three regions of the disc -- inner, solar, outer -- and by returning a similar number of stars per bin.

\begin{figure*}
\centering
\includegraphics[scale=0.5]{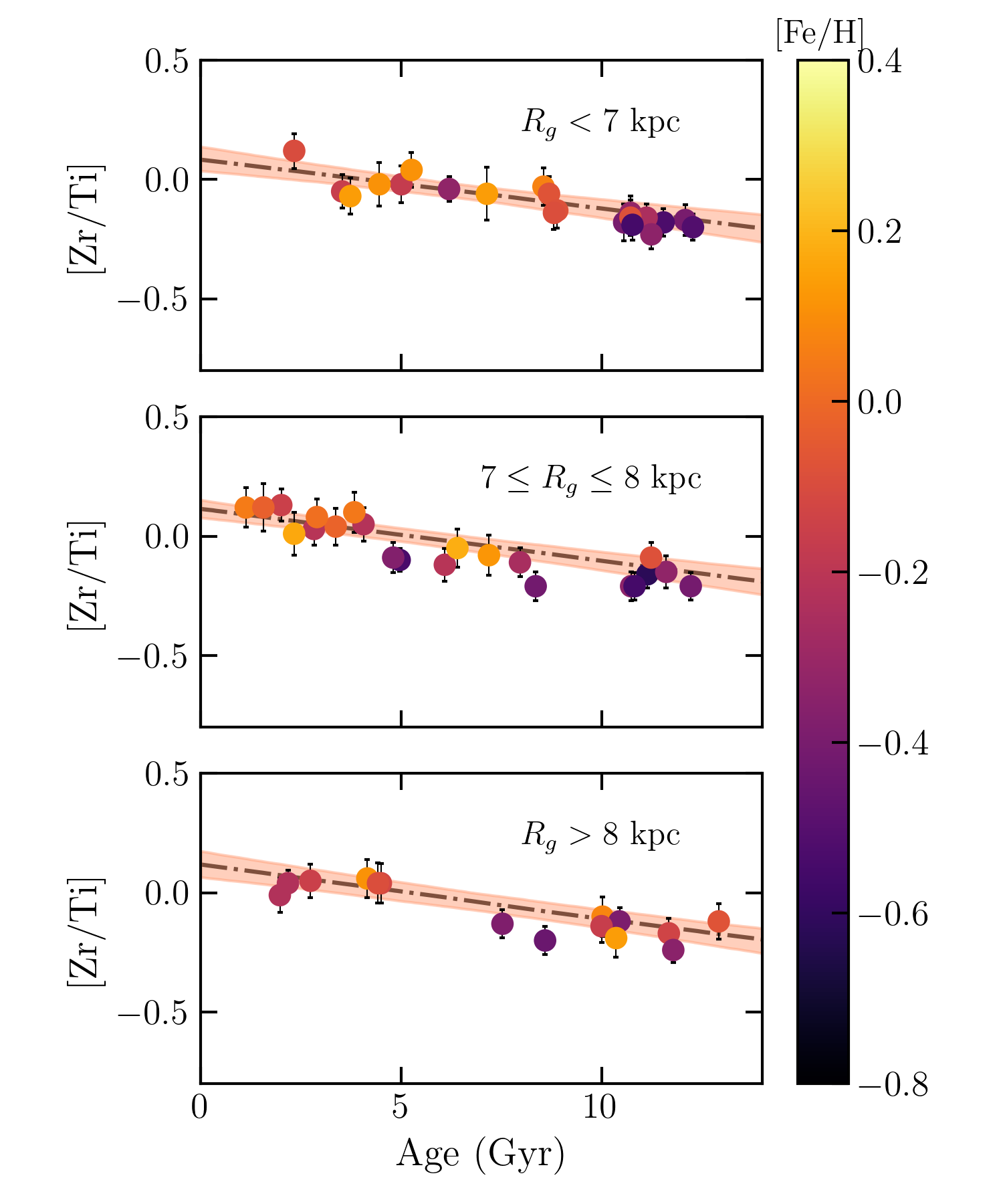}
\includegraphics[scale=0.5]{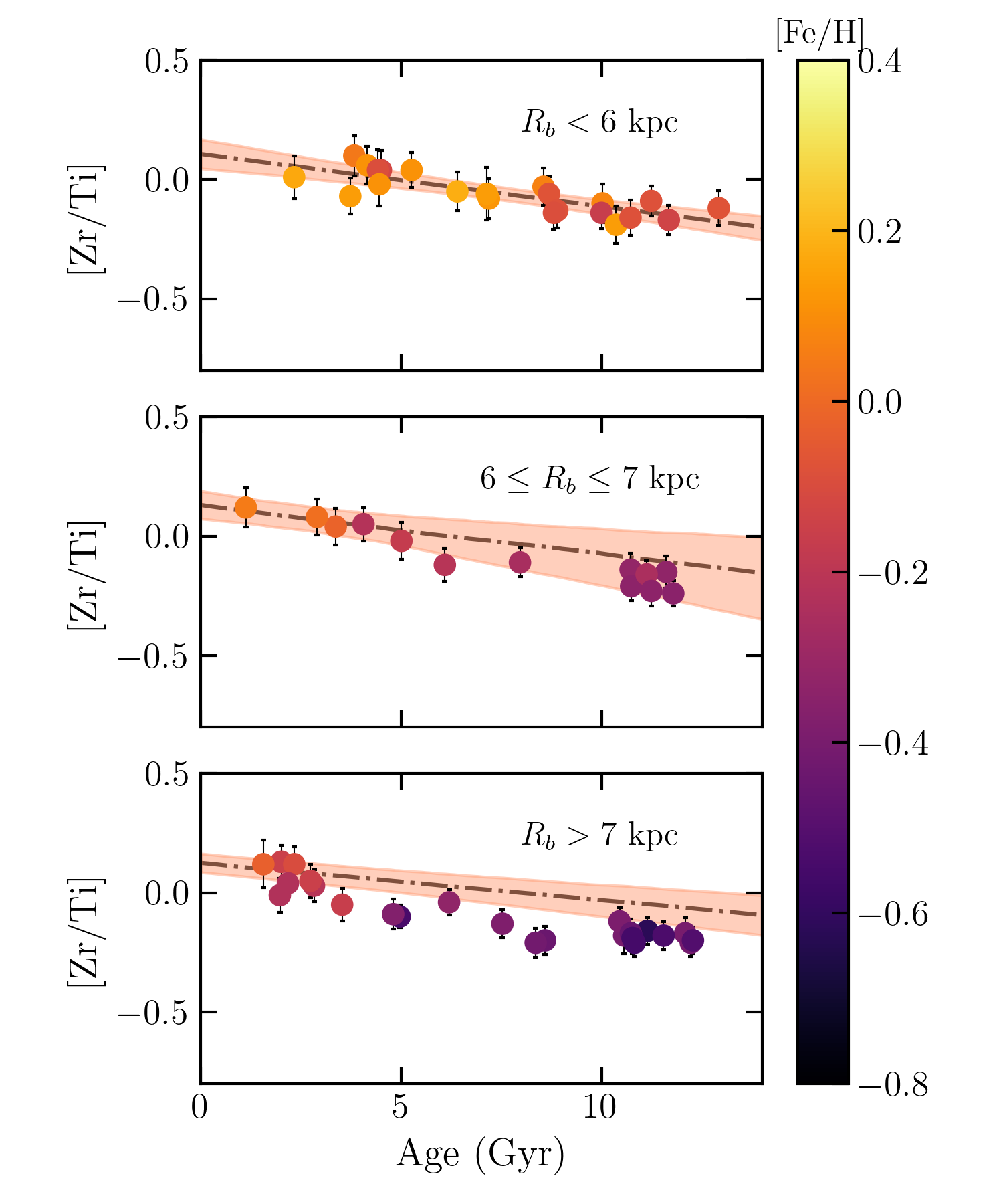}
\caption{[Zr/Ti] vs. stellar age for the \kepler sample in different bins of \rl (left panels) and \rb (right panels). The data are colour-coded by metallicity. The dash-dotted line represents the best fit, and the shaded area the 68\% confidence interval plus intrinsic scatter. Although our sample of stars spans a range of metallicities, the fits in the plots are shown for [Fe/H] = 0, which may not accurately represent the metallicity of some stars that appear discrepant from the fit.
\label{fig:zrti}}
\end{figure*}

\begin{figure*}
\centering
\includegraphics[scale=0.5]{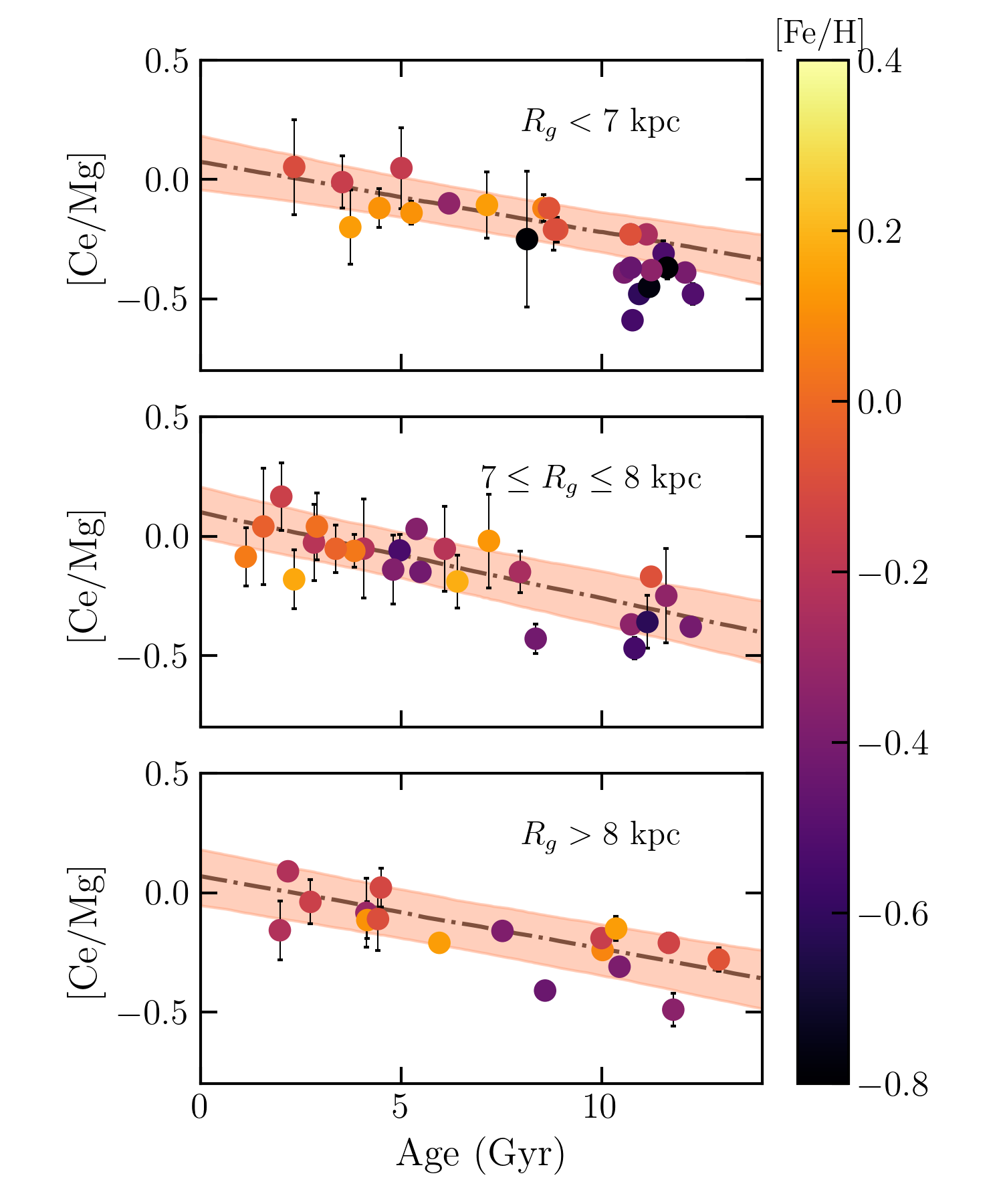}
\includegraphics[scale=0.5]{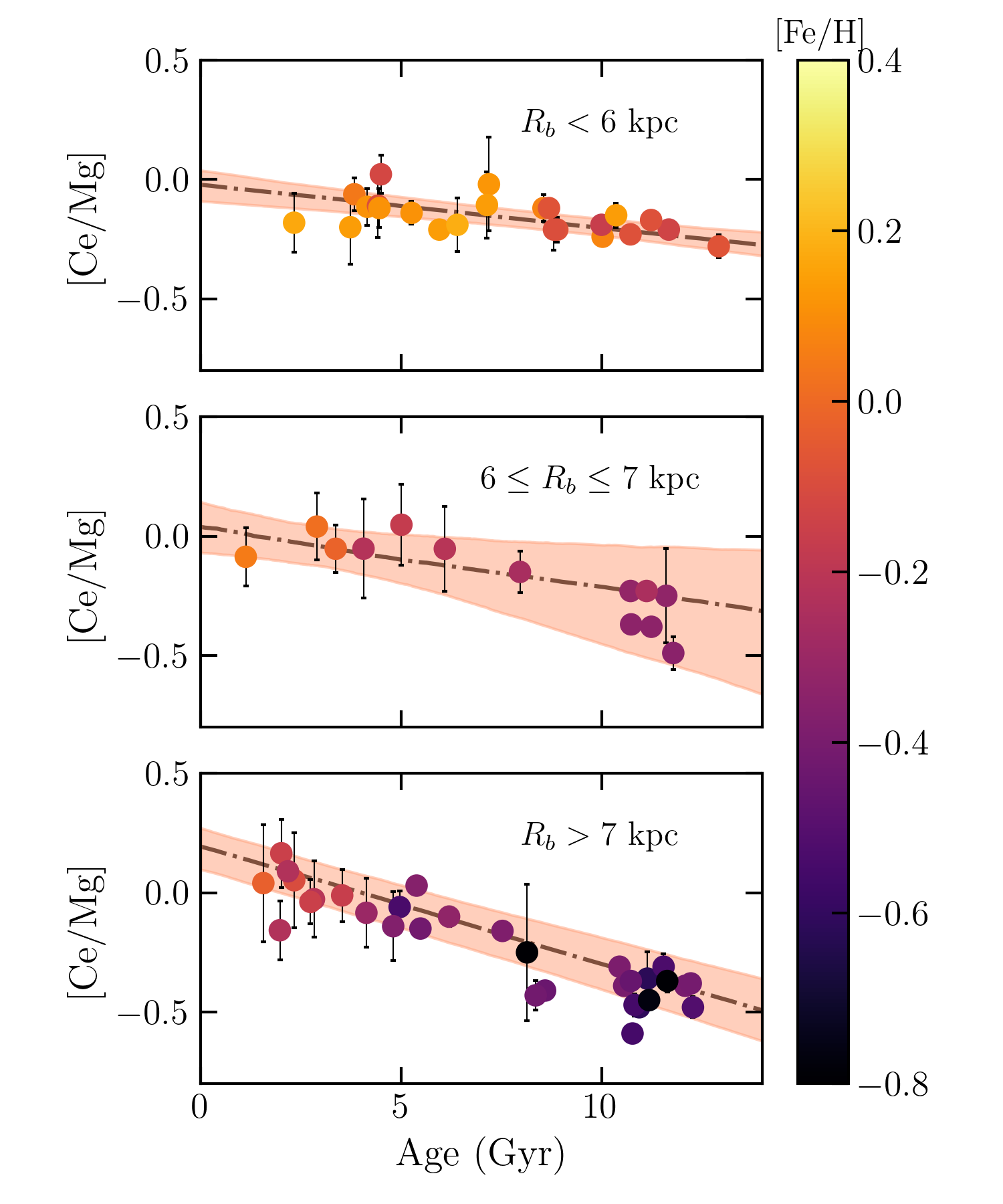}
\caption{[Ce/Mg] vs. stellar age for the \kepler sample in different bins of \rl (left panels) and \rb (right panels). The data are colour-coded by metallicity. The dash-dotted line represents the best fit, and the shaded area the 68\% confidence interval plus intrinsic scatter. Although our sample of stars spans a range of metallicities, the fits in the plots are shown for [Fe/H] = 0, which may not accurately represent the metallicity of some stars that appear discrepant from the fit. 
\label{fig:cemg}}
\end{figure*}

In our calculation, we took into account the uncertainties on [s/\alfa], [Fe/H], and stellar age. 
The best fits with the spread (68\% confidence interval plus intrinsic scatter) of the models resulting from the posteriors are represented in Fig.~\ref{fig:zrti} for [Zr/Ti] and Fig.~\ref{fig:cemg} for [Ce/Mg] with a black line and a red shaded area respectively for the bins in \rl and $R_{b}$ (relations at [Fe/H]=0 are shown). The values of all relations are shown in Table~\ref{tab:mcmc_zrti_cemg}.
The [Fe/H] colour-coding of these stars show a dependence on metallicity. We can see that stars with lower metallicity have higher [s/\alfa] at a given age.  This behaviour was already shown in \citet{feltzing17}, \citet{delgado19}, \citet{casali20}, \citet{magrini21}, \citet{casali23}, suggesting that the metallicity dependence is due to the different star formation history and the non-monotonic dependence of s-process yields on [Fe/H] \citep{busso01, vescovi20, cristallo15}. 

Figure~\ref{fig:zrti} shows the chemical clock [Zr/Ti] vs. stellar age divided in bins of \rl (left panels) and \rb (right panels). 
Regarding the \rl bins, the slopes with ages are consistent within the errors (see $m_{1}$ in Table~\ref{tab:mcmc_zrti_cemg}). 
This could be attributed to the \rl distribution, which has fewer stars at the smallest and largest radii ($R_{g} < 6$ and $R_{g} > 8.5$ kpc). Most stars are concentrated near to the solar region (around 6-8 kpc), suggesting that the stars in the inner and outer bins have radii that are not significantly different from those in the central bin. Moreover, as we discussed in Sect.~\ref{sec:clocks}, Zr is an element of the first peak as Y, which does not show differences at different \rl \citep[see also Fig.~2 of ][]{ratcliffe23}.

When comparing the sample in terms of birth radii and guiding radii, it is observed that more metal-rich stars tend to have smaller birth radii than guiding radii: metal-rich stars are more concentrated in the inner disc, whereas the metal-poor stars in the outer disc. 
Although the \rb distribution is broader than the \rl distribution, extending to both very small and very large radii, the slopes in the three \rb bins remain still consistent with each other within the errors. They are also consistent with those obtained for the \rl bins.
In conclusion, the relations in the three bins in \rl and \rb do not show variation in their slopes, implying no differences in [Zr/Ti]-age relations among inner, solar and outer regions. 
For this reason, we also computed the relation taking all the stars together, without dividing them in three spatial bins. The parameters (named "no binning") are listed in Table~\ref{tab:mcmc_zrti_cemg}. This relation shows similar parameters to those for the binned relations within the errors; however, the uncertainties on the parameters and the intrinsic scatter are smaller. Indeed, by increasing number of samples we get a better measurement of the sample variance following the central limit theorem.


Figure~\ref{fig:cemg} shows the relations for the chemical clock [Ce/Mg]. The slopes in the three \rl regions are consistent within errors as we saw for [Zr/Ti] (the central bin shows a more negative slope), due to the narrow \rl distribution: stars are more concentrated around the solar regions, with their radii not differing significantly from those in the solar bin when compared to the outer and inner bins.
However, when we look at the distribution in $R_{b}$, the situation is different. The inner region is quite flat, whereas the slopes of the relations become steeper moving towards the outer region. 
This behaviour is already seen in \citet{casali23} and \citet{ratcliffe23} for Ce, implying a larger dependency on the radial birth position.
The intrinsic scatter for the relations binned in \rb is smaller than the intrinsic scatter for the relations binned in $R_{g}$, unlike [Zr/Ti]. 



\begin{table*}
 \caption{Parameters of the MCMC fitting.}
\begin{center}
\begin{tabular}{cccccc}
\hline
Ratio & Bin & $m_{1}$~(dex/Gyr) & $m_{2}$~(dex) & $c$~(dex) & $\epsilon$~(dex)  \\
\hline
$\rm [Zr/Ti]$ &  $R_{g} < 7$~kpc          & ${-0.020} \pm {0.006}$& ${0.072} \pm {0.094}$ & ${0.085} \pm {0.048}$ & ${0.014} \pm {0.013}$ \\
$\rm [Zr/Ti]$ &  $7 \le R_{g} \le 8$~kpc  & ${-0.022} \pm {0.005}$& ${0.150} \pm {0.078}$ & ${0.115} \pm {0.033}$ & ${0.016} \pm {0.014}$ \\
$\rm [Zr/Ti]$ &  $R_{g} > 8$~kpc          & ${-0.023} \pm {0.005}$& ${0.173} \pm {0.112}$ & ${0.119} \pm {0.048}$ & ${0.019} \pm {0.017}$ \\
\hline
$\rm [Zr/Ti]$ &  $R_{b} < 6$~kpc          & ${-0.023} \pm {0.007}$& ${-0.066}\pm {0.188}$ & ${0.112} \pm {0.057}$ & ${0.017} \pm {0.015}$ \\
$\rm [Zr/Ti]$ &  $6 \le R_{b} \le 7$~kpc  & ${-0.021} \pm {0.013}$& ${0.283} \pm {0.377}$ & ${0.130} \pm {0.050}$ & ${0.019} \pm {0.018}$ \\
$\rm [Zr/Ti]$ &  $R_{b} > 7$~kpc          & ${-0.016} \pm {0.005}$& ${0.287} \pm {0.138}$ & ${0.127} \pm {0.037}$ & ${0.013} \pm {0.012}$ \\
\hline
$\rm [Zr/Ti]$ &  No binning               & ${-0.022} \pm {0.003}$& ${0.123} \pm {0.048}$ & ${0.104} \pm {0.022}$ & ${0.009} \pm {0.008}$ \\
\hline
\hline
$\rm [Ce/Mg]$ &  $R_{g} < 7$~kpc          & ${-0.028} \pm {0.010}$& ${0.264} \pm {0.093}$ & ${0.068} \pm {0.084}$ & ${0.073} \pm {0.018}$ \\
$\rm [Ce/Mg]$ &  $7 \le R_{g} \le 8$~kpc  & ${-0.036} \pm {0.009}$& ${0.080} \pm {0.136}$ & ${0.095} \pm {0.063}$ & ${0.080} \pm {0.025}$ \\
$\rm [Ce/Mg]$ &  $R_{g} > 8$~kpc          & ${-0.030} \pm {0.008}$& ${0.174} \pm {0.146}$ & ${0.068} \pm {0.069}$ & ${0.087} \pm {0.025}$ \\
\hline
$\rm [Ce/Mg]$ &  $R_{b} < 6$~kpc          & ${-0.018} \pm {0.006}$& ${-0.234} \pm {0.159}$ & ${-0.026} \pm {0.053}$ & ${0.034} \pm {0.014}$ \\
$\rm [Ce/Mg]$ &  $6 \le R_{b} \le 7$~kpc  & ${-0.030} \pm {0.020}$& $ {0.202} \pm {0.497}$ &  ${0.099} \pm {0.087}$ & ${0.062} \pm {0.030}$ \\
$\rm [Ce/Mg]$ &  $R_{b} > 7$~kpc          & ${-0.049} \pm {0.008}$& $ {0.110} \pm {0.161}$ &  ${0.188} \pm {0.054}$ & ${0.069} \pm {0.019}$ \\
\hline
$\rm [Ce/Mg]$ &  No binning        &       ${-0.032} \pm {0.005}$& ${0.194} \pm {0.059}$ & ${0.092} \pm {0.037}$ & ${0.078} \pm {0.011}$ \\ 

\hline
\end{tabular}
\end{center}
\label{tab:mcmc_zrti_cemg}
\end{table*}

It is important to note that the high precision ages and abundances derived in this work significantly reduce the scatter found in \citet{casali23} when trying to calibrate [Ce/Mg] using APOGEE data with ages from global seismic parameters. This is a sobering example of the power of chemical clocks when high precision data is available. Leveraging on the improved precision of the [Ce/Mg] calibration derived here, we thus apply these relations to the APOGEE sample where the large number of stars available allow a meaningful focus on average age trends (see Sec.~\ref{sec:appl}).

We estimated the accuracy and precision of each relation in recovering the input age of our sample of stars. We defined the accuracy as the mean average of the relative errors, $\rm (Age_{input} - Age_{output})/Age_{input}$, between input and output ages obtained from our relations, while the precision is its standard deviation (see also Sec.~\ref{sec:testage}). The relations between [Zr/Ti]/[Ce/Mg]-[Fe/H]-age are able to reproduce the asteroseismic ages of our sample with an overall accuracy and precision of 7\% and 60\% for [Zr/Ti], and 7\% and 70\% for [Ce/Mg]. 
However, it is worth pointing out that accuracy and precision change considerably as a function of age (since the denominator of the relative error is small for young ages and large for old ages), whereby above 5 Gyr typical accuracy and precision for chemical clocks is of the order of 0.3\% and 25\% for [Zr/Ti] and 3\% and 35\% for [Ce/Mg], respectively. 
Typical input uncertainties to build these relations are 0.06 dex in [Zr/Ti], 0.08 dex in [Ce/Mg], 0.04 in [Fe/H] and 8\% in stellar age. 

In summary, it is important to emphasise that chemical ages are inherently statistical. They can be used to estimate age distributions or trends across stellar populations, but their usability for individual stars is limited by the precision. Further tests on the accuracy and precision of chemical ages using data of different quality are discussed in the next Section. 


%

\subsection{High precision for calibrators}
\label{sec:testage}
To highlight the significance of the high-precision abundances and ages used in this work to calibrate chemical clocks, we examined how the tightest of these relations would degrade when using input data of lower quality. Using the same technique discussed in Sec.~\ref{sec:relations}, here we derived the [Ce/Mg]-[Fe/H]-age relations for the sample of 68 \kepler stars, using ages from individual mode frequencies, but Ce and Mg abundances from APOGEE DR17 instead\footnote{Albeit the relation with [Zr/Ti] is even tighter than the one using [Ce/Mg], we focus on the latter since the only survey overlapping with our sample is APOGEE for which Zr is not available.}. These relations have an average intrinsic scatter of $\sim0.15$~dex, against the $\sim 0.08$~dex when using our abundances instead. 

\begin{figure}
\includegraphics[trim={3cm 0cm 0cm 0cm}, clip,scale=0.45]{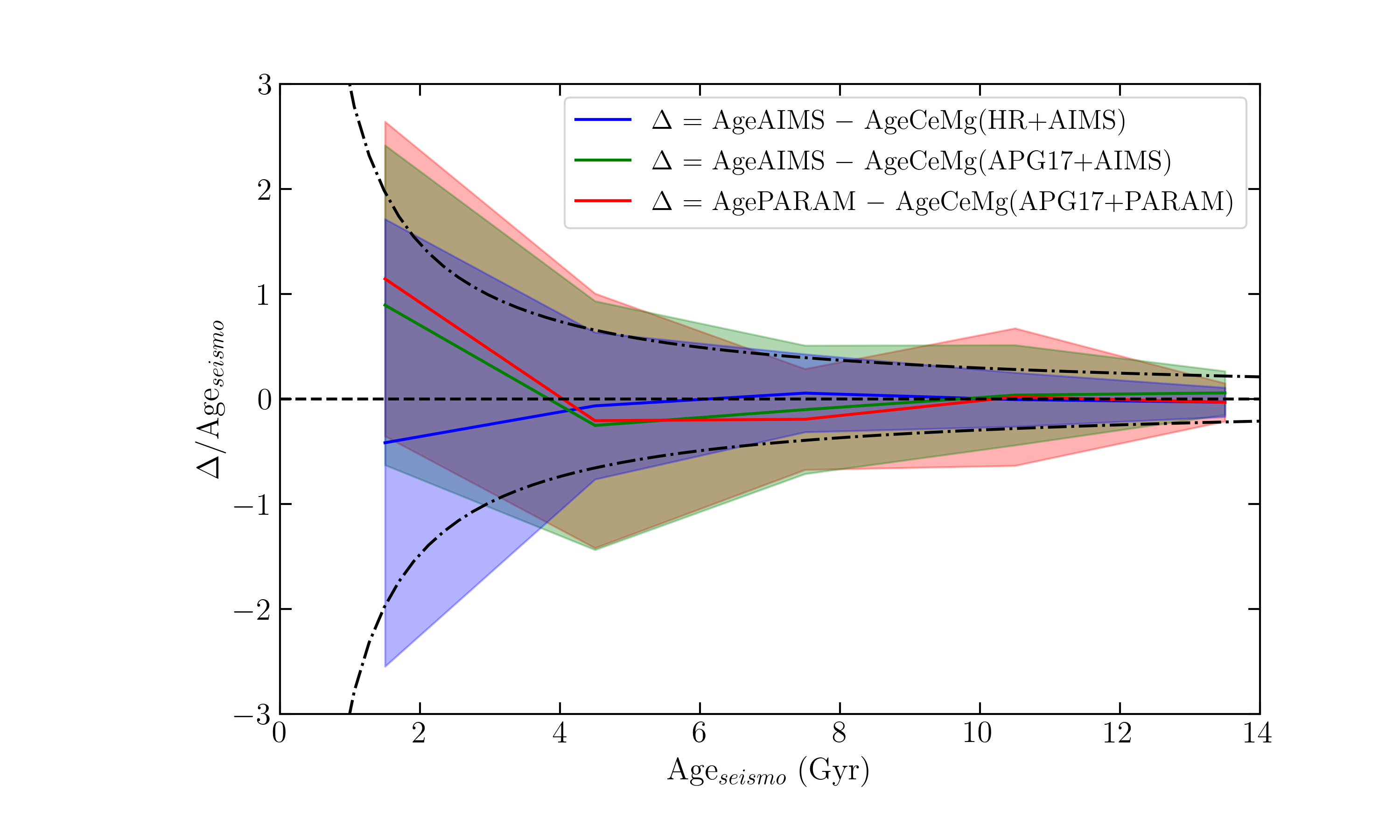}
\caption{The mean (solid line) and standard deviation (shaded area) of $\Delta/\rm Age_{seismo}$ in different age bins, where $\Delta$ is explained in the legend. $\rm Age_{seismo}$ is AgeAIMS or AgePARAM depending on the left term of the $\Delta$. The dot-dashed lines indicate the relative error in ages for an absolute uncertainty of 2.95 Gyr.} \label{fig:confrontoeta}
\end{figure}

Finally, for the same 68 \kepler stars we performed a recovery test comparing chemical ages inferred from [Ce/Mg] against input asteroseismic ages used to build the relations. These results are shown in Figure~\ref{fig:confrontoeta}, where the mean and standard deviation of the recovered ages are shown for the following cases:
\vspace{-\topsep}
\begin{itemize}
\item[--] high-precision abundances from this work with high-precision 
ages from individual mode frequencies (blue line and shade. Hereafter: chemical ages HR+AIMS);
\item[--] abundances from APOGEE DR17 with high-precision 
ages from individual mode frequencies (green line and shade. Hereafter: chemical ages APG17+AIMS);
\item[--] abundances from APOGEE DR17 with ages from global asteroseismic parameters, \numax and $\Delta \nu$ (red line and shade. Hereafter: chemical ages APG17+PARAM).
\end{itemize}
\vspace{-\topsep}
To estimate the accuracy and precision of the chemical ages, we calculated the mean and standard deviation of the relative errors of the chemical ages compared to the seismic ages:

\begin{equation}
\rm \frac{Age_{seismo} - Age_{chemical}}{Age_{seismo}},
\end{equation}
where $\rm Age_{chemical}$ are chemical ages (from HR+AIMS, APG17+AIMS or APG17+PARAM), and $\rm Age_{seismo}$ are ages from AIMS or from PARAM, depending on the set used to calibrate the relations.
While the HR+AIMS relation clearly outperforms the other two, the comparison between APG17+AIMS or APG17+PARAM suggests that using ages from individual frequencies instead of global seismic parameters have a moderate impact on calibrating chemical clocks. This is likely due to the fact that relations for chemical clocks are averaged over ages, with individual frequencies mostly improving upon precision as Fig.~\ref{fig:agecomp} already indicated. On the other hand, the quality of input abundances can have a considerable impact on the precision at which chemical ages can be derived. 

We interpret the mean of these relative errors as a measurement of the accuracy, while the standard deviation as a measurement of the precision of chemical ages. While above we quote the relative dispersion, the absolute one, $\rm Age_{seismo} - Age_{chemical}$, is overall independent of ages and of order of 2.95, 4.84 and 4.93 Gyr for HR+AIMS, APG17+AIMS and APG17+PARAM respectively. This is highlighted by the dot-dashed line in Fig.~\ref{fig:confrontoeta} which shows the relative error in ages for an absolute uncertainty of 2.95 Gyr. Because of the inverse age trend, it is worth noticing that for ages older than about 5 Gyr relative errors are quite constant. In particular, chemical clocks calibrated onto ages from individual mode frequencies are able to deliver ages of old stars with a precision of a few tens of percent. 


\section{Application to field stars}
\label{sec:appl}
Our final aim is to validate our relations and study their ability to provide a reliable age estimate for a larger sample of stars. We seek to verify whether the chemical ages can well reproduce key features already studied in literature, such as the age dissection of low-$\alpha$ and high-$\alpha$ sequences, old metal-rich stars, old metal-poor low-$\alpha$ stars and disc flaring as discussed in the following sections.

The first step is to apply our relations to the stars present in the large spectroscopic surveys APOGEE \citep{apogeedr17} and \emph{Gaia}-ESO \citep[][hereafter GES]{randich22}. 
We applied the relations at different \rl for [Ce/Mg] to APOGEE because Ce is the only s-process element measured in this survey, and the relations for [Zr/Ti] to the GES survey because Ce is measured in fewer stars than Zr. We also tested that the results do not qualitatively change if instead we apply the relations without binning in $R_{g}$. We would like to remind the reader that we cannot use the relations binned in \rb since we do not have information about stellar age for these stars, necessary to compute the $R_{b}$.
Finally, since our sample has metallicity in the range \ensuremath{[-0.8,0.2]} dex, the chemical ages for stars out of this metallicity range are extrapolated. In the following sections, we also address the ages extrapolated for stars beyond the range of our sample, noting that the resulting chemical ages remain within a reasonable values.

Regarding the APOGEE survey, we removed stars with the following flags: ASPCAPFLAG = STAR\_WARN, STAR\_BAD, STARFLAG = VERY\_BRIGHT\_NEIGHBOR, LOW\_SNR, PERSIST\_HIGH, PERSIST\_JUMP\_POS, PERSIST\_JUMP\_NEG, SUSPECT\_RV\_COMBINATION, and ELEMFLAG = GRIDEDGE\_BAD, CALRANGE\_BAD, OTHER\_BAD, FERRE\_FAIL for Ce.
Since our observed stars are also present in the APOGEE survey, we computed the deference between our abundances and the ASPCAP ones for Ce and Mg and [Fe/H], and we applied them to the APOGEE abundances to computed the chemical ages. 
Regarding the GES survey, instead, we removed stars with Zr uncertainties larger than 0.2 dex.

\subsection{[\alfa/Fe] vs. [Fe/H] plane}

The [\alfa/Fe] vs. [Fe/H] plane can be explained with the early, rapid contribution to the Galactic chemical enrichment by Type II supernovae, and later, long contribution by Type Ia supernovae \citep{tinsley80,matteucci89,kobayashi06,Imig2023,Patil2023}. The bimodality seen in the [\alfa/Fe] vs. [Fe/H] diagram reflects distinct formation and evolutionary histories for the two \alfa-sequences \citep{chiappini97,chiappini09,Grisoni2017,Mackereth2019b,Warfield2021}. This can be translated into different age distributions.

\begin{figure}
\includegraphics[trim={2cm 0cm 0cm 0cm}, clip,scale=0.45]{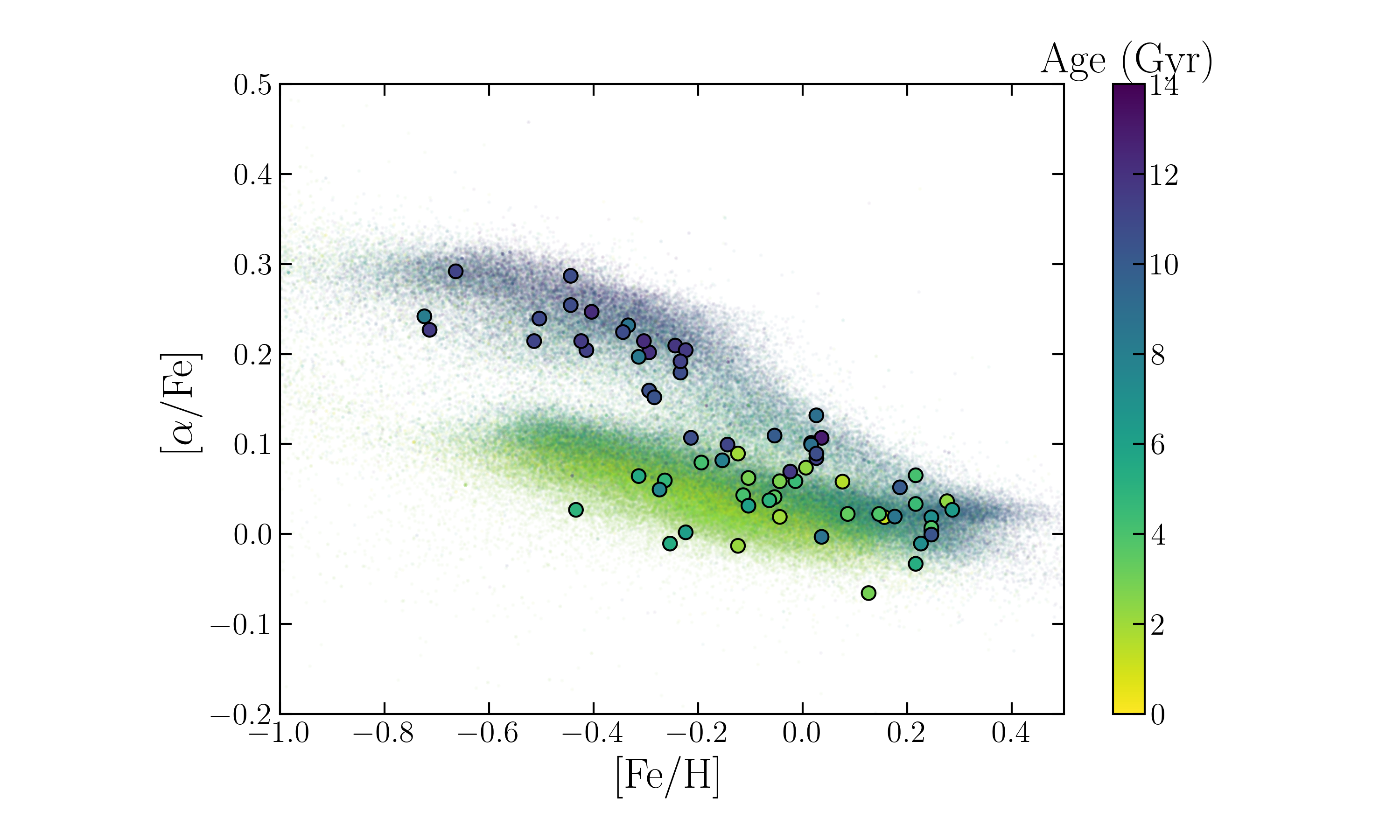}
\includegraphics[trim={2cm 0cm 0cm 0cm}, clip,scale=0.45]{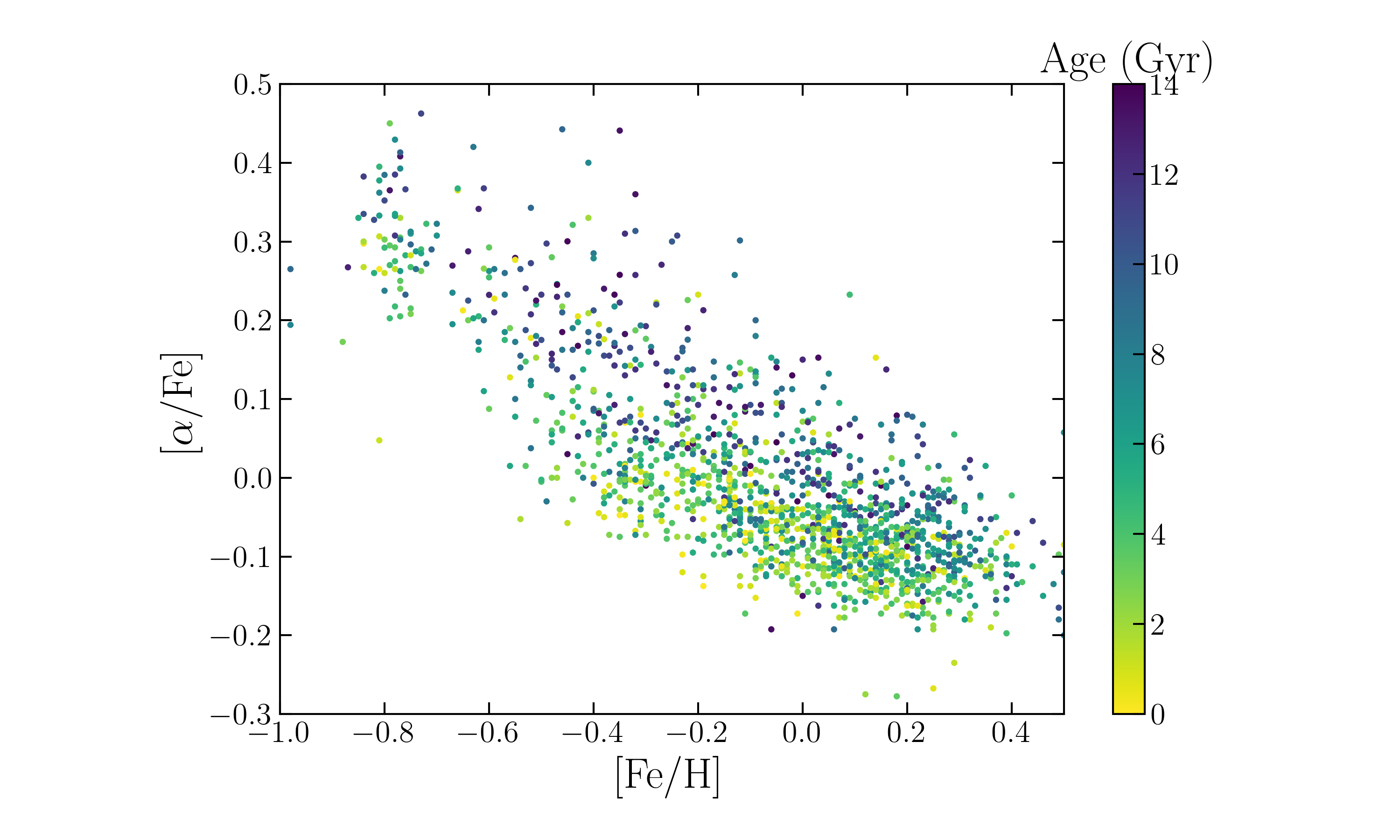}
\caption{[$\alpha$/Fe] vs. [Fe/H] for field stars in the APOGEE (top) and \emph{Gaia}-ESO (bottom panel) surveys. The stars are colour-coded by chemical ages computed using the relations for [Ce/Mg] and [Zr/Ti] respectively for the two surveys. The circles are the 68 \kepler stars. 
\label{fig:appl}}
\end{figure}

Figure~\ref{fig:appl} displays the [\alfa/Fe] vs. [Fe/H] planes for the APOGEE and GES surveys where stars in each panel are colour-coded using the chemical ages. The number of stars for which we found chemical ages is $\sim 270,000$ and $1,500$, respectively. This is one of the largest sample with ages so far, along with the catalogues with APOGEE abundances with ages from astroNN\footnote{\url{https://data.sdss.org/sas/dr17/env/APOGEE_ASTRO_NN/}}, from XGBoost \citep{anders23} and a catalogue with \emph{Gaia} DR3 abundances with ages from {\sc StarHorse} \citep{nepal24}.
The panels show the separation in age for the low- and high-$\alpha$ sequences for both surveys: low-$\alpha$ stars are younger than the high-$\alpha$ ones and show a gradient in age increasing with $\alpha$. 

Furthermore, in APOGEE the separation in low-\alfa~and high-\alfa~sequence is more evident than in the GES sample. However, chemical age can help us to distinguish the two sequences, even when the separation is less clear based solely on \alfa~abundances.

In the following subsections, we will just focus on the APOGEE survey since the number of stars is larger than GES. Additionally, APOGEE stars are better suited to test chrono-chemo-dynamical trends across the Galaxy, whereas GES mainly comprises stars located in the solar neighbourhood \citep[see][for a comparison of the volumes covered by different spectroscopic surveys]{queiroz23}.



\subsection{[Fe/H] and [$\alpha$/Fe] vs. stellar age}
The age-metallicity relation is a fundamental observational constraint for understanding how the Galactic disc formed and evolved chemically in time. Its existence has been a matter of debate in the past. 
Although a tight age-metallicity relation is known to be absent among disc stars \citep[e.g.][]{edvardsson93,nordstrom04,haywood13,bergemann14,casagrande16,Mackereth2019b,Kruijssen2019}, the increasing availability of large-scale survey data and asteroseismic ages allows us to probe its scatter and possible substructures in greater detail.

\begin{figure}
\includegraphics[trim={1cm 0cm 0cm 0cm}, clip,scale=0.36]{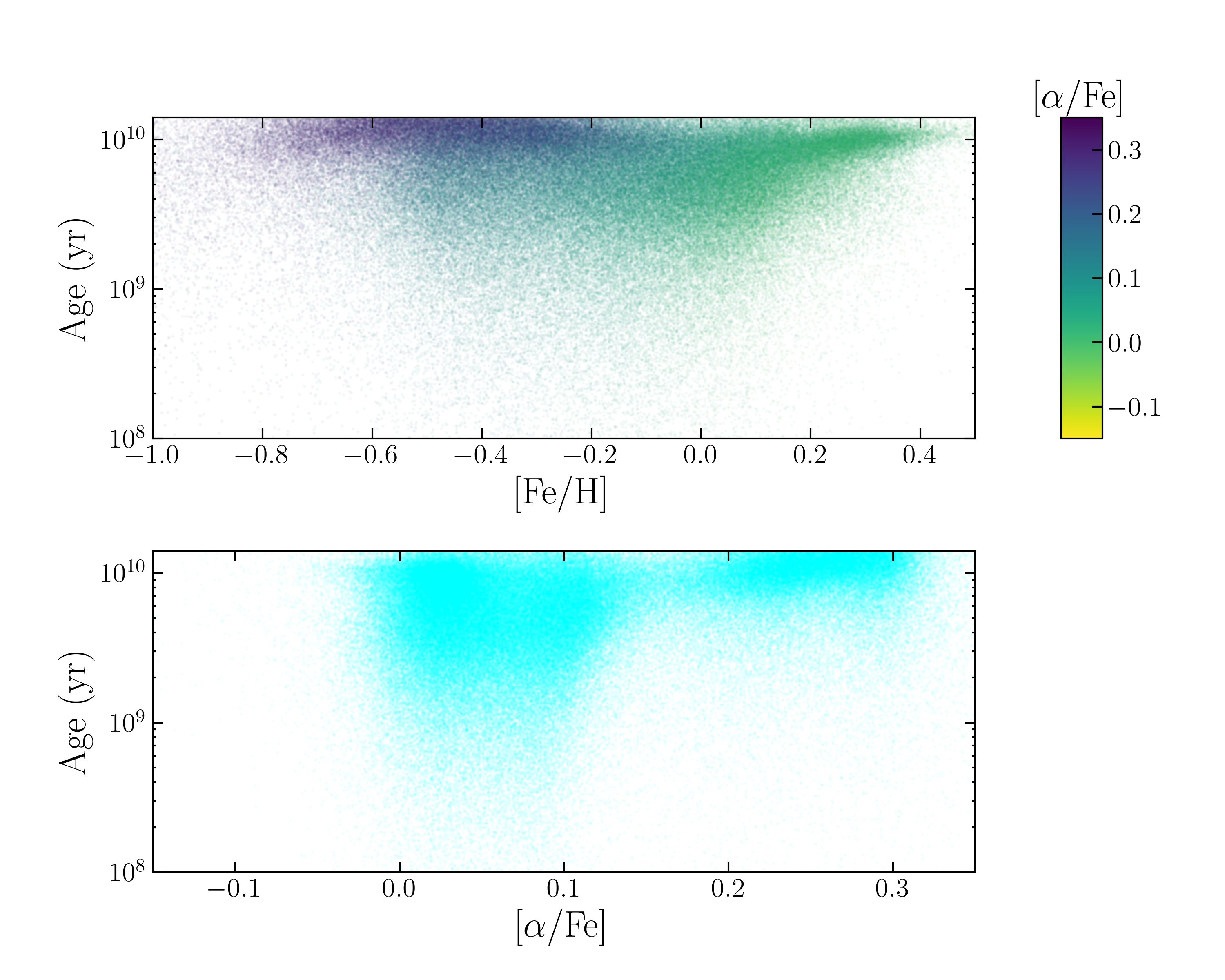}
\caption{Top panel: logarithmic chemical age vs. metallicity for the APOGEE survey where the stars are colour-coded by [\alfa/Fe]. Bottom panel: logarithmic chemical age vs. [\alfa/Fe] for the APOGEE survey. \label{fig:apo_amr}}
\end{figure}

In our case, we have a large number of stars with chemical age to our advantage. This is also complementary, as previous works have focused on isochrone ages of main sequence turnoff (MSTO) and sub-giant stars, while our analysis concentrated on giant stars,  which are more luminous and allow the study of more distant regions. We studied, indeed, the age-metallicity relation for the APOGEE sample. 
The top panel of Fig.~\ref{fig:apo_amr} shows the logarithmic chemical age vs. metallicity where the stars are colour-coded with [$\alpha$/Fe]. 
With increasingly older ages, stars show a broader range of metallicities, whereby stars with $\rm [Fe/H] < -0.6$ and $> 0.2$ dex are consistently old. This picture is consistent with the idea that young stars form from the local ISM at roughly solar metallicity \citep[e.g.,][]{nieva12}, whereas dynamical processes are responsible for moving more metal-poor and metal-rich stars from other locations in the Galaxy on Gyr timescales \citep[e.g.,][]{schonrich09,casagrande11,minchev13,minchev14,casagrande14,miglio21}. 

The bottom panel of Fig.~\ref{fig:apo_amr} shows the logarithmic chemical age vs. [\alfa/Fe] plane. It suggests that the chemical evolution of stars in the low-$\alpha$ sequence (mainly stars with $\rm [\alpha/Fe]<0.1$ dex) happened on much longer timescales compared to the high-$\alpha$ sequence. In fact, the high-$\alpha$ stars reveal an average age of $\sim 11$ Gyr with a narrow distribution \citep[see also][]{miglio21,queiroz23,nepal24}. The low-$\alpha$ stars, instead, unveil a broad age range, reaching ages as young as $\sim 2$ Gyr and as old as the high-$\alpha$ sequence. This behaviour is already seen in several works, e.g. \citet{hawkin14,bensby14,miglio21,nepal24} and others. 

\subsection{Old metal-rich stars}
In the [$\alpha$/Fe] vs. [Fe/H] panels, there is a bulk of old metal-rich stars (age $\gtrsim$ 8 Gyr, [Fe/H] $\gtrsim$ 0 dex), particularly evident for the APOGEE survey. 
The existence of old and metal-rich stars has been confirmed by several works: e.g., \citet{trevisan11} and \citet{casagrande11} using isochrone fitting; \citet{anders17}, \citet{miglio21} and \citet{puls23} using asteroseismic ages. 
These stars, often found in the thin disc, have higher metallicities despite their advanced ages, suggesting that they formed in regions of the Galaxy where star formation had already enriched the interstellar medium with heavy elements. Their presence challenges simple models of Galactic evolution, indicating a more complex history of star formation and migration within the disc. Studies such as \citet{chen03,haywood08,trevisan11,adibekyan11,puls23,nepal24a} explored the connection among age, kinematics and metallicity, revealing that these stars could be the result of radial migration -- where stars formed in the inner, metal-rich regions of the Galaxy migrate outward over time. 
They can be kinematically subclassified in samples of thick and thin disc stars: some of them have high eccentricity and low maximum height above the Galactic plane like thin-disc stars, while others have space velocity $V - V_{LSR} < -50$~km s$^{-1}$ \citep[see][]{trevisan11}, which is more typical of a thick disc, but show solar [$\alpha$/Fe] ratios that are more compatible with thin disc.

We assigned the probability of each star belonging to the kinematically-defined thin disc, kinematically-defined thick disc, or kinematically-defined halo using the method described in \citet{reddy06}. This assumes that the sample is a mixture of the three populations and their space velocities follow a Gaussian distribution. We considered that a probability larger than 70\% in one of the three components assigns the corresponding star to that population. 

\begin{figure}
\includegraphics[trim={1.5cm 0cm 0cm 0cm}, clip,scale=0.4]{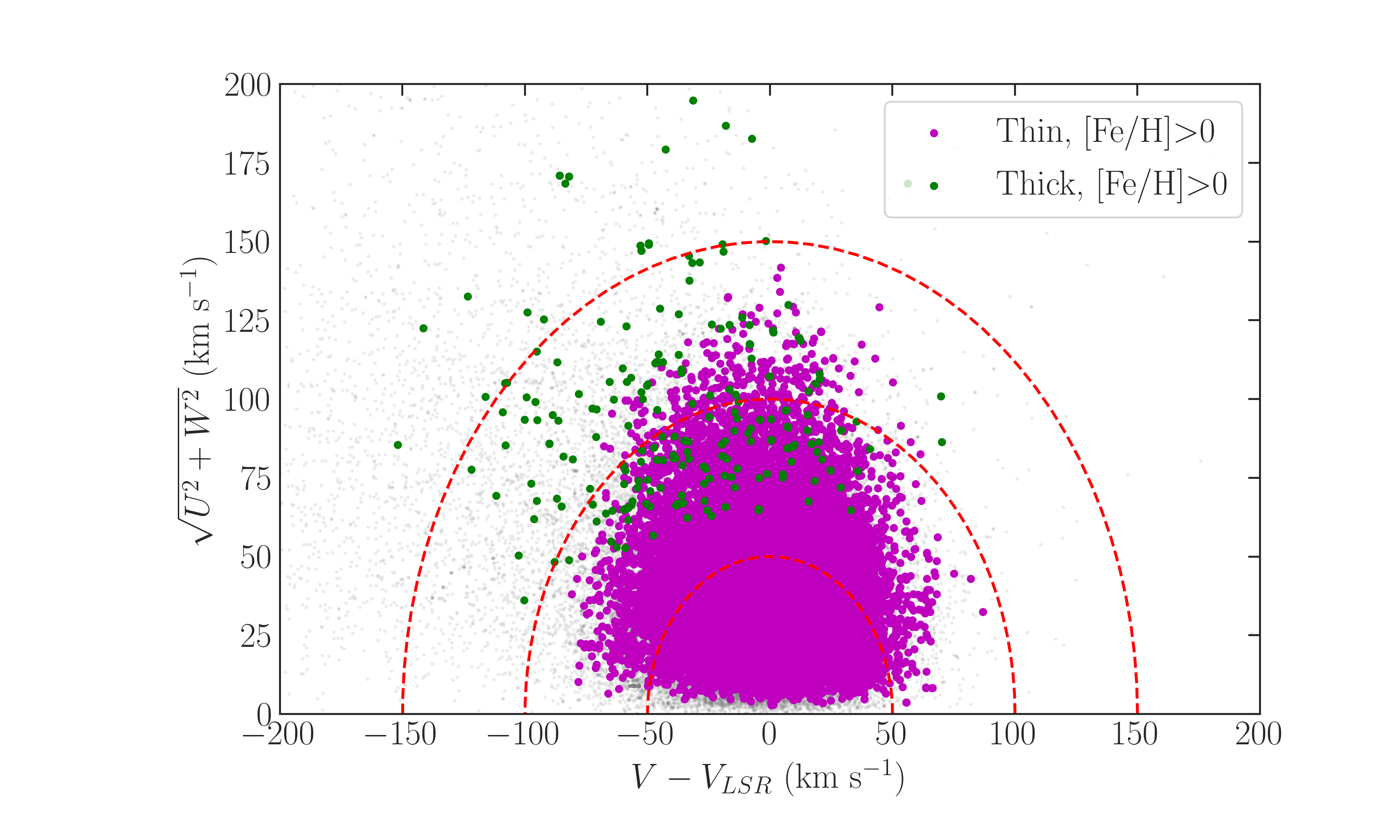}
\caption{Toomre diagram for the APOGEE stars with [Fe/H] > 0. The kinematically-defined thin-disc, thick-disc stars \citep[determined using the probability by][]{reddy06} are represented by green, magenta dots, respectively. The dashed lines indicate the total space velocity, $v_{\rm total} = \sqrt{U^2 + V^2 + W^2}$, in steps of 50 km/s. The grey dots on the background are stars with uncertainties on $U$, $V$, and $W$ larger than 30\%. \label{fig:toomre_omr}}
\end{figure}

We selected stars with $\rm [Fe/H]>0$ dex for the APOGEE sample and plotted them following their probabilities in the Toomre diagram as in Fig.~\ref{fig:toomre_omr}. Velocities have been computed using {\sc GalPy} as described in Sec.~\ref{sec:spatial}. 
In Fig.~\ref{fig:toomre_omr}, we kept just stars with uncertainties on $U$, $V$ and $W$ lower than 30\%.  Among these metal-rich stars, most have thin-disc kinematics with a mean age of 6.8 Gyr, while only 3\% have thick-disc kinematics with a mean age of 7 Gyr. Moreover, they have a solar [$\alpha$/Fe] ratio, belonging to the low-$\alpha$ sequence.
This means that in a thick disc defined by kinematics, part of the stars can be low-\alfa~\citep[see \text{[$\alpha$/Fe]-[Fe/H]} planes at different positions in the Galaxy in, e.g.,][]{anders14,hayden15,queiroz23}, underlining how the kinematically-defined thick disc and chemically-defined thick disc are different \citep[see, e.g.,][]{hayden17}.

\begin{figure*}
\hspace*{-1cm}
\includegraphics[scale=0.5]{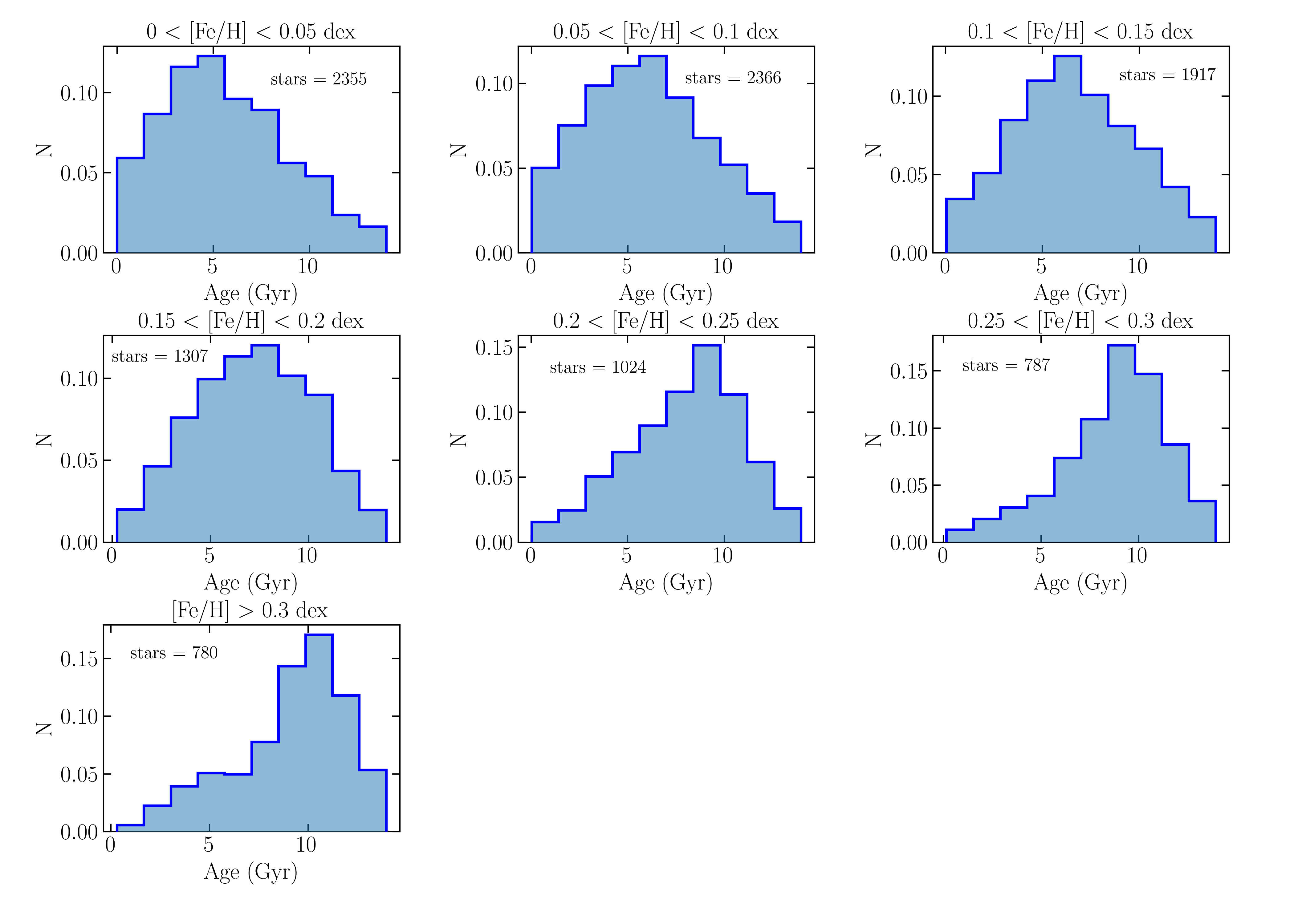}
\caption{Chemical age distribution of APOGEE stars in different bin of metallicity ($\rm [Fe/H]>0$). N is a probability density as in Fig.~\ref{fig:agedist}. The number of stars per distribution is shown in each panel. \label{fig:hist_omr}}
\end{figure*}

A better visualisation of their distribution in age at different bins of metallicity is presented in Fig.~\ref{fig:hist_omr}. The peak of the age distribution moves from $\sim 5$ Gyr in the most metal-poor bin to $\sim 10$ Gyr in the most metal-rich bin. We noticed the majority of the super metal-rich stars \citep{miglio21} -- $\rm [Fe/H] > 0.2$~dex -- are old. This behaviour is already seen in other works, such as \citet{miglio21} for giants and \citet{nepal24a} for MSTO and sub-giants stars. 

\begin{figure}
\hspace*{-1cm}
\includegraphics[trim={3cm 0cm 0cm 0cm}, clip,scale=0.45]{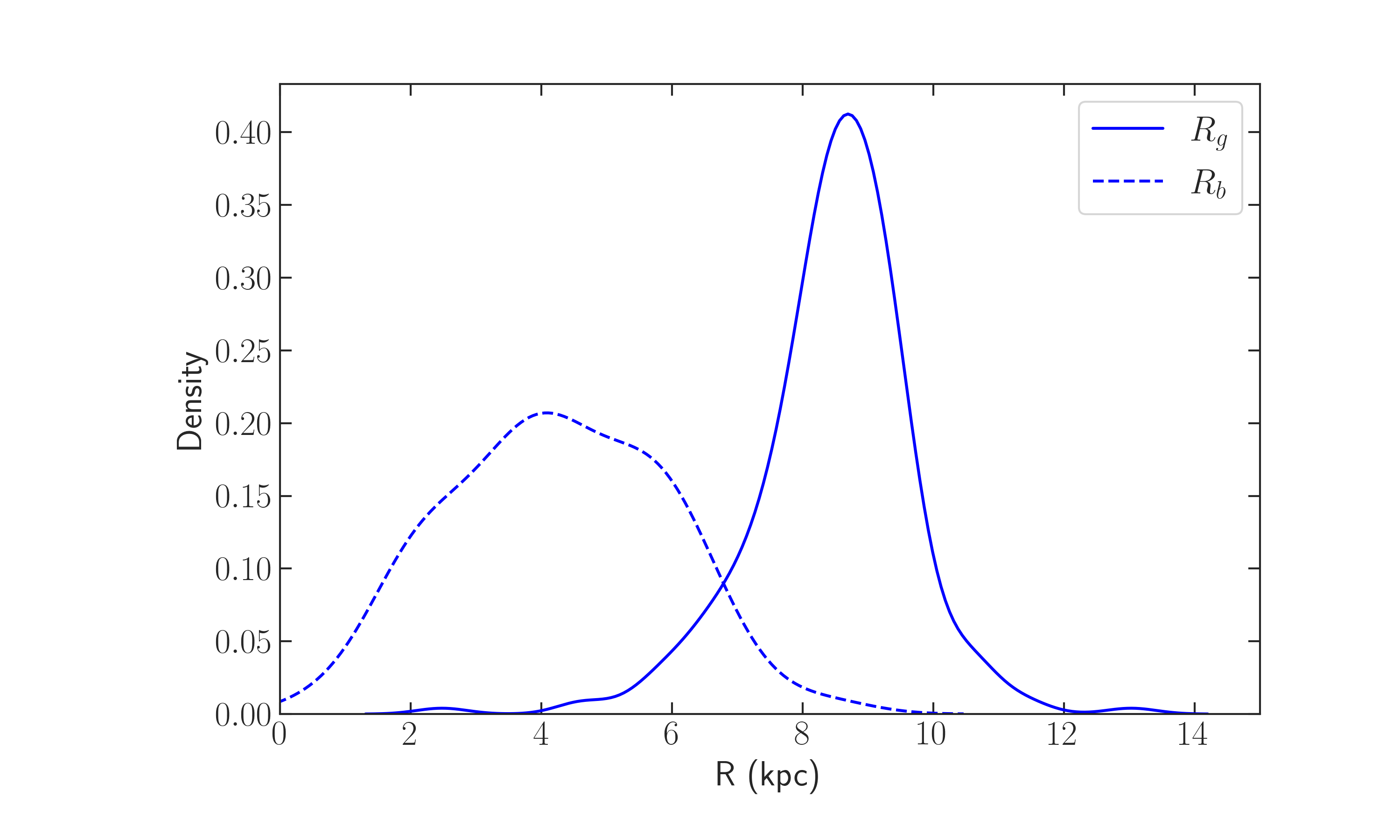}
\caption{Spatial distribution of the APOGEE super metal-rich stars ($\rm [Fe/H]>0.2$) in \rl and \rb (computed using the chemical age from [Ce/Mg]). \label{fig:rl_rb_omr}}
\end{figure}

The spatial distribution of the old super metal-rich stars peaks at the solar neighbourhood in $R_{g}$, whereas their birth radii \rb (using the \citet{lu24}'s method and the chemical ages from [Ce/Mg]) show their origin from the inner disc (see Fig.~\ref{fig:rl_rb_omr}).  
Indeed, they are too metal-rich to be a result of the star formation history of the solar vicinity, and cannot be explained by chemical evolution models which predict a maximum metallicity of $\rm [Fe/H] \sim 0.2$ dex in the solar neighbourhood. 

\subsection{Old metal-poor low-$\alpha$ stars}


Old metal-poor low-$\alpha$ stars trace early accretion events in our Galaxy such as Gaia-Enceladus \citep{helmi18,belokurov18} and Sagittarius \citep{ibata01,limberg23} among others \citep[][for a review]{helmi20}. These stars are characterised by their low metallicity and low [$\alpha$/Fe] ratios, indicating that they likely formed from gas that had undergone limited enrichment from Type II supernovae, which produce $\alpha$-elements, but possibly significant contributions from Type Ia supernovae, which mainly contribute to iron. Their chemical signatures suggest that they formed in environments that were chemically less enriched, potentially in smaller, accreted satellite galaxies or in isolated pockets of the early Galactic disc. Several studies have used APOGEE data to study possible accretion scenarios, such as \citet{hasselquist21}, \citet{horta23}, \citet{queiroz23} and references therein.

\begin{figure}
\includegraphics[trim={2cm 0cm 0cm 0cm}, clip,scale=0.45]{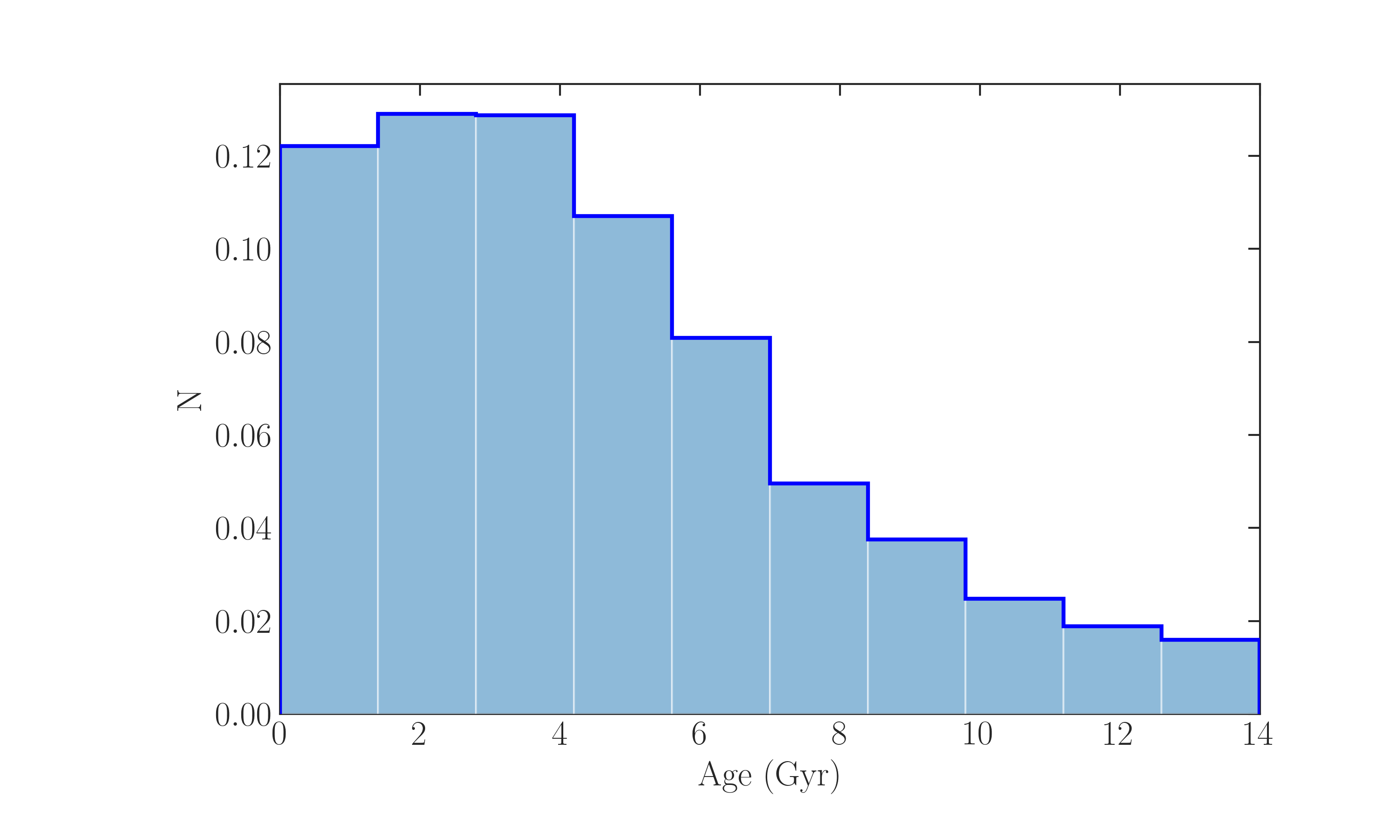}
\caption{Chemical age distribution for the low-$\alpha$ $\rm [Fe/H]<-0.5$ for the APOGEE stars.  N is a probability density as in Fig.~\ref{fig:agedist}. \label{fig:omp_hist}}
\end{figure}

The age distribution of low-$\alpha$ $\rm [Fe/H]<-0.5$ stars in Fig.~\ref{fig:omp_hist} is composed mainly of young stars, but there is a tail towards old ages.  We also identified old, low-$\alpha$ stars in our small sample of \textit{Kepler} stars, with six of them exhibiting ages older than 8 Gyr at sub-solar metallicity.  

\begin{figure}
\includegraphics[trim={2cm 0cm 0cm 0cm}, clip,scale=0.45]{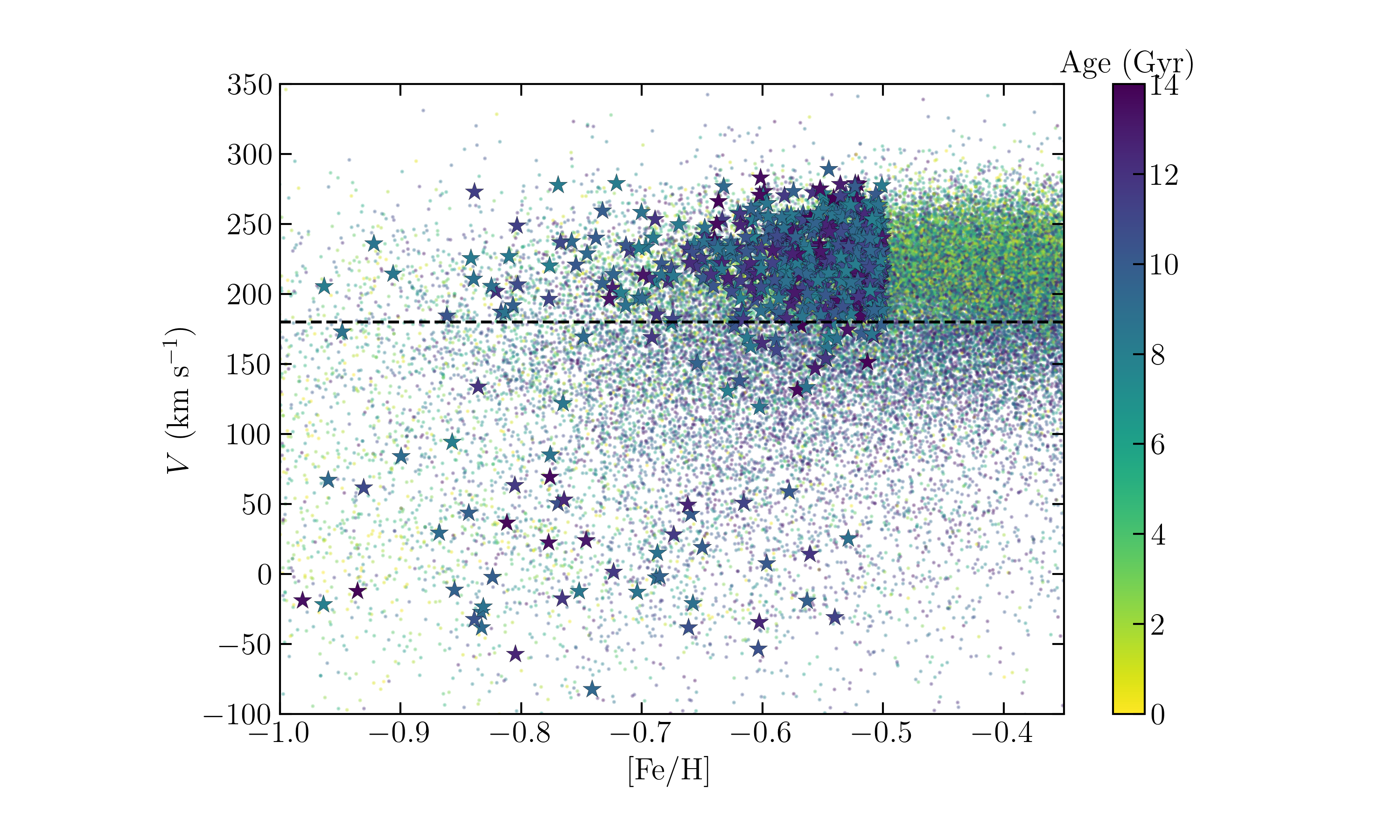}
\caption{$V$ vs. [Fe/H] plane for the APOGEE stars colour-coded by chemical age. The star symbols are the low-$\alpha$ $\rm [Fe/H]<-0.5$ APOGEE stars with chemical age older than 8 Gyr. \label{fig:omp_vfeh}}
\end{figure}

We selected low-$\alpha$ $\rm [Fe/H]<-0.5$ stars from this age distribution with age older than 8 Gyr and we plotted them in the $V$ vs. [Fe/H] plane in Fig.~\ref{fig:omp_vfeh} with star symbols \citep[see discussion in][and references therein]{nepal24}. 
Most of them show a thin-disc kinematics ($V > 180$ km/s), while the others have a thick-disc kinematics ($V < 180$ km/s) or they are in the $V- \rm [Fe/H]$ space designated as "Splash" \citep[\text{$-0.7 < \rm [Fe/H] < 0.2$~dex, $V <100$~km/s}, see][]{belokurov20}. 
The presence of the old low-\alfa~metal-poor stars suggest an early formation scenario for the Milky Way thin disc with respect to to the estimates around 8–9 Gyr, when the beginning of the formation of thin disc is dated.

\subsection{Disc flaring with chemical ages}

One aspect of particular interest is the phenomenon of disc flaring of our Galaxy, where young disc stars show an increasing scale height moving towards larger Galactocentric distances. This feature has been studied by several works: e.g., \citet{miglio13} and \citet{stokholm23}, using seismic data; \citet{ness16} used [C/N] as proxy of age; works that transferred the age labels from \citet{miglio21}, such as \citet{leung23}, \citet{anders23}; and finally also works as \citet{minchev15} and \citet{yu21}, who have explored its implications for the structure and evolution of the Galactic disc. By applying ages from chemical clocks, we aim to contribute to this topic of interest, providing an understanding of how disc flaring relates to the age distribution of stars across different regions of the Milky Way.

\begin{figure}
\hspace*{-1cm}
\includegraphics[scale=0.45]{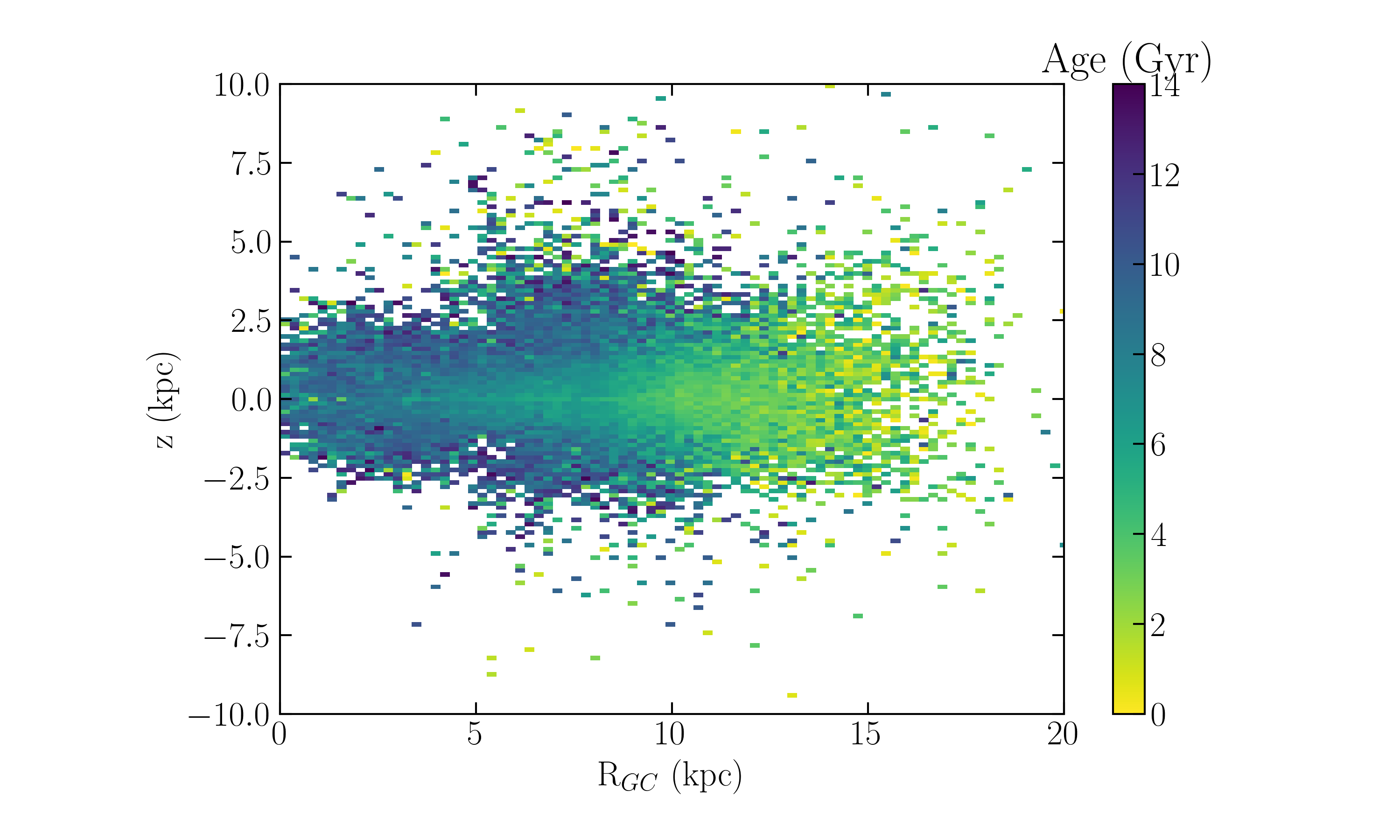}
\caption{$z$ vs. $R_{GC}$ plane for the APOGEE stars with $\rm [Fe/H] > -1$. The distribution is colour-coded following the chemical ages computed using the [Ce/Mg]-[Fe/H]-age relations. \label{fig:flaring}}
\end{figure}

Figure~\ref{fig:flaring} illustrates the $R_{GC}$ vs. $z$ plane (with $z$ height on the Galactic plane), where the distribution of the stars is colour-coded with the chemical ages, as indicated by the [Ce/Mg] ratio. The values for $R_{GC}$ and $z$ are obtained using the {\sc GalPy} package of Python as in Sec.~\ref{sec:spatial}.
The figure clearly reveals the disc flaring of the Milky Way. 
It shows that stars in a narrow $|z|$ range near the Galactic plane are predominantly young, with this range broadening as one moves toward the outer regions. In contrast, older stars are found at smaller $R_{GC}$ values and are typically located farther from the plane than their younger counterparts.

The presence of the youngest stars, particularly around $z \sim 0$, suggests an ongoing star formation in the gas-enriched regions of the Galaxy. Their extension to larger radii further supports an inside-out formation scenario for the Milky Way \citep{matteucci89,chiappini01,Grisoni2018}, where star formation initially occurred in the central regions and progressively expanded outward. This inside-out growth model is consistent with the idea that the Milky Way’s disc evolved over time, with older stars remaining closer to the Galactic centre while younger stars formed in the more distant regions as the disc expanded. Through this analysis, we aim to enhance our understanding of how stellar populations and the structural evolution of the Milky Way are interconnected.




\section{Conclusions}
\label{sec:conclusions}
In this work, we analysed high-resolution spectra of a sample of 68 RGB \kepler stars, collected with HARPS-N@TNG and FIES@NOT spectrographs. We obtained precise estimates of their atmospheric parameters ($T_{\rm eff}$, $\log g$, [Fe/H] and $\xi$) and abundances of 25 elements. 
We confirmed the stellar parameters of these 68 stars found by APOGEE DR17 with optical spectra.
We derived their ages through individual radial-mode frequencies, reaching an unprecedented precision ($\sim 8\%$ for this sample). 

We used this sample of stars as calibrators for chemical clock relations. 
We studied mainly two chemical clocks: [Ce/Mg] and [Zr/Ti]. They show the smallest scatter related to stellar age among all abundance ratios explored in this work, maximising the correlation with time. 
We investigated the relations chemical clock-[Fe/H]-age in different bins of guiding radius \rl and birth radius $R_{b}$, finding a difference less than 1~$\sigma$ among the parameters obtained for the inner, solar and outer regions for both radii. 

The advantages of using these stars as calibrators is due to their high precision in spectroscopy and stellar age. We tested the accuracy and precision of the chemical ages derived by chemical clock-[Fe/H]-age relations, using in input both our precise abundances and the APOGEE ones. 
The relationships derived using the APOGEE abundances exhibit greater intrinsic scatter ($\sim 0.15$ dex) compared to those based on the abundances measured in this work ($\sim 0.08$ dex). As a result, these relationships yield chemical ages that differ more significantly from the asteroseismic ages, which are used as input, than when the abundances from this study are used.
The result is even worst 
if we use in input the APOGEE abundances and less precise ages such as those derived from \numax and $\Delta \nu$. However, the less precise abundances affect more the relations than the less precise ages.
This work demonstrates how high-precision spectroscopy and high-precision stellar age 
are crucial to obtain relations precise enough to be applied to infer ages of field stars in large spectroscopic surveys. The precision present in large spectroscopy surveys such as APOGEE is not sufficient to use their abundances to obtain these empirical relations.
Indeed, having chemical clock-[Fe/H]-age relations with an intrinsic scatter less than 0.1 dex and abundances with uncertainties less than 0.08 dex, as in our case, is fundamental to achieve the precision necessary to estimate chemical ages for a huge sample of stars.  This level of precision is necessary if we want to have relative error in stellar ages lower than 3 Gyr across the entire Galactic chronochemical history as we showed in the Sec.~\ref{sec:testage} and Fig.~\ref{fig:confrontoeta}.

Finally, we applied these relations to field stars present in the large spectroscopic surveys APOGEE and \emph{Gaia}-ESO. Our aim was to validate this age-dating method reproducing key features of Galactic archaeology.
The ages derived from chemical clocks show good agreement with results already present in the literature, such as the dichotomy in age between the high- and low-$\alpha$ sequence, age-metallicity relation, old metal-rich stars, old low-\alfa~stars, and disc flaring, where more precise methods have been employed. While our estimates may be less precise than, e.g., asteroseismology, they still provide a valuable contribution, particularly by increasing the statistical sample, along with other works in literature that use machine learning techniques, such as \citet{anders23} with XGBoost, or \citet{leung23} with astroNN. 
Moreover, these methods are especially useful when other dating techniques cannot be applied or are less effective. Therefore, the ages obtained through chemical clocks represent a valuable tool for deepening our understanding of the chemical and dynamical evolution of the Galaxy.
This immediate application confirms the potential power of chemical clocks to improve our knowledge of stellar ages.



\section*{Acknowledgements}
We thank Dr Alessio Mucciarelli, Dr Valentina D'Orazi and Dr Marica Valentini for helpful discussions during the preparation of this paper.
GC, AM, JS, MM, KB, VG, MT, AS, EW acknowledge
support from the European Research Council Consolidator Grant funding scheme (project ASTEROCHRONOMETRY, G.A. n. 772293, \url{http://www.asterochronometry.eu}. 
LM and GC acknowledge support from INAF with the Grant "Checs, (CHEmical ClockS) Seeking a theoretical foundation for the use of chemical clocks".
 LM  thanks INAF for the support (Large Grant 2023 EPOCH) and the financial support under the National Recovery and Resilience Plan (NRRP), Mission 4, Component 2, Investment 1.1, Call for tender No. 104 published on 2.2.2022 by the Italian Ministry of University and Research (MUR), funded by the European Union – NextGenerationEU– Project ‘Cosmic POT’  Grant Assignment Decree No. 2022X4TM3H  by the Italian Ministry of Ministry of University and Research (MUR).
 AS acknowledges the support from the European Research Council (ERC) under the European Union’s Horizon 2020 research and innovation programme (CartographY; grant agreement ID 804752).
 VG acknowledges financial support from INAF under the program “Giovani Astrofisiche ed Astrofisici di Eccellenza - IAF: Astrophysics Fellowships in Italy" (Project: GalacticA, "Galactic Archaeology: reconstructing the history of the Galaxy") and INAF Mini Grant 2023. 
This article is based on observations made in the Observatorios de Canarias del IAC with the Telescopio Nazionale Galileo operated on the island of La Palma by INAF in the Observatorio del Roque de los Muchachos. This article is also based on observations made with the Nordic Optical Telescope, owned in collaboration by the University of Turku and Aarhus University, and operated jointly by Aarhus University, the University of Turku and the University of Oslo, representing Denmark, Finland and Norway, the University of Iceland and Stockholm University at the Observatorio del Roque de los Muchachos, La Palma, Spain, of the Instituto de Astrofisica de Canarias.

\section*{Data Availability}
Relevant data underlying this work is available in the article. All other data will be shared upon reasonable request to the corresponding author.



\bibliographystyle{mnras}
\bibliography{Bibliography} 


\bsp	
\label{lastpage}
\end{document}